\documentclass[twocolumn]{aastex63}
\usepackage{amsmath}
\usepackage{dsfont}

\submitjournal{ApJ}

\graphicspath{{./}{figures/}}

\begin{document} 

\title{Temporally and Spatially Extended Star Formation in the \\ Brightest Cluster Galaxy of MACS\,J0329.7$-$0211 at $z=0.45$: \\ Implications for Stellar Growth}

\correspondingauthor{Juno Li}
\email{junolee@connect.hku.hk}

\author[0000-0002-8184-5229]{Juno Li}
\affiliation{Physics Department, the University of Hong Kong, Pokfulam Road, Hong Kong}

\author[0000-0002-5899-3936]{Leo W.H. Fung}
\affiliation{Institute for Advanced Study/Department of Physics, Hong Kong University of Science and Technology, Clear Water Bay, Hong Kong}

\author[0000-0003-4220-2404]{Jeremy Lim}
\affiliation{Physics Department, the University of Hong Kong, Pokfulam Road, Hong Kong}

\author[0000-0001-9490-3582]{Youichi Ohyama}
\affiliation{Institute of Astronomy and Astrophysics, Academia Sinica, 11F of Astronomy-Mathematics Building, No. 1, Sec. 4, Roosevelt Rd., Taipei 10617, Taiwan}

\begin{abstract}
Brightest cluster galaxies (BCGs), particularly those at the centers of cool-core clusters, can exhibit star formation over spatial extents of up to $\gtrsim$100\,kpc at inferred rates of up to $\gtrsim100\rm\,M_\sun\,yr^{-1}$. Is their star formation also extended over time, as might be expected if fuelled by cooling of the surrounding hot intracluster gas -- a residual cooling flow -- as demonstrated hitherto only for the BCG in the Perseus cluster? Here, to infer the formation history of relatively young stars in the BCG of MACS\,J0329.7$-$0211, we fit model single-stellar-populations to the spectral energy distributions (spanning near-UV to near-IR) measured along different sightlines towards its young stellar population. Employing a Markov Chain Monte Carlo method, we show that star formation in this BCG has persisted at a relatively constant rate of $\sim2{\rm\,M_\sun\,yr^{-1}}$ (factors of 10--40 below the rates previously inferred using simpler methods and/or ad hoc assumptions) over the past $\sim$400\,Myr, beyond which any star formation falls below the observational detection threshold. Such persistent star formation from a residual cooling flow can contribute up to $\sim$10\% of the original stellar mass of this BCG if its progenitor was among the most massive red nuggets known at $z\sim$2 having masses of $\sim1\times10^{11}\rm\,M_\sun$, but only a few percent of its overall growth in stellar mass to $\sim8\times10^{11}\rm\,M_\sun$ at $z=0.45$. Although constituting only a minor pathway for the stellar growth of this BCG, persistent star formation from a residual cooling flow can nevertheless contribute significantly to the enormous number of globular clusters found around BCGs in the local Universe.
\end{abstract}



\section{Introduction}\label{sec:intro}
Elliptical galaxies are generally red (dominated by an old stellar population that formed in the early Universe) and dead (no detectable ongoing or recent star formation) -- apart from a small minority that, in most cases, have been temporarily revived by the accretion of cool gas through interactions or mergers with other galaxies thus fueling a brief episode of star formation.  One might therefore expect giant elliptical galaxies at the centers of massive galaxy clusters -- brightest cluster galaxies, henceforth BCGs -- to be especially lifeless, as ram-pressure stripping by the hot intracluster medium removes much of the gas in cluster member galaxies that move at high speeds on predominantly radial orbits \citep[e.g.,][]{Jaffe2015} by the time dynamical friction slows these galaxies down sufficiently to be captured by the BCG.
 
Indeed, it has been argued since \citet{White1976} and \citet{Ostriker1977} that BCGs grew over time in stellar mass and, especially, stellar size primarily through dry mergers with cluster member galaxies (i.e., those lacking cool gas).  Such mergers are especially effective in growing galaxies in stellar size, as stars from the more strongly disrupted companion (or, if they have comparable masses and sizes, both disrupted galaxies) are deposited primarily at the outskirts of the BCG (or merger remnant).  In wet mergers, by comparison, cool gas accreted from the disrupted companion sinks dissipatively into the center of the BCG (or merger remnant) to fuel star formation; although this cool gas is eventually dispersed by stellar winds, radiation pressure, and supernova explosions associated with the newly-formed stars, along possibly by actions associated with an Active Galactic Nucleus (AGN) fueled by the cool gas, the contraction of the BCG (or merger remnant) in response to its central change in gravitational potential (owing to the newly-formed stars) partially mitigates its overall increase in stellar size owing to the accretion of stars from its disrupted companion.  Further underpinning this picture, BCGs have since been postulated to descend from red nuggets, the oldest and among the most massive galaxies discovered (by \citealt{Daddi2005} and \citealt{Trujillo2006}) at high redshifts ($z \sim 2$).  These galaxies have large stellar masses of $\sim$$10^{10}$--$10^{11} {\rm \, M_\sun}$, but remarkably small sizes with effective radii of just $\lesssim 1$\,kpc.  They must therefore grow perhaps just modestly in mass (by a factor of a few) but spectacularly in size (by about an order of magnitude) to resemble present-day BCGs, which in the local Universe have stellar masses spanning $\sim$$10^{11}$--$10^{12} {\rm \, M_\sun}$ (typically several $10^{11} \, M_{\sun}$) and effective radii $\gtrsim 5$\,kpc.  The manner by which this growth actually occurs continues to pose an especially big challenge to our understanding of galaxy assembly, and constitutes a key test of galaxy formation models.

In contrast to the expectations laid out above, an increasing number of BCGs at relatively low redshifts -- mostly $z \lesssim 0.1$ where searches have been concentrated \citep{Edge2001,Salome2003}, but recently up to $z \sim 0.6$ (\citealt{Castignani2020}, and references therein) and now even up to $z \sim 1.2$ \citep{Dunne2021} -- have been found to contain large quantities of molecular gas as traced in CO, the reservoir for star formation.  The inferred masses span $\sim$$10^9$--$10^{11} {\rm \, M_\sun}$; to place such large gas masses into context, the upper end of this range in gas mass closely approaches the typical present-day stellar masses of BCGs.  In addition, by contrast with theoretical models \citep{DeLucia2007, Laporte2013} that invoke very little if any star formation following an initial gas-rich dissipational collapse onto massive dark matter halos fuelling star formation at $z \gtrsim 2$ \citep{Oser2010} thereby giving rise (presumably) to red nuggets, a significant fraction of BCGs at $z \lesssim 1$ have been found to exhibit ongoing or recent star formation with inferred rates as high as $\gtrsim 100 {\rm \, M_\sun \ yr^{-1}}$ (e.g., \citealt{Donahue2015,McDonald2016}).  Appreciable star formation in BCGs is far from exceptional: between one-third and one-half of BCGs in massive X-ray-selected clusters at $z \lesssim 0.1$ exhibit luminous optical emission-line nebulae, for which their line strengths in H$\alpha$ are correlated (albeit with substantial scatter) with UV emission from newly-formed stars (\citealt{Donahue2010}, and references therein).  Even if star formation at rates of $\sim$$100 {\rm \, M_\sun \ yr^{-1}}$ persist for just 100\,Myr (or lower rates over commensurately longer time intervals) to consume $\sim$$10^{10} {\rm \, M_\sun}$ of molecular gas, the newly-formed stars would add $\sim$10\%--100\% to the original stellar masses of red nuggets.  As a fiducial comparison, theoretical models \citep[e.g.,][]{Laporte2013} postulate, and observations (\citealt{Lidman2012}; see also discussion and references in \citealt{Lidman2013}) find, that the stellar masses of BCGs have nearly doubled (i.e., increased by nearly 100\%) since $z \sim 1$ (over the past $\sim$8\,Gyr).    

Does in-situ star formation since $z \lesssim 2$ therefore play a significant role in the stellar growth of  BCGs since they were red nuggets?  Can such star formation contribute to, or at least not too severely mitigate by deepening the central gravitational potential of the burgeoning galaxy, the overall growth in their stellar sizes so as to become present-day BCGs? To address these questions, we need to better understand the star-formation history of BCGs -- both temporally and spatially -- since $z \lesssim 2$.

A first step towards addressing the nature of star formation in BCGs is to address the origin of their molecular gas, the reservoir for star formation.  A strong relationship has long been established between BCGs that contain cool gas as traced in optical emission lines and the physical properties of their surrounding hot intracluster gas that emits in X-rays.  At low to moderate redshifts ($z \lesssim 0.5$), such BCGs are found exclusively in cool-core clusters\footnote{A designation introduced by \citet{Molendi2001} for galaxy clusters that exhibit a temperature decrement in their intracluster X-ray emitting gas at the cluster core.  The X-ray gas permeating such clusters exhibits a surface brightness that is strongly centrally peaked, indicating prodigious radiative loss at the cluster core.  The resulting loss in pressure support was predicted to result in an inflow of cooling intracluster gas, referred to as an X-ray cooling flow \citep{Cowie1977,Fabian1977}.  \citet{Molendi2001} found, as demonstrated more robustly by \citet{David2001} and \citet{Peterson2003}, that the mass-deposition rate from any such flow is much lower than had previously been inferred.  Today, it is widely recognized that X-ray cooling flows are strongly quenched by AGNs in BCGs \citep[e.g., review by][]{Fabian2012}.   At low redshifts, cool-core clusters constitute the majority ($\gtrsim 70\%$) of X-ray selected samples of galaxy clusters (e.g., \citealt{Edge2002}; \citealt{Edge1992,Hudson2010}), and $\sim$30\% of essentially unbiased samples of galaxy clusters \citep{Eckert2011,McDonald2013}}, specifically those in which the entropy of the hot intracluster gas at the cluster core lies below a well-defined threshold \citep{Cavagnolo2008}; the same threshold separates BCGs that exhibit star formation from those that do not \citep{Rafferty2008}.  This relationship implies that the cool gas in BCGs must originate from cooling of the hot intracluster gas \citep{Pulido2018}, albeit counteracted in large part by re-heating owing to the actions of their AGN jets \citep[e.g., review by][]{Fabian2012}.  The manner by which a residual cooling nevertheless occurs is not fully understood, although likely to involve a complex interplay between AGN jets and the surrounding intracluster gas \citep[e.g.,][]{Qiu2020} rather than a simple inflow of cooling quiescent gas.  Despite these uncertainties, the common entropy threshold found for the onset of cool gas and star formation implies that the residual -- but still large -- amounts of cooled gas (nearly) always fuels star formation.  

The deposition of cool gas in BCGs owing to a residual cooling of their surrounding intracluster gas -- hereafter a residual cooling flow, no matter how this cooling actually occurs -- differs from any cool gas accreted through wet mergers with cluster member galaxies in several important ways: (i) the mass of cool gas in BCGs can be, as appears to be observed, substantially higher than what would be expected from mergers with cluster member galaxies that have had much of their gas removed by ram-pressure stripping; (ii) the spatial distribution of cool gas can be highly extended owing to an interplay with the AGN jets, rather than dissipatively accumulating at the centers of BCGs as would be expected in wet mergers; and (iii) the continuous deposition (albeit perhaps at a time-varying rate) and therefore replenishment of cool gas can sustain star formation at a high rate over an indefinite period, by contrast with wet mergers whereby gas is accreted in a single episode to spark a brief period of star formation.  Consistent with these expectations, the optical emission-line nebulae of BCGs can, and indeed quite often, extend over several 10\,kpc if not over 100\,kpc \citep[e.g.,][]{Lynds1970,Conselice2001,Tremblay2015}, as do their molecular gas as traced in CO \citep{Salome2011,McNamara2014,Russell2014,Russell2016,Vantyghem2016,Russell2017a,Russell2017b} and star formation as traced in UV emission \citep{Tremblay2015,Donahue2015}.  In the best studied example, NGC\,1275 at the center of the Perseus cluster, its filamentary optical emission-line nebula spans $\sim$140\,kpc \citep{Lynds1970,Conselice2001} compared with an effective (optical) radius for this galaxy of $\sim$25\,kpc \citep{Smith1990}.  The nebula is multi-phase, containing not just atomic/ionic gas seen in optical emission lines, but also ionized gas seen in X-rays, and a counterpart in molecular gas seen in the rotational-vibrational line of H$_2$ in the IR \citep{Lim2012} and as traced in CO \citep{Salome2006,Lim2008,Salome2008,Ho2009,Salome2011}.  NGC\,1275 is the only star-forming BCG sufficiently close for individual star clusters to be observable (and even marginally resolved): this galaxy has formed thousands of star clusters having masses ranging from $\sim$$10^4 {\rm \, M_\sun}$, imposed by the observational detection threshold, up to $\sim$$10^6 {\rm \, M_\sun}$, thus spanning the same mass range as globular clusters, at an essentially constant rate of $\sim0.1 {\rm \, M_\sun \ yr^{-1}}$ over at least the past 1\,Gyr, beyond which the relatively young star clusters cannot be distinguished from the even more numerous globular clusters in this galaxy \citep{Lim2020}.  Although the bulk of the molecular gas is located at a projected radius of $\lesssim 10$\,kpc from the center of NGC\,1275, the majority of the newly-formed star clusters are located much further out in close association with the outer gas filaments.  Beginning their free-fall from far beyond the strong tidal fields at the inner region of the BCG, the more massive star clusters ($\gtrsim 10^5 {\rm \, M_\sun}$) are likely to survive a Hubble time, whereas the more numerous star clusters having lower masses are more easily tidally disrupted to help grow the galaxy in, especially, stellar size (see a more detailed discussion in \citealt{Lim2020}, as well as earlier arguments by \citealt{Conselice2001}).

\begin{figure*}[ht]
\centering
\includegraphics[width=18cm]{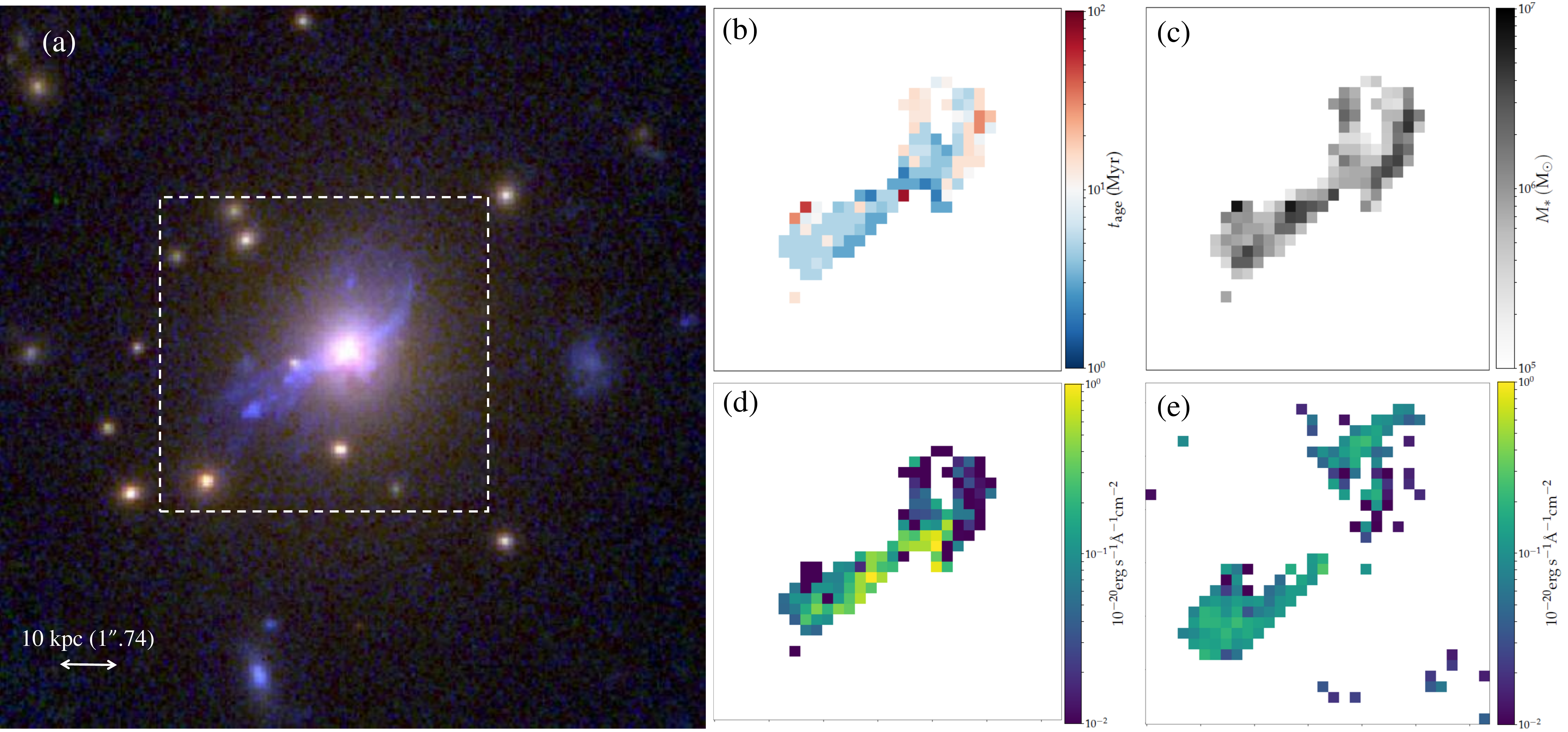}
\caption{(a) RGB image of the central region of the cluster MACS\,J0329.7$-$0211 composed from images in F435W (for B), F775W (for G) and F105W (for R).  A young and spatially complex (blue), as well as an old and spheroidal (red), stellar population is apparent in the BCG.  The dashed box encompasses the region over which we determined the physical properties of the young stellar population in the BCG, and includes cluster members as a sanity check of our methodology.  The remaining panels have the same size as this box.   Nominal (b) ages and (c) total stellar masses at birth inferred using a MCMC approach (Section\,\ref{MCMC}) for the young stellar population in the BCG by adopting $Z = 0.4\,\rm Z_{\odot}$, the approximate metallicity of the intracluster gas.  Our model predicted (d) maximal H$\alpha$+[NII] emission from HII regions associated with the young stellar population, and (e) minimal H$\alpha$+[NII] emission from a line-emitting nebula not excited by the young stellar population (Section\,\ref{Nebula}).}
\label{fig:intro}
\end{figure*}

Despite the numerous examples of spatially extended star formation as cited above, NGC\,1275 was -- until the work described here -- the only BCG definitively demonstrated to show not just spatially but also temporally extended star formation: implying that star formation in BCGs can be both spatially widespread and sustained indefinitely by residual cooling flows.  Nonetheless, there remain concerns that NGC\,1275 may be exceptional and therefore its gas properties and star formation extraordinary.  This perception stems in large part to the spectacular appearance of its optical emission-line nebula, which in the past has made NGC\,1275 unusual, but is owed in part to no more than its relative proximity (at $z=0.01756$) making its parent cluster the X-ray brightest cluster in the sky.  Here, we make use of the Cluster Lensing And Supernova survey with Hubble (CLASH) program \citep{Postman2012}, a Hubble Space Telescope (HST) treasury program in which 25 massive galaxy clusters were imaged in 16 filters spanning UV to IR wavelengths, to study the star formation of the BCG in MACS\,J0329.7$-$0211.  At $z = 0.45$, this BCG is seen at an epoch dating back to about two-thirds the present age of the Universe.  From the CLASH images of this galaxy, \citet{Donahue2015} found that its UV emission has a complex morphology, as shown in Figure\,\ref{fig:intro}, with a maximal projected spatial extent of $\sim$30\,kpc.  \citet{Castignani2020} report a mass in molecular hydrogen gas for this BCG of $3.4 \pm 1.0 \times 10^{10} \rm \, M_\sun$ as traced in CO.  In our work, we extract the spectral energy distributions (SEDs) of the young stellar population in this galaxy along all spatially-separated sightlines to reconstruct its recent star-formation history.  The smaller panels in Figure\,\ref{fig:intro} provide a preview of our results, showing the ages of the young stellar population and their total stellar masses at birth along different sightlines (upper row), as well as the maximal possible H$\alpha +$[NII] emission from HII regions and the minimal H$\alpha +$[NII] emission from a nebula not associated with HII regions (lower row).

In Section \ref{sec:method}, we explain how we first removed, from the images in each filter, the old stellar population in the BCG so as to isolate the young stellar population along with the emission-line nebula in this galaxy.  We present images of the young stellar population and the emission-line nebula combined in Section \ref{SED young stellar pop}, and explain how we extracted their SEDs along different sightlines for model fitting.  Then, in Section \ref{sec:analysis}, we describe how we fit model SEDs generated from a stellar population synthesis code to the measured SEDs of the young stellar population along each sightline, taking care to avoid wavelengths in the measured SEDs containing a significant contribution from the emission-line nebula (which, as we show, is largely not associated with HII regions) in this galaxy.  We present the results for the ages and masses of the young stellar population along each sightline, and compute the star-formation history -- star-formation rate as a function of time -- of this population in Section \ref{sec:results}.  In this section, we also present the morphology of the line-emission nebula, and examine the extent to which its emission may correspond to HII regions.   In Section \ref{sec:discussion}, we examine and correct for selection biases -- to the degree possible -- in the inferred star-formation history.
We then consider the implications of the temporally and spatially extended star formation on the growth of this BCG.  Finally, in Section \ref{sec:conclusions}, we provide a summary of the most important points of this paper.  We adopt a distance to MACS\,J0329.7$-$0211 of 2497.6\,Mpc based on a standard flat $\Lambda CDM$ cosmology with $\Omega_m=0.3$, $\Omega_{\Lambda}=0.7$, and $H_0 = 70 \rm \, km \, s^{-1} \, Mpc^{-1}$, so that $1\arcsec = 5.76\,\rm kpc$.


\section{Removing old stellar population}\label{sec:method}
\begin{figure*}[htb]
\centering
\includegraphics[width=\linewidth]{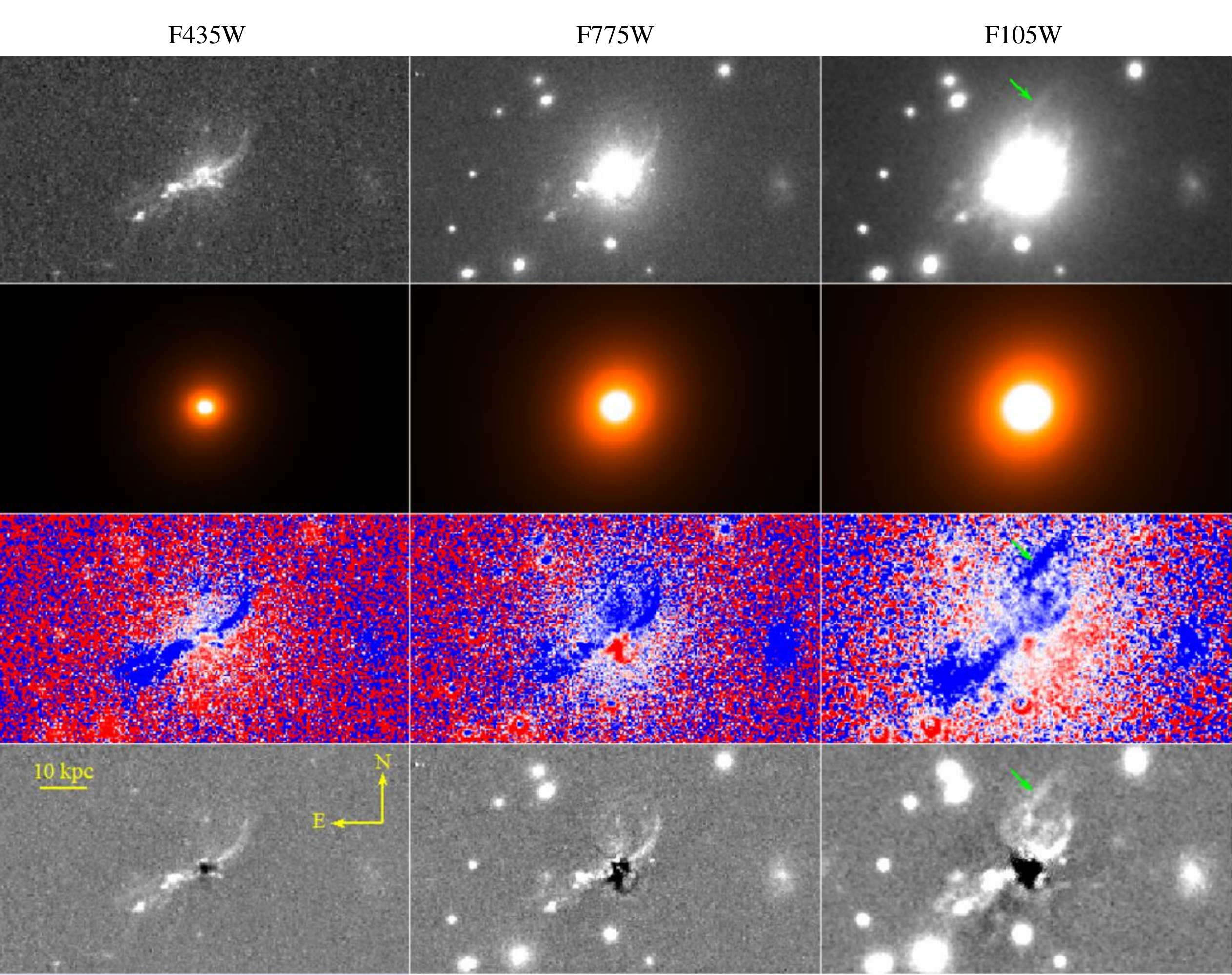}
\caption{Images in or involving the F435W (first column), F775W (middle column), and F105W (right column) filters.  First row are the original CLASH images extracted from the archive.  Second row are our best-fit 2-dimensional models for the projected light distribution of old stellar population.  Third row are color images corresponding to the ratio in intensity between the corresponding filter and the longest wavelength filter, F160W in the near-IR, such that blue corresponds to a higher intensity ratio (i.e., bluer colors) than red.  Fourth row are images derived by subtracting images in the second row from those in the first row, thereby removing the old stellar population.  Green arrows point to H$\alpha$+[NII] emitting gas not spatially coincident with any detectable young stellar population in F435W or F775W (nor in the remaining filters not encompassing the H$\alpha$+[NII] lines).}
\label{fig:ratios} 
\end{figure*}

We downloaded images having a pixel size of 65\,mas (those having a smaller pixel size of 30\,mas sometime show artefacts owing to the manner in which they were reconstructed from the dithered images) from the CLASH archive.  Figure\,\ref{fig:ratios} (first row) shows cropped images centered on the BCG in three representative filters -- rest-frame UV (F435W), optical (F775W), and near-IR (F105W) -- revealing: (i) a spatially complex feature extending from the south-east to the north-west with a projected longer dimension of $\sim$6\arcsec ($\sim$30\,kpc) that dominates the light at ultraviolet to short optical wavelengths (F225W to F475W filters); and (ii) a spatially simple and nearly circularly-symmetric feature that increasingly contributes and eventually dominates the light at longer wavelengths (e.g., F105W filter).  As might be supposed and as we shall demonstrate, the former is produced by a young stellar population and the latter by an old stellar population.  Their different spatial distributions permits model fits to the projected 2-dimensional (2-D) light distribution of the old stellar population after masking out the light from the young stellar population (and other features as described below).  Subtracting the fitted light profiles of the old stellar population leaves only light from the young stellar population (subject to any internal dust extinction, along with a contribution from an emission-line nebula), thus permitting the spectral energy distributions (SEDs) of just this population to be isolated and modelled.  In the remainder of this section, we describe the procedure used to carefully derive the 2-D light distribution of the old stellar population.

\subsection{Masking}\label{masking}
Owing to their simple morphologies, analytical functions can provide an excellent description of the two-dimensional (2-D) light distribution of the old stellar population in elliptical galaxies.  Before we are able to fit analytical models to the 2-D light distribution of the old stellar population in the BCG of MACS\,J0329.7$-$0211, however, we first have to mask out sources unrelated to this population.  Such sources include the young stellar population, dust visible in silhouette, a gaseous emission-line nebula, along with neighbouring galaxies. 

To aid in identifying and/or clarifying the spatial extents of the aforementioned sources, we constructed color images by dividing the image in a given filter with that in the longest-wavelength filter (F160W).  Example color images involving a filter in the rest-frame UV (F435W), optical (F775W), and near-IR (F105W) are shown in Figure\,\ref{fig:ratios} (third row); note that the F105W filter spans the H$\alpha$+[NII] line at the redshift of the BCG.  As can be seen, the light extending from the south-east to the north-west across the center of the BCG is relatively blue, characteristic of a young stellar population, whereas the nearly circularly-symmetric light in the BCG is relatively red, characteristic of an old stellar population.  Silhouette dust is apparent as a reddish patch and curved filament just south of the BCG center, especially in the (F775W$-$F160W) image.  Finally, a prominent blue filament north-north-west of the BCG center (indicated by an arrow) can be seen in the (F105W$-$F160W) color image but not the (F435W$-$F160W) nor (F775W$-$F160W) color images; this feature corresponds to line emission from a gaseous nebula, which is commonly found in BCGs at the centers of cool-core clusters.  Such emission-line nebulae are not always spatially coincident with young stars \citep[e.g.,][]{Canning2010,Canning2014,Lim2020}, if any are indeed present.  Because features such as the emission-line nebula appear only in a restricted number of filters, we did not impose an intensity threshold for computing the color images; as a consequence, regions with neighbouring pixels that very randomly between blue and red correspond to noise.

\subsubsection{Young stellar population}
The young stellar population dominates if not produces all the light of the BCG in the UV filters (F225W to F390W).  Thus, to mask out the young stellar population in these filters, we simply masked all pixels with intensities above 3$\sigma_{\rm noise}$, where $\sigma_{\rm noise}$ is the root-mean-square (rms) noise fluctuation of the image in a given filter (Section \ref{error}).  At longer wavelengths, the old stellar population increasingly contributes and eventually dominates the BCG light.  In these filters, we manually masked the young stellar population guided by their spatial distribution both in the UV images and in the color images (i.e., relatively blue regions) constructed at optical and near-IR wavelengths (e.g., third row of Fig.\,\ref{fig:ratios}).  This approach ensures that the entire young stellar population, even if detectable only in the optical and/or near-IR filters, is masked out.

\subsubsection{Neighbouring Galaxies}
We used the source catalog generated by \cite{Postman2012} using SExtractor to mask all sources (galaxies and stars) in the field apart from the BCG.  Although the detectable BCG light is confined to a relatively small area (first row of Fig.\,\ref{fig:ratios}) where the only cataloged sources are galaxies, extensive masking of all cataloged sources is necessary for determining the rms noise of the image in each filter as explained in Section \ref{error}.  In the vicinity of the BCG, we carefully tuned the sizes of the circular or elliptical masks for the cataloged sources in each image to best match their detectable sizes, leaving as much of the old stellar population in the BCG as possible for fitting models to its 2-D light distribution.

\subsubsection{Silhouette dust}
Silhouette dust is visible as a dark filamentary patch in the images shown in Figure\,\ref{fig:ratios} (top row), although its full spatial extent is best revealed as the reddest regions in color images such as those shown in Figure\,\ref{fig:ratios} (third row).  By selecting pixels with an intensity ratio below $1.1$ within a central radius of $\sim$15\,kpc in the (F775W-F160W) color image (Fig.\,\ref{fig:ratios}, third row, middle column), where we find the silhouette dust to have the greatest contrast, we created a common mask for this feature in all the images.

\subsubsection{Emission-line nebula}\label{emission-line nebula}
As mentioned earlier, an emission-line nebula is clearly detected as relatively blue regions away from the young stellar population in color images involving filters that span the H$\alpha+$[NII] lines, such as the (F105W$-$F160W) image shown in Figure\,\ref{fig:ratios} (third row, right column).  To mask the nebula in these filters, we selected relatively blue pixels in the color images having intensity ratios above 1.5 within a radius of $\sim$15\,kpc centered on the BCG.

\subsection{Model 2-D light distribution}\label{model old stellar population}

\begin{figure}[ht]
\gridline{\fig{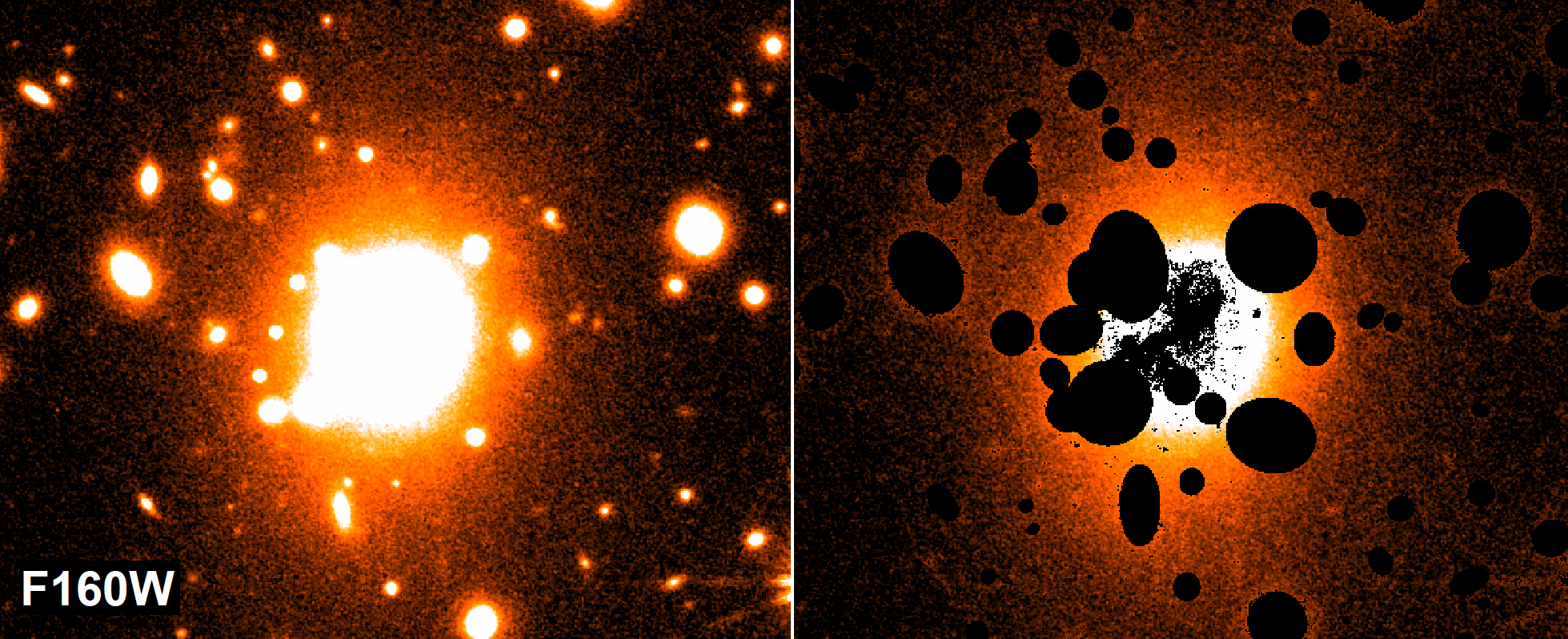}{0.45\textwidth}{}}
\gridline{\fig{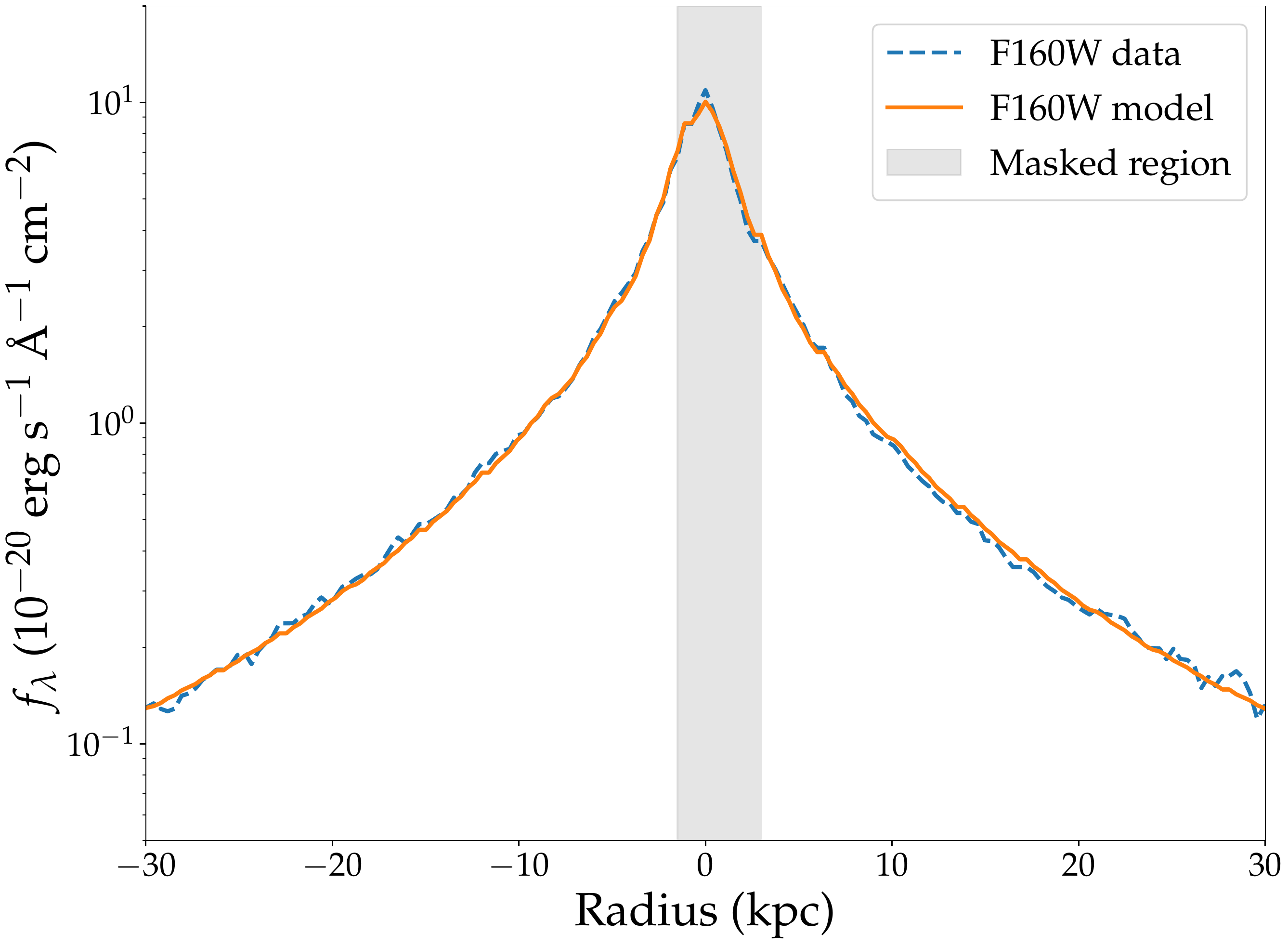}{0.45\textwidth}{}}
\caption{Upper row: Original CLASH image of the central region of MACS\,J0329.7$-$0211 in F160W (left panel) and the same image after masking (right panel) to enable a fit for the light distribution of the old stellar population in the BCG.  Only an area of size $280 \times 220$\,kpc is shown centered on the BCG.   Lower row: Radial light profile of the original image as indicated by the blue dashed curve, and of the best-fit model for the old stellar population as indicated by the red solid curve, at a position angle of $150^\circ$ chosen to best avoid sources unrelated to the old stellar population.  The model is in good agreement with the data except within the masked region, shaded gray, owing to light from a young stellar population and extinction from dust.  
}
\label{fig:image_light_profile}
\end{figure}

An example of the original image and the same image after masking, both in the longest-wavelength filter (F160W), is shown in Figure\,\ref{fig:image_light_profile} (upper panels) over an area of size $180 \times 220$\,kpc centered on the BCG.  To model the remaining light in the masked image produced solely by an old stellar population in the BCG, we fitted 2-D S$\rm \acute{e}$rsic functions to the masked images over a generous area of size $750 \times 750$\,kpc.  We found that a combination of three S$\rm \acute{e}$rsic functions having different power-law slopes ($n$), ellipticities ($e$), position angles ($PA$), and effective radii ($R_{\rm e}$) are required to fully capture the 2-D light distribution of the old stellar population.  The different $e$ and $PA$ of the three S$\rm \acute{e}$rsic functions reflect the triaxial shapes of elliptical galaxies projected onto the plane of the sky, resulting in twists in the major axis and changes in the ellipticity of their light isophotes with radius.  We found that the parameters of the fitted S$\rm \acute{e}$rsic components vary little among the images in the different filters, providing a measure of their robustness.  Because the old stellar population is brightest in the near-IR images, we took the mean of the model parameters for each S$\rm \acute{e}$rsic component in the three longest-wavelength filters (F125W, F140W, and F160W), as listed in Table\,\ref{table:sersic}, to represent the optimal model parameters for the 2-D light distribution of the old stellar population.  Thus fixing the values of these model parameters, we then fitted for the light intensity at $R_{\rm e}$ for each of the three S$\rm \acute{e}$rsic components in the individual images. Reassuringly, the relative intensities of the three fitted S$\rm \acute{e}$rsic components at their respective $R_{\rm e}$ was similar among all the filters.  Examples of the model 2-D light distribution for the old stellar population are shown in Figure\,\ref{fig:ratios} (second row).

\begin{deluxetable}{c c c c}[ht]
\tablecaption{S$\rm \acute{e}$rsic Parameters\label{sersic-parameters}}
\tablehead{\colhead{} & \colhead{Outer S$\rm \acute{e}$rsic} & \colhead{Middle S$\rm \acute{e}$rsic} & \colhead{Inner S$\rm \acute{e}$rsic} }
\startdata
$n$  & 0.788 & 1.816 & 1.004 \\
$e$  & 0.26 & 0.13 & 0.22 \\
$PA$ ($\deg$)  & 179.1 & 149.9 & 80.3 \\
$R_e$ (kpc)  & 55.0 & 11.7 & 2.3 \\
\enddata
\end{deluxetable}\label{table:sersic}

Figure\,\ref{fig:image_light_profile} (lower panel) compares the 1-D radial light profile of the BCG in the F160W filter (in which the old stellar population most dominates the light) along a position angle of $150^{\circ}$ (to best avoid, where possible, sources unrelated to the old stellar population), against that extracted from the model image in the same filter and along the same position angle.  Except at the inner regions where a young stellar population is evident, the radial light profile of the model image closely follows that of the observed image, providing a measure of the accuracy of our model fit.  

\begin{figure}[htbp]
\centering
\fig{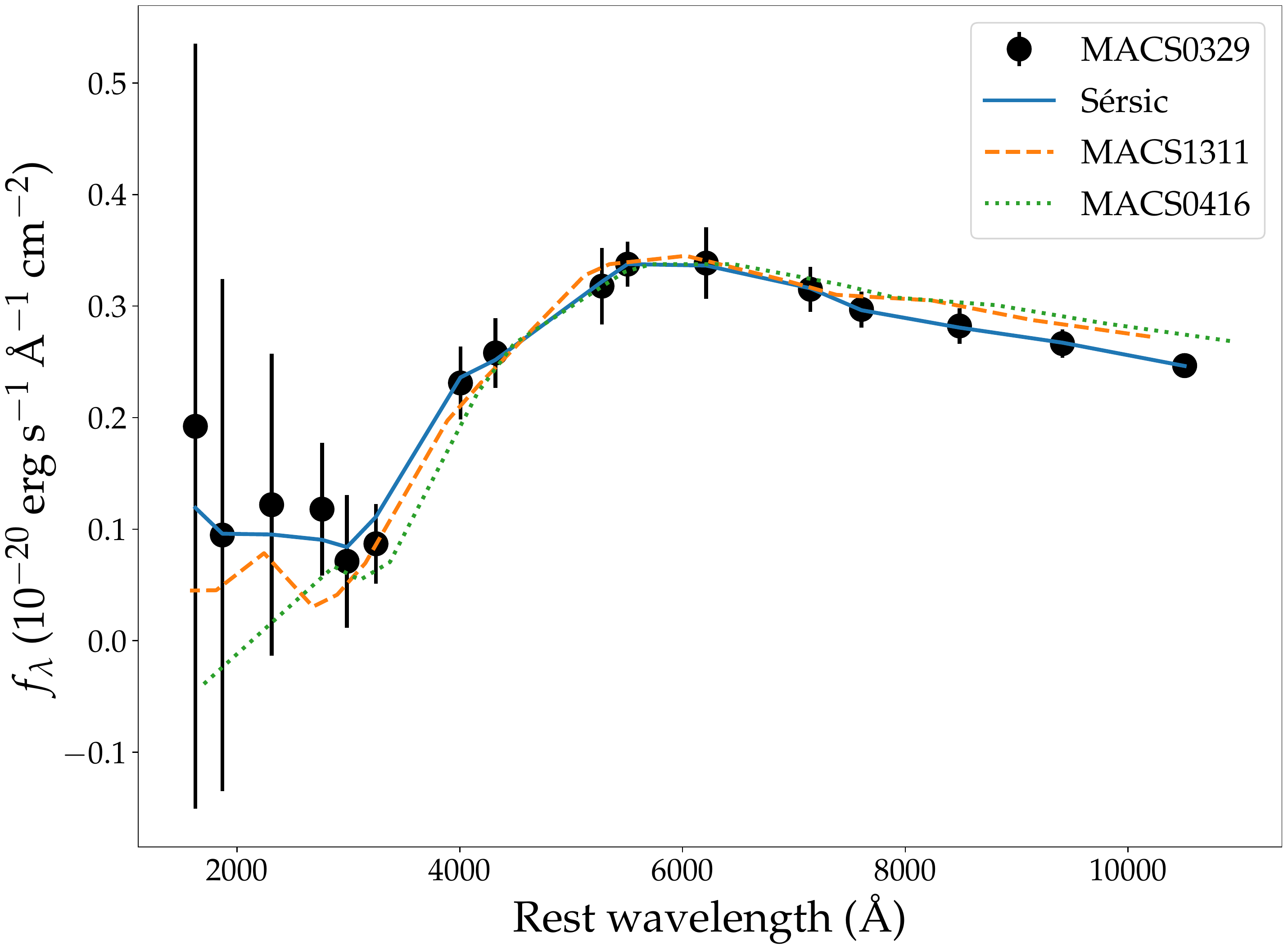}{0.95\linewidth}{}
\caption{
Measured SED indicated by filled circles with $\pm1\sigma_{\rm noise}$ error bars (Section\,\ref{error}) extracted from a region of the BCG in MACS\,J0329.7$-$0211 that avoids all sources (the young stellar population, silhouette dust, and line-emitting gas) apart from the old stellar population.  SED extracted over the same region from our best-fit triple S$\rm \acute{e}$rsic model (Table\,\ref{table:sersic}) for the old stellar population in this BCG is indicated by the blue curve.  For comparison, measured SED of the BCG in MACSJ0416.1$-$2403 at $z=0.395$ indicated by the dotted green curve, and that of the BCG in MACSJ1311.0$-$0311 at $z=0.494$ indicated by the dashed orange curve.  These BCGs, which do not exhibit any detectable young stellar populations, have redshifts straddling that of MACS\,J0329.7$-$0211.
}
\label{plot-SED-model}
\end{figure}

Figure\,\ref{plot-SED-model} compares the SED measured over a region in the BCG that avoids all other sources apart from its old stellar population, and the SED extracted over the same region from our model images for the old stellar population.  Once again there is an excellent match, providing another measure of the accuracy of our model fit.  Finally, the measured SEDs of the BCGs in, respectively, MACSJ0416.1$-$2403 ($z=0.395$) and MACSJ1311.0$-$0311 ($z=0.494$) are shown also in Figure\,\ref{plot-SED-model}, scaled so as to be superposed against the measured SED of the BCG in MACS\,J0329.7$-$0211 for easy comparison.  These galaxy clusters, which were also observed in the CLASH program, have redshifts that straddle the redshift of the galaxy cluster studied here.  The BCGs in these clusters show no detectable evidence for a young stellar population or a gaseous nebula (Levitsky et al., in preparation); nonetheless, we masked the respective centers of these BCGs before extracting their SEDs.  As is apparent, the measured SED of the old stellar population in the BCG of MACS\,J0329.7$-$0211 closely matches the measured SEDs of the BCGs in MACSJ0416.1$-$2403 and MACSJ1311.0$-$0311, indicating that we are genuinely tracing the old stellar population in the target BCG.

\subsection{Subtracted Images}
Figure\,\ref{fig:ratios} (last row) shows examples of the resulting images after subtracting our model images of the old stellar population in the BCG from the observed images.  In the subtracted images, the young stellar population now stands out in clear relief, as does the silhouette dust, the emission-line nebula (in the F105W filter), and of course projected neighbouring galaxies.  A final stage of image processing is necessary before extracting the SEDs of the young stellar population for model fitting, as is explained next.


\section{SED\lowercase{s} of Young Stellar Population}\label{SED young stellar pop}

\subsection{PSF homogenization}
\begin{figure*}[ht]
\fig{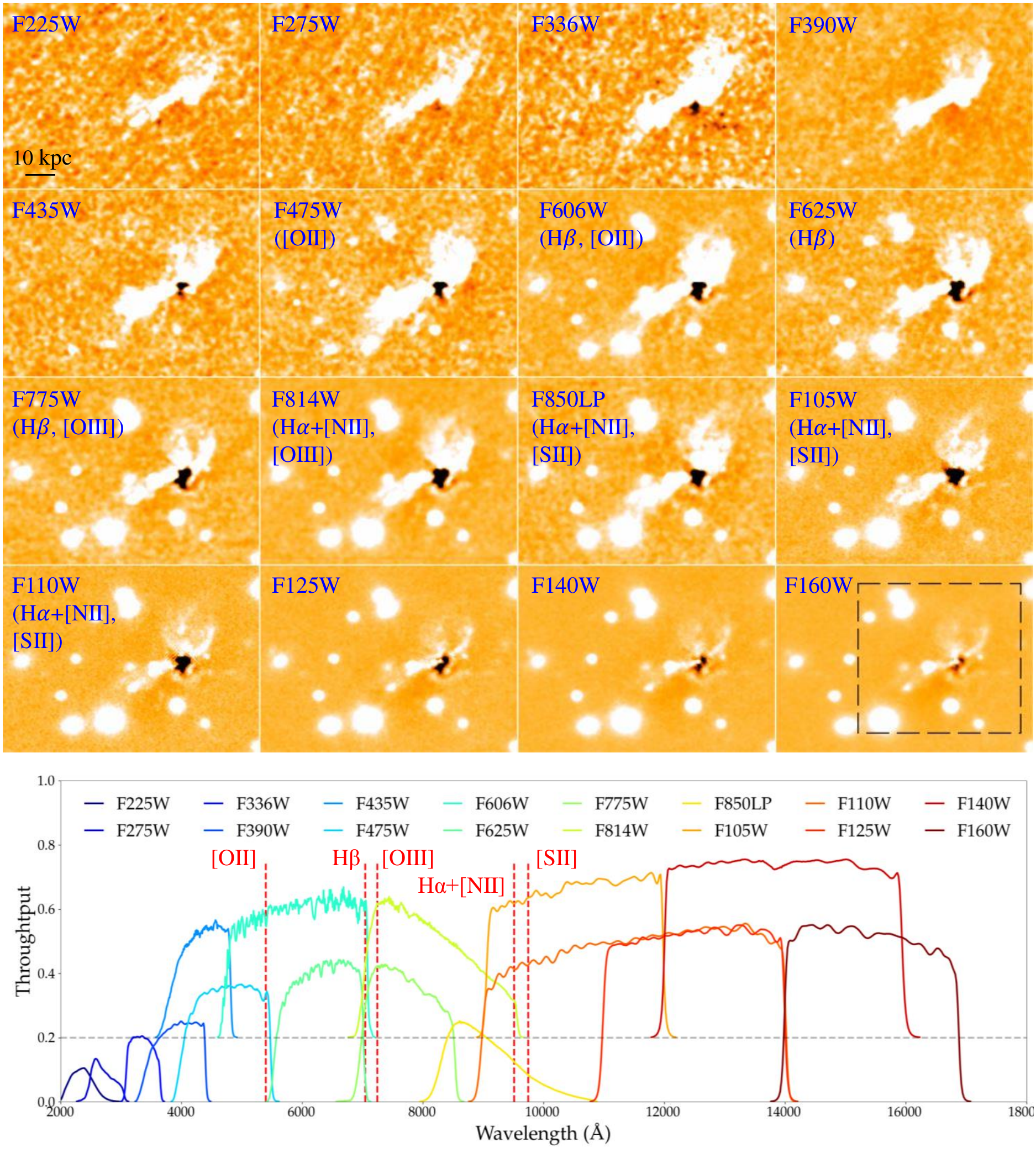}{0.8\linewidth}{}
\caption{Upper panels: subtracted images in all 16 filters of the CLASH program after PSF homogenization.  Filter names are indicated in each panel, along with selected lines in brackets that might be encompassed in each filter.    Dashed box in the F160W panel encloses the region over which we extracted SEDs for modelling.  Bottom panel is the throughput of all the CLASH filters plotted with python package {\it stsynphot} \citep{stsynphot2020}, with wavelengths of the same selected lines at the redshift of the BCG in MACS\,J0329.7$-$0211 indicated by dashed vertical lines. Note some filters are translated up for clarity.}
\label{mosaic-residuals}
\end{figure*}

The images in the different filters shown in Figure\,\ref{fig:ratios} all have different point spread functions (PSFs).  Before extracting the SEDs of the young stellar population, we therefore convolved the subtracted images to a common angular resolution set by the broadest PSF among the different filters.  From measurements of the 2-D radial profiles of field stars visible in each image, we found that the PSFs in all the images are well represented in their cores by a circularly-symmetric Gaussian profile, with the broadest at full-width half-maximum (FWHM) belonging to that of the image in the F110W filter.  We therefore convolved every other image to a FWHM  of $\sim$0\farcs23, corresponding to that in the F110W filter.  The final subtracted images after PSF homogenization are shown in Figure\,\ref{mosaic-residuals}.  Also shown in this figure are the passbands of the individual filters used in the CLASH program, along with where the H$\alpha$, H$\beta$, and [OIII] doublet (which, along with the [NII] doublet closely straddling H$\alpha$, are the brightest emission lines from HII regions over the wavelength range encompassed by these filters) would appear given the redshift of the BCG.  We note that the feature indicated by an arrow in Figure\,\ref{fig:ratios} appears in all filters containing the H$\alpha$ line but not in any of those containing the [OIII] doublet, further indicating that it is not related to HII regions (see Section\,\ref{masking}).

\subsection{Noise level}\label{error}
The rms noise fluctuation cannot be directly measured from the individual images as each image is constructed from multiple dithered images in a given filter.  In the process, the images were sub-sampled so that their pixel sizes (65\,mas) are smaller than the pixel sizes of the camera (either the Wide-field Camera 3, WFC3, or the Advanced Camera for Surveys, ACS) used in the observations.  As a consequence, neighbouring pixels in these images are not statistically independent.

\begin{figure}[ht]
\centering
\fig{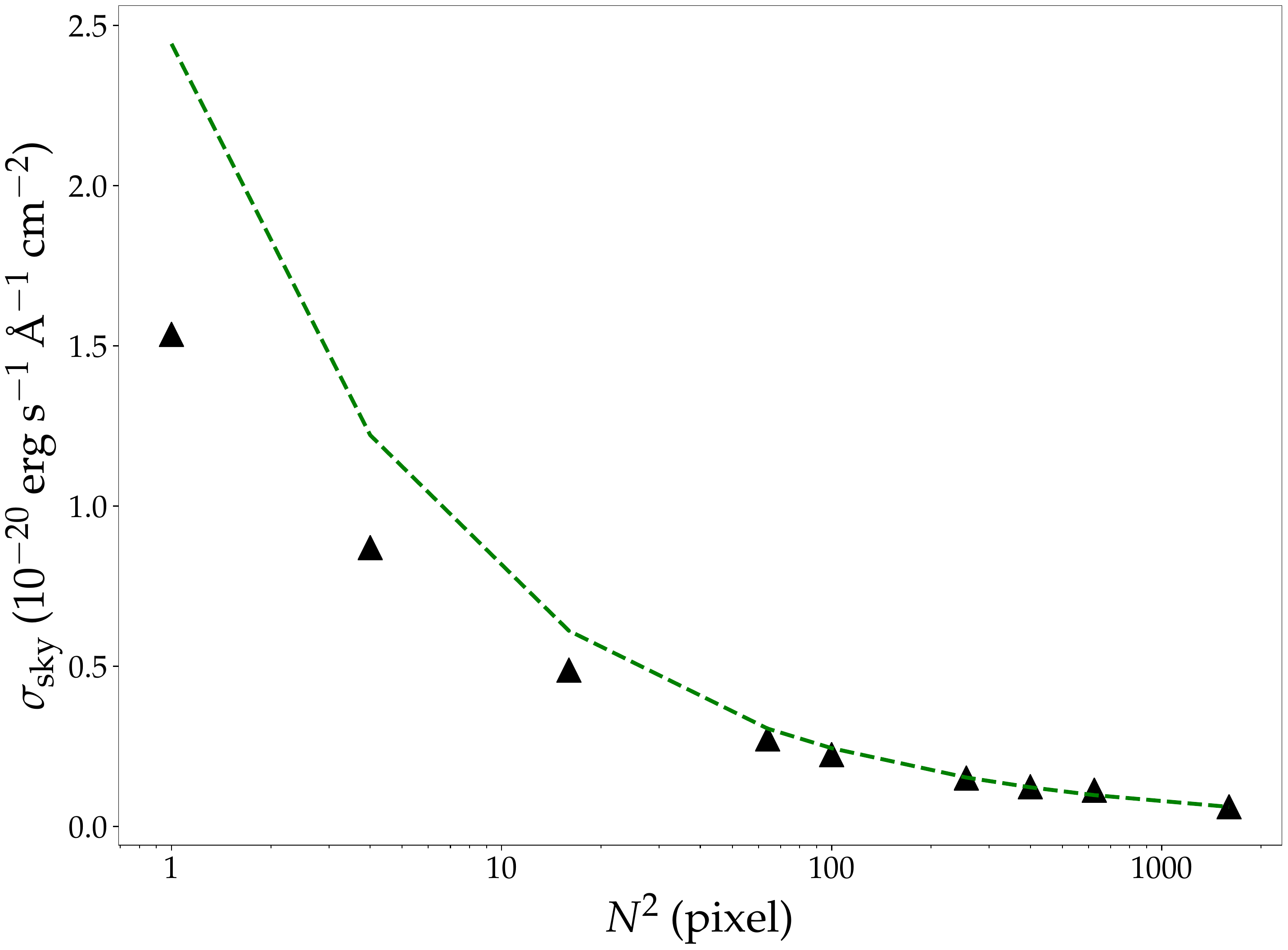}{0.95\linewidth}{}
\caption{Measured sky noise, $\sigma_{\rm sky}$, in the F225W image indicated by triangles, computed from images masked of all sources and binned over $N$ pixels on a side.  An extrapolation of $\sigma_{\rm sky}$ from $N^2 \ge 256$ to smaller $N^2$ based on a $1/\sqrt{N^2}$ relation is indicated by a dashed line, demonstrating that the sky noise would be underestimated at $N^2 \lesssim 100$ if directly computed from the archival image owing to dithering.  We determine the true sky noise for the measured SEDs, extracted over $4 \times 4$ pixels, from the extrapolation to $N^2 = 4 \times 4$ for the image in each filter.}
\label{fig:gauss-fit-errors}
\end{figure}

The rms noise fluctuation in each pixel, $\sigma_{\rm noise}$, can be separated into Poisson noise from light sources, $\sigma_{\rm source}$, and that contributed by the CCD (read and dark noise) along with the diffuse astronomical sky both lumped together (as they cannot be distinguished in the archival images) into $\sigma_{\rm sky}$.  The latter (i.e., $\sigma_{\rm sky}$) can be measured from apparently blank regions of the archival images in the following manner.  First, we masked out all the cataloged sources (including the BCG) in the original images so as to leave only apparently blank regions of sky, and then convolved the images to the same (common) FWHM for their PSFs like in the final subtracted images.  After that, we binned the images over $N$ pixels on a side so that each rebinned pixel has an area of $N^2$, and for each such rebinned image computed $\sigma_{\rm sky}$ as a function of $N^2$ as shown in Figure\,\ref{fig:gauss-fit-errors} for the image in the F226W filter.   As can be seen, for $N^2 \ge 256$, the rms noise decreases with increasing number of pixels binned as $1/\sqrt{N^2}$, as is expected for a Gaussian noise distribution.  At smaller $N^2$, however, the rms noise departs from this dependence, reflecting the fact that neighbouring pixels (at the number of pixels binned) are not statistically independent.  Thus, to estimate $\sigma_{\rm sky}$, we extrapolated the rms noise measured from images binned to $N^2 \ge 256$ pixels to smaller $N^2$ using a $1/N$ dependence, as is shown by the dashed line in Figure\,\ref{fig:gauss-fit-errors}.  The extrapolated value at $N^2 = 4 \times 4$, the area over which we extracted measured SEDs as described next, is taken as $\sigma_{\rm sky}$.  By contrast with the calculation for $\sigma_{\rm sky}$, the value of $\sigma_{\rm source}$ in a given pixel can simply be directly computed from the count rate in that pixel.  In the SED plots shown below, the measurement uncertainty associated with each filter is therefore $\sigma_{\rm noise} = \sqrt{\sigma_{\rm source}^2+\sigma_{\rm sky}^2}$.

\subsection{Extracted SEDs}\label{sec:extracting SEDs}
From the subtracted images after PSF homogeneization as shown in Figure\,\ref{mosaic-residuals}, we extracted SEDs summed over $4 \times 4$ pixels corresponding therefore to $0\farcs26 \times 0\farcs26$ ($1.5 \times 1.5$\,kpc); i.e., with sides just slightly larger than the common FWHM of the images.  The area over which we extracted SEDs spans $128 \times 128$\,pixels, corresponding to $8\farcs3 \times 8\farcs3$ ($48 \times 48$\,kpc), centered on the BCG as indicated by the dashed box in the last panel of Figure\,\ref{mosaic-residuals}.  This area encloses a number of neighbouring galaxies, which proved useful for providing a sanity check on our model SED fits.  All the SEDs were corrected for Galactic dust extinction by adopting $A_V=0.165$ \citep{SF11} and $R_V=3.1$ in the dust extinction curve of \citet{CCM1989}.


\section{Analyses}\label{sec:analysis}

\subsection{Model Single Stellar Populations}\label{model SSPs}
We started with the hypothesis that the measured SEDs of the young stellar population contained within individual apertures of size $4 \times 4$ pixels can each be characterised by a coeval stellar population sharing the same metallicity; i.e., a single stellar population (SSP).  The task then is to determine which model SED generated for SSPs having different ages and metallicities, and if justified also extinctions, best matches a particular measured SED.  As the measured SEDs are each extracted over a region of size $1.5 \times 1.5$\,kpc, it would not be surprising if a mixture of different SSPs (i.e., separate star clusters), including those that overlap along the sightline, contribute to the individual measured SEDs.  In such cases, the best-fit model SED will either reflect the SSP deemed to dominate a particular measured SED or still poorly reproduce the measurements, or both.

To generate the model SEDs, we used the publicly available software code YGGDRASIL\footnote{\url{https://www.astro.uu.se/~ez/yggdrasil/yggdrasil.html}} \citep{YGGnebula,YGGmodel}, which employs model SSPs from Starburst99 \citep{Leitherer1999,Vazquez2005} based on a Kroupa initial mass function \citep{Kroupa2001,Kroupa2002} over the stellar mass interval $0.1$--$100 \, \rm M_{\odot}$ along with Padova stellar evolutionary tracks.  At sufficiently young ages for the model SSPs ($\lesssim$\,10\,Myr), surrounding gas that is photoionized to form a HII region can contribute line along with continuum emission, the latter from recombination, free-free, and 2-photon emission.  The model brightnesses of such HII regions are parameterised by the fraction of Lyman continuum photons from internal (hot and massive) stars that is absorbed by the surrounding gas, $f_{\rm cov} = 1 - f_{\rm esc}$, where $f_{\rm esc}$ is the fraction that escapes; i.e., $f_{\rm cov} = 0$ corresponding to none, and $f_{\rm cov} = 1$ corresponding to a maximal amount, of the surrounding gas being photoionized.  Because stars that produce the required Lyman continuum photons have largely vanished by an age of $\sim$10\,Myr, we found that the selected value of $f_{\rm cov}$ only strongly affects the model spectra at younger ages.  Specifically, after convolving over the bandpasses of the individual filters used in the CLASH program so as to provide a direct comparison with the measured SEDs, we found that emission lines do not contribute appreciably to the (convolved) model SEDs beyond an age of 5\,Myr.  Nonetheless, up to ages of $\sim$100\,Myr, the intensities in all filters are slightly elevated for $f_{\rm cov} = 1$ compared with $f_{\rm cov} = 0$.  We do not know whether this excess is produced by leftover photoionized gas or, as we suspect, small numerical errors as the reverse can happen at ages beyond 100\,Myr.

We considered two different metallicities, $Z$, for the model SSPs of $Z = \rm Z_{\sun}$, the solar metallicity, and $Z = 0.4 \, \rm Z_{\sun}$, approximately that of the intracluster gas around the BCG \citep{Maughan2008}.  For reasons we shall explain, we also subjected the model spectra to varying amounts of ``internal" dust extinction -- that within the BCG to the foreground of the SSP (recall that the measured SEDs have been corrected for Galactic dust extinction) -- based on the dust extinction curve of \citet{CCM1989}, for which we adopted $R_V=3.1$.  Finally, the template spectra were Doppler shifted to the redshift of the BCG and then convolved with the spectral responses of the filters used in the CLASH program, thus generating model SEDs that can be directly compared with the measured SEDs.  In this way, we generated model SEDs at logarithmically space intervals beginning at an age of 1\,Myr, the youngest permitted by the code, up to an age of 3\,Gyr (far exceeding the oldest age found for the young stellar population; see below), except for a denser sampling between 10\,Myr and 15\,Myr in steps of 1\,Myr.  

\subsection{Color-Color Diagrams}\label{color-color}

Before evaluating which model SEDs best fit the measured SEDs, we found it instructive to construct color-color diagrams from the measured SEDs to compare against evolutionary tracks generated from the model SEDs.   As we shall see, this comparison yielded crucial insights on the overall range of physical parameters -- in particular, ages and (internal) dust extinctions -- spanned by the young stellar population, thus constraining the parameter space over which we needed to explore to find best-fit model SEDs.  Just as importantly, this comparison revealed bright line emission in filters encompassing H$\alpha$+[NII] on sightlines towards the young stellar population where they have inferred ages $\gg 5$\,Myr, by which time line emission from HII regions is predicted to be undetectable.  
An emission-line nebula unrelated to HII regions is therefore present along most if not all sightlines towards the young stellar population, in addition to sighliness away from this population as pointed out earlier in Figure\,\ref{fig:ratios}.  Our model SEDs are not able to account for the contribution from such an emission-line nebula, making it necessary to exclude filters encompassing H$\alpha$+[NII] when fitting model SEDs to the measured SEDs as described later.

\begin{figure*}
    \fig{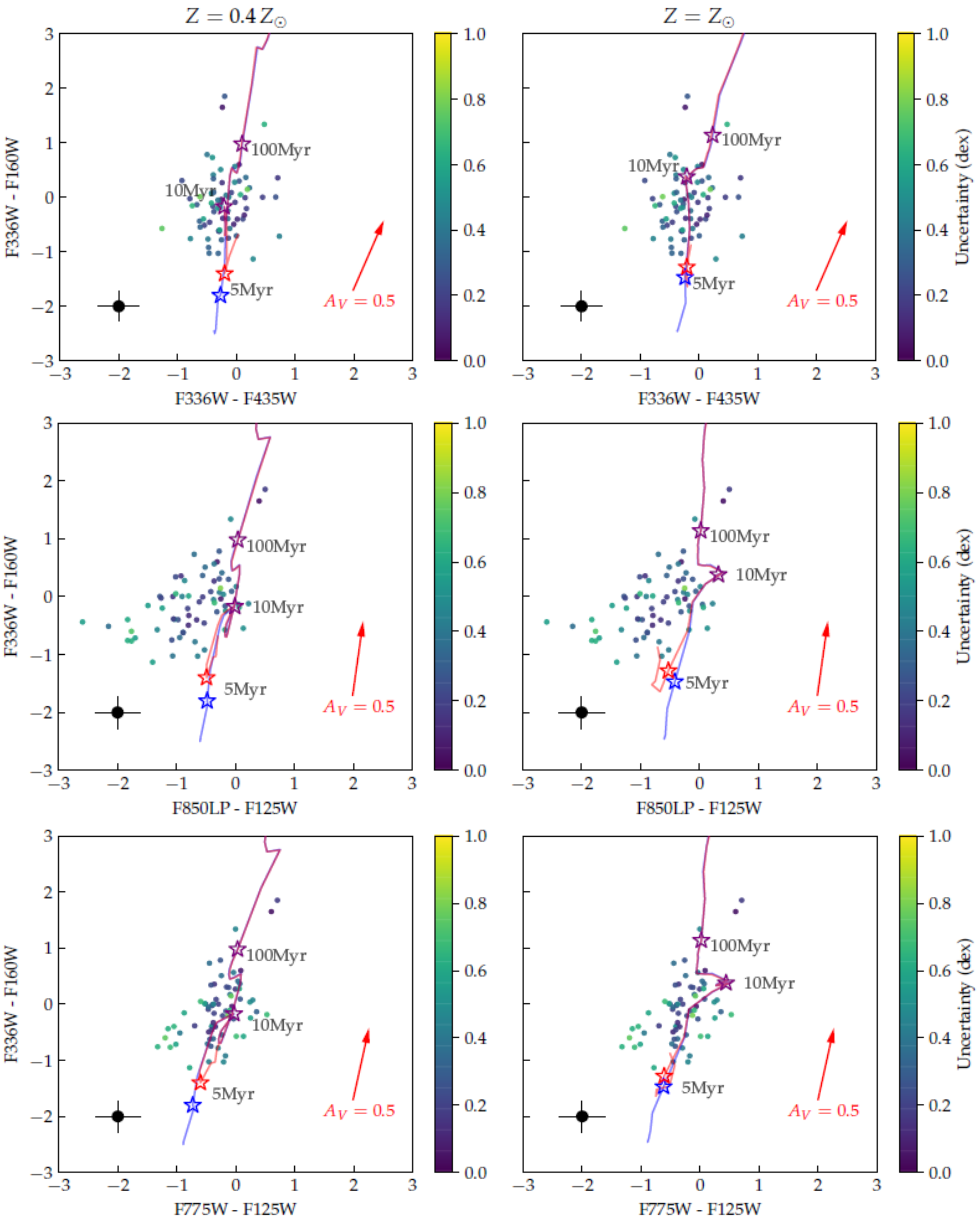}{0.75\textwidth}{}
    \caption{
    Color-color diagrams selected to highlight a relatively broad range of ages (top row), H$\alpha$+[NII] emission in excess of that predicted from HII regions (middle row), and little if any appreciable H$\beta$ and [OIII] emission (bottom row).  Measured SEDs along individual sightlines are indicated by filled circles, with colors corresponding to their measurement uncertainties as indicated in the color bar accompanying each panel.  The median $\pm 1\sigma_{\rm noise}$ uncertainty of the measured SEDs is indicated at the lower left corner of each panel.  Evolutionary tracks are plotted in each panel, with the blue tracks corresponding to $f_{\rm cov} = 0$ (i.e., no HII region) and the red tracks to $f_{\rm cov} = 1$ (i.e., maximal predicted emission from a HII region) for adopted metallicities of $0.4 \rm Z_\sun$ in the left column and $\rm Z_\sun$ in the right column.  The red arrow in each panel indicates the expected color change owing to dust extinction as large as $A_V=0.5$.
    }
    \label{fig:CC}
\end{figure*}

To construct the color-color diagrams presented in Figure\,\ref{fig:CC}, we selected sightlines along which the signal-to-noise ratio (S/N) of the young stellar population exceeds $1\sigma_{\rm noise}$ in each of the 16 filters.  Through trial-and-error, we found this criterion to select a representative range of ages exhibited by the young stellar population while preserving an adequate S/N for their colors.  In each panel of Figure\,\ref{fig:CC}, evolutionary tracks derived from the model SSPs are plotted for metallicities of $Z = 0.4 \, \rm Z_{\sun}$ (left column) and $Z = \rm Z_{\sun}$ (right column), for which we chose the extreme limits of $f_{\rm cov} = 0$ (no HII region; red track) and $f_{\rm cov} = 1$ (maximal emission from a HII region; blue track) to illustrate the maximal possible differences in colors for the evolutionary tracks at a given metallicity.

Figure\,\ref{fig:CC} (top row) shows color-color diagrams selected to best differentiate between age and (internal) dust extinction.  As can be seen, the evolutionary tracks (spanning ages from 1\,Myr to 400\,Myr) are nearly vertical in this color-color diagram, reflecting a rapid change in colors between (rest-frame) UV (F336W) and near-IR (F160W) along the ordinate, but a nearly constant color in UV (F336W and F435W) along the abscissa, as age increases.  By contrast, internal dust extinction displaces the colors of the model SSPs along a more tilted direction as indicated by extinction vectors having a length of $A_V = 0.5$, thus permitting a distinction between age and appreciable dust extinction for a color spread among the young stellar population.  The measured colors of the young stellar population straddle the model evolutionary tracks, with no apparent systematic shift to the right of these tracks as might be expected if extinction is appreciable along many sightlines.  Their overall distribution in color-color space suggests a collection of un-extincted SSPs having a range of ages spanning a few Myr to a few 100\,Myr.  The lack of apparent dust extinction is consistent with that observed for young star clusters over the same age range in NGC\,1275 \citep{Lim2020}, where the vast majority of star clusters lie next to rather than within the emission-line nebula associated with that galaxy.  Note the weak dependence on the colors of the evolutionary tracks with metallicity: as a consequence, our best-fit model SEDs to the measured SEDs do not permit a strong constraint on metallicity (over, at least, the range 0.4--1.0\,$\rm Z_{\sun}$).

Figure\,\ref{fig:CC} (middle row) shows color-color diagrams selected to highlight, along the abscissa, any line emission in the F850LP filter, which encompasses the H$\alpha$ line, [NII] doublet, and [SII] doublet at the redshift of the BCG (see bottom panel of Fig.\,\ref{mosaic-residuals}).  The other filter involved in the abscissa, F125W in the near-IR, does not encompass any bright emission lines that we are aware of.  The ordinate involves the same two filters as before in Figure\,\ref{fig:CC} (first row).   
The extinction vector is nearly parallel to the evolutionary tracks in this color-color diagram, and so the spread in measured colors along the ordinate can in principle reflect either an age spread or spatially-variable extinction (or both); as just demonstrated in Figure\,\ref{fig:CC} (top row), however, any extinction is too low to produce a detectable systematic shift in the color-color diagram, and so the color spread along the ordinate must primarily reflect an age spread.  A large systematic shift to the left leaving relatively few points to the right of the evolutionary tracks is clearly apparent in Figure\,\ref{fig:CC} (middle row), indicating bright emission lines in the F850LP filter even along sightlines where the young stellar population have inferred ages of $\gg 5$\,Myr.  Along many sightlines, the total line emission in the F850LP filter is much stronger than the maximal predicted line emission from HII regions (i.e., for $f_{\rm cov} = 1$).
At this point, we note that such a systematic shift is observed not only for colors involving the F850LP and F125W filters, but also for colors involving either the F814W, F105W, or F110W -- which also encompass the H$\alpha$ line+[NII] lines --  and F125W filters.

To further emphasize that emission lines in the F850LP filter are predominantly unrelated to HII regions, Figure\,\ref{fig:CC} (bottom row) shows color-color diagrams selected to highlight, along the abscissa, any line emission in the F775W filter, which encompasses the [OIII] doublet and H$\beta$ at the redshift of the BCG (see bottom panel of Fig.\,\ref{mosaic-residuals}; notice that H$\beta$ lies at the edge of the passband, where the throughput is only $\sim$50\% that of the [OIII] doublet).  The ordinate involves the same two filters as before.  By contrast with Figure\,\ref{fig:CC} (middle row), there is, at best, a weak systematic shift to the left of the evolutionary tracks.  The contribution by emission lines to the F775W filter is therefore much weaker if at all appreciable by comparison with that to the F850LP filter.  This behavior is inconsistent with expectations for HII regions, whereby emission in the individual [OIII] doublets is typically about as bright as that in H$\alpha$, but consistent with expectations for emission-line nebulae in BCGs, whereby [OIII] is, where at all detectable, much dimmer than H$\alpha$ \citep[e.g.,][]{Hatch2006}.  

\subsection{Model SED Fitting}\label{model sed fitting}

\begin{figure*}[htb]
    \fig{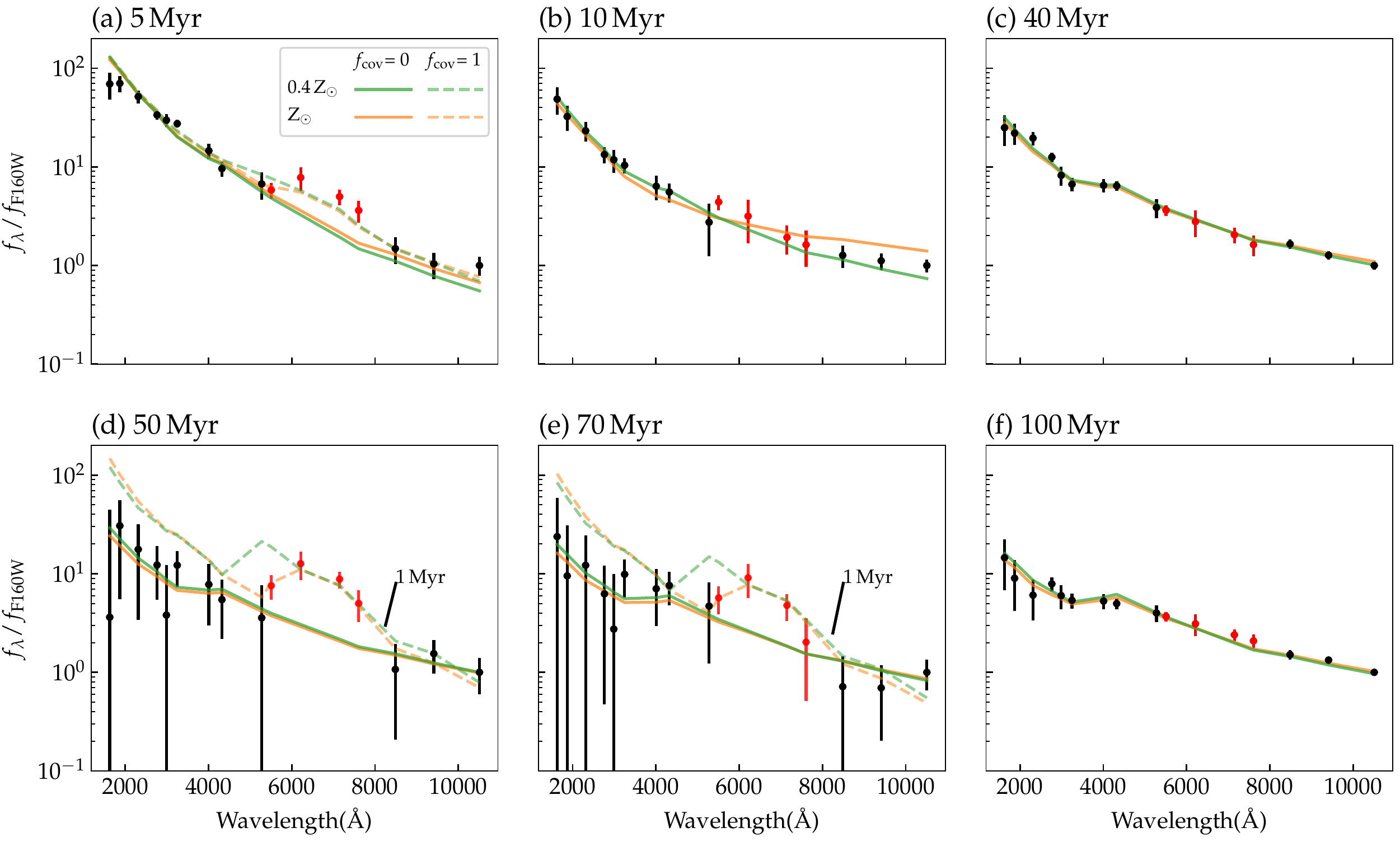}{\linewidth}{}
    \caption{
    Measured SEDs indicated by filled circles selected to highlight a broad range of ages along different sightlines to the young stellar population.  All have $\pm 1\sigma_{\rm noise}$ error bars indicated by vertical bars; those colored red encompass the H$\alpha$+[NII] lines.  Model SEDs having ages as indicated above each panel or for an age of 1\,Myr as indicated are plotted as solid curves for $f_{\rm cov} = 0$ (no HII region) and dashed curves for $f_{\rm cov} = 1$, and in green for $0.4 \, \rm Z_\sun$ and orange for $\rm Z_\sun$.  The dashed curves in panel (a) underpredict the red data points, suggesting that only a portion of the H$\alpha$+[NII] emission along this sightline may be generated by HII regions, wheres none of the same line emission in panels (d) and (e) can be generated by HII regions.}
    \label{fig:ssp-model}
\end{figure*}

The color-color diagrams presented in Figure\,\ref{fig:CC} provide encouragement that each of the measured SEDs, despite encompassing an area of size $1.5 \times 1.5$\,kpc, can be closely reproduced by model SEDs for SSPs.  To assess how well these model SEDs actually fit the measured SEDs, Figure\,\ref{fig:ssp-model} shows six representative examples of measured SEDs fitted by model SEDs having different SSP ages -- corresponding to the approximate ages inferred from the color-color diagram of Figure\,\ref{fig:CC} (top row) -- as indicated at the top of each panel, except for the model SEDs indicated by dashed lines in Figure\,\ref{fig:ssp-model}$d$--$e$ that have ages of only 1\,Myr.  Model SEDs having $Z = 0.4 \, \rm Z_{\sun}$ are plotted in green, and those having $Z = \rm Z_{\sun}$ are plotted in orange.  The solid and dashed lines are for model SEDs having, respectively, $f_{\rm cov} = 0$ and $f_{\rm cov} = 1$.
The SEDs are plotted at the rest frame of the BCG, with each data point centered at the effective wavelength (i.e., weighted by the spectral response) of the corresponding filter after correcting for the Doppler shift of the BCG.  The data points colored red correspond to the four filters (F814W, F850LP, F105W, and F110W) that encompass the (redshifted) H$\alpha$+[NII] lines.

Apart for the model SEDs plotted as dashed lines in Figure\,\ref{fig:ssp-model}$d$--$e$, the remaining model SEDs plotted fit the measured SEDs over all filters (panels $b$, $c$ and $f$), or over all filters except those encompassing the H$\alpha$+[NII] lines as indicated by the red data points (panels $a$, $d$ and $e$), remarkably well -- even though these model SEDs need not necessarily represent the very best fits to the measured SEDs (see Section \ref{chi^2} and Section \ref{MCMC}) as their SSP ages were inferred solely and only approximately from the color-color diagram of Figure\,\ref{fig:CC} (top row).  Over the SSP age range 5--100\,Myr encompassed by this figure, both the measured and model SEDs change with age in two important ways: (i) their slope, especially from UV to optical wavelengths, becomes shallower with age; (ii) a discontinuity in the slope near 4000\,\AA, owing predominantly to the Balmer break at 3646\,\AA\ (rather than absorption by ionized metals that create the 4000\,\AA\ break, a feature that dominates at ages older than the range relevant here), becomes increasingly prominent with age.  
The young stellar population must therefore encompass a relatively broad range of ages, confirming this aspect of the color-color diagrams shown earlier (Fig.\,\ref{fig:CC}, top row).  At a given age, the model SEDs exhibit little change with metallicity, except at an age of 10\,Myr whereby the slope of the SED at near-IR wavelengths is different for the two different metallicities plotted -- owing to the different timescales over which massive stars evolve to become red supergiants as a function of metallicity.  As a consequence, the measured SEDs do not provide strong constraints on metallicity, at least over the range 0.4--$1.0 \rm \, Z_\sun$. 

In Figure\,\ref{fig:ssp-model}$a$, the model SEDs at an age of 5\,Myr formally under-predict the measurements in filters containing H$\alpha +$[NII] (red data points) even for $f_{\rm cov}= 1$ (dashed lines), albeit well reproducing the measurements in the other filters.  Even more glaringly, in Figure\,\ref{fig:ssp-model}$d$--$e$, the model SEDs (solid lines) far under-predict the measurements in filters containing H$\alpha +$[NII] albeit well reproducing the measurements in the other filters.  The measured SEDs shown in Figure\,\ref{fig:ssp-model}$d$--$e$ exhibit a prominent Balmer break, and must therefore have ages far exceeding 10\,Myr by which time any HII regions would have vanished.
The dashed lines in the same panels show model SEDs having $f_{\rm cov}= 1$ and an age of just 1\,Myr (the youngest allowed by the SED code used), thereby giving the maximal possible H$\alpha +$[NII] emission from an associated HII region.   Although these model SEDs can reproduce the measurements in filters containing H$\alpha +$[NII] as well as those at longer wavelengths, they far over-predict the measured brightnesses in filters at short optical and UV wavelengths.  These examples illustrate the contribution by an emission-line nebula unrelated to HII regions to filters encompassing the H$\alpha +$[NII] lines, confirming this aspect of the color-color diagram shown earlier (Fig.\,\ref{fig:CC}, middle row).  As a consequence, the inclusion of these filters complicates if not renders impossible satisfactory fits to the measured SEDs in instances where the emission-line nebula makes a detectable contribution.  Because the emission-line nebula contributes detectably along many sightlines as demonstrated in Figure\,\ref{fig:CC} (middle row), we omitted the four filters encompassing H$\alpha +$[NII] from the model fits as described next.

\subsubsection{$\chi^2$ minimization}\label{chi^2}

As a first step towards evaluating quantitatively how well the model SEDs match the measured SEDs as well as the age range over which individual measured SEDs can be constrained, we computed the reduced $\chi^2$ (henceforth, $\chi_{\rm red}^2$) of the model SEDs fitted to the individual measured SEDs.  As explained earlier, model SEDs were generated over the age range of 1\,Myr to 3\,Gyr for metallicities of either $Z = 0.4 \, \rm Z_\sun$ or $Z = \rm Z_\sun$.  Because we omitted the four filters containing H$\alpha +$[NII] from consideration owing to contamination by an emission-line nebula, we henceforth consider only model SEDs having $f_{\rm cov} = 0$.  Finally, as the color-color diagram of Figure\,\ref{fig:CC} (top row) indicates negligible internal dust extinction, we consider for now only model SEDs having $A_V = 0$.  For each model SED, we computed the optimal normalization in brightness -- thus yielding the total stellar mass at birth (henceforth, birth mass) for the corresponding model SSP -- that provides the best fit to the measured SED as judged by the lowest $\chi_{\rm red}^2$.

\begin{figure*}[htb]
    \fig{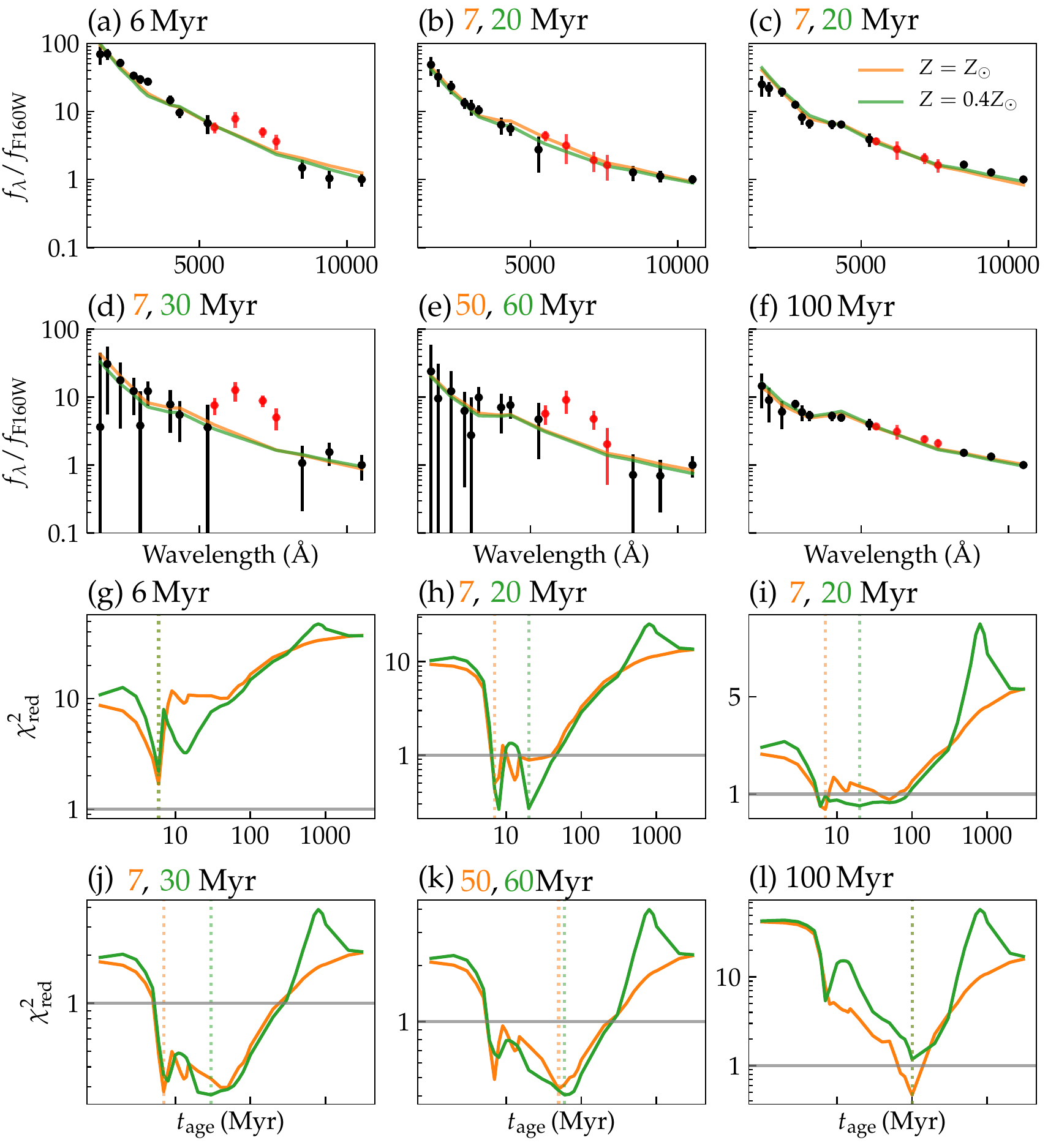}{0.75\linewidth}{}
    \caption{
    Measured SEDs in first and second rows are the same as those in Fig.\,\ref{fig:ssp-model}.  Model SEDs having the same color coding as Fig.\,\ref{fig:ssp-model} are plotted at SSP ages, $t_{\rm age}$, indicated at the top of each panel, corresponding to the lowest $\chi_{\rm red}^2$ among the model SEDs fitted to the individual measured SEDs as indicated by the vertical dotted lines in the $\chi_{\rm red}^2$ versus $t_{\rm age}$ plots in the third and fourth rows.  The measurements in red, which encompass the H$\alpha$+[NII] lines, were excluded when making model SED fits to the measured SEDs for reasons described in the text.}
    \label{fig:chisq-plot}
\end{figure*}

The lowest two rows of Figure\,\ref{fig:chisq-plot} shows example plots of the $\chi_{\rm red}^2$ versus model SSP age, $t_{\rm age}$, of the model SEDs fitted to the measured SEDs shown in the upper two rows of Figure\,\ref{fig:chisq-plot}.  The measured SEDs are the same as those shown in Figure\,\ref{fig:ssp-model}.  Like before, the green lines are for model SEDs having $Z = 0.4 \, \rm Z_\sun$ and the orange lines for those having $Z = \rm Z_\sun$.  Satisfactory fits having $\chi_{\rm red}^2 \le 1$ can be found for all the examples shown except Figure\,\ref{fig:chisq-plot}$a$, where the lowest $\chi_{\rm red}^2$ is between 1 and 3 (depending on the metallicity).
As can be seen, the $\chi_{\rm red}^2$ versus $t_{\rm age}$ can exhibit either: (i) a single narrow minimum, indicating a narrow (in logarithmic scale) range of best-fit ages (e.g., orange in panels $g$ and $l$); (ii) two narrow local minima whereby one is significantly deeper than the other, indicating a relatively narrow range of best-fit ages (e.g., green in panel $a$); (iii) two narrow and comparably deep local minima, indicating two relatively narrow ranges of best-fit ages (e.g., both orange and green in panel $h$); and most commonly (iv) a broad minimum, indicating a relatively broad range of best-fit ages (e.g., green in panel $i$, both orange and green in panels $j$ and $k$).  The model SEDs overlaid on the measured SEDs in the upper two rows of Figure\,\ref{fig:chisq-plot} correspond to the $t_{\rm age}$ at the lowest $\chi_{\rm red}^2$ as indicated by the dotted vertical green and orange lines (which overlap in panels $g$ and $l$).  
Despite the sometimes broad range of ages spanned by the model SEDs that provide satisfactory fits to the measured SEDs, the measured SEDs shown in Figure\,\ref{fig:chisq-plot} clearly span about two decades in ages.

\subsubsection{Markov Chain Monte Carlo}\label{MCMC}

To fully explore and quantify the likelihood of a given model SED in matching an individual measured SED, we used a Markov Chain Monte Carlo (MCMC) method.  For this purpose, we developed an algorithm that yields the likelihood surface of individual model SSP parameters used to generate the suite of model SEDs fitted to an individual measured SED, as described in full in Appendix \ref{app:mcmc}.  The free parameters explored are SSP age, $t_{\rm age}$, over the range 1\,Myr to 1\,Gyr (although model SEDs were generated from 1\,Myr to 3\,Gyr as mentioned in Section\,\ref{model SSPs}, from the work described in Section\,\ref{chi^2} we found that all sightlines towards the young stellar population have very large $\chi_{\rm red}^2$ for ages beyond several hundreds of Myr), and birth mass $M_*$ (corresponding to the sum over the stellar mass interval 0.1--100\,$\rm M_\sun$ with a Kroupa initial mass function).  We set the metallicity at either $Z = 0.4 \rm \, Z_\sun$ or $Z = \rm Z_\sun$, and set $f_{\rm cov} = 0$ as all four filters encompassing H$\alpha +$[NII] are omitted from the fit.  We also set $A_V = 0$ as the color-color diagram indicates negligible (internal) dust extinction, although we later also explored how our results would change if we allow $A_V \leq 0.5$\,mag.  We assumed flat priors (i.e., uniform initial likelihood) for each free parameter, and extracted the maximum a posteriori estimate of each parameter from the peak of the inferred multi-dimensional likelihood surface -- thus deriving parameter values for the most probable model SED (among those considered) that matches a given measured SED.  Owing to discrete sampling of the multi-dimensional likelihood surface, the maximum a posteriori estimate may not coincide exactly with -- but is always close to -- the set of parameter values that give the lowest $\chi_{\rm red}^2$, except possibly in relatively rare cases where the multi-dimensional likelihood surface exhibits two well-separated peaks having comparable heights such that the parameter value having the lowest $\chi_{\rm red}^2$ would be found close to the somewhat lower peak if the sampling was continuous.
We henceforth refer to the maximum a posteriori estimates derived from the smoothed multi-dimensional likelihood surface as the nominal parameter values.

For each free parameter, we compute its one-dimensional likelihood function by marginalizing over all the other free parameters.  The uncertainty, $\widetilde{\sigma}$, of a given free parameter is defined, by analogy with the variance in a Gaussian distribution, as the area under its posterior distribution that encompasses $68.2\%$ (i.e., corresponding to 2$\widetilde{\sigma}$) of the total area over which this distribution has the highest values\footnote{We use $\widetilde{\sigma}$ rather than $\sigma$ as a reminder to the reader that the uncertainty inferred for a given parameter from our MCMC method should not be regarded in the same way as the root-mean-square deviation of a Gaussian distribution.}.  For those not familiar with the MCMC approach, we emphasize here caution in making too close an analogy between the uncertainty as computed using the MCMC approach with the variance as computed for a Gaussian distribution: even if the posterior distribution has a single high and narrow peak but broad wings, the uncertainty can be much broader than the visual width of the peak so as to encompass a broad portion of the wings.  Furthermore, because the posterior distribution need neither be unimodal nor symmetric, the absolute values of $\widetilde{\sigma}$ need not be the same on opposite sides of the peak in the posterior distribution.  If the posterior distribution is multimodal, the range of parameter values encompassed within the uncertainties may include those having very low if not near zero probabilities; e.g., those examples in Figure\,\ref{fig:chisq-plot} where the $\chi_{\rm red}^2$ as a function of age exhibits two local minima having $\chi_{\rm red}^2 \leq 1$ separated by a plateau having $\chi_{\rm red}^2 > 1$, such that the parameter values encompassed within the uncertainties span the entire range between two local minima.  Finally, we note that even the most probable SED does not necessarily guarantee a satisfactory fit to a given measured SED, so that it is important to check the $\chi_{\rm red}^2$ of any fit.

\begin{figure*}[ht]
\centering
\gridline{\fig{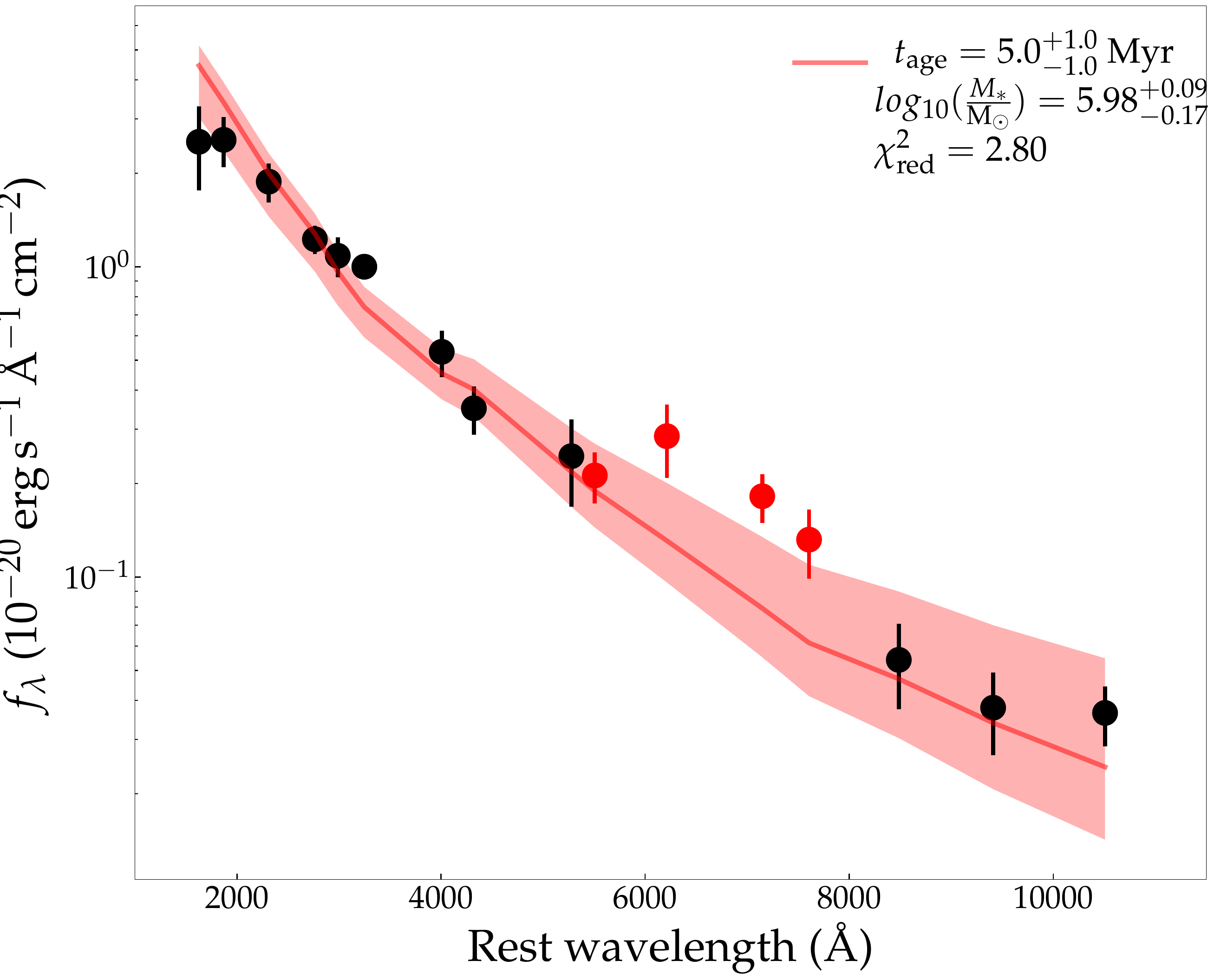}{0.48\textwidth}{}
          \fig{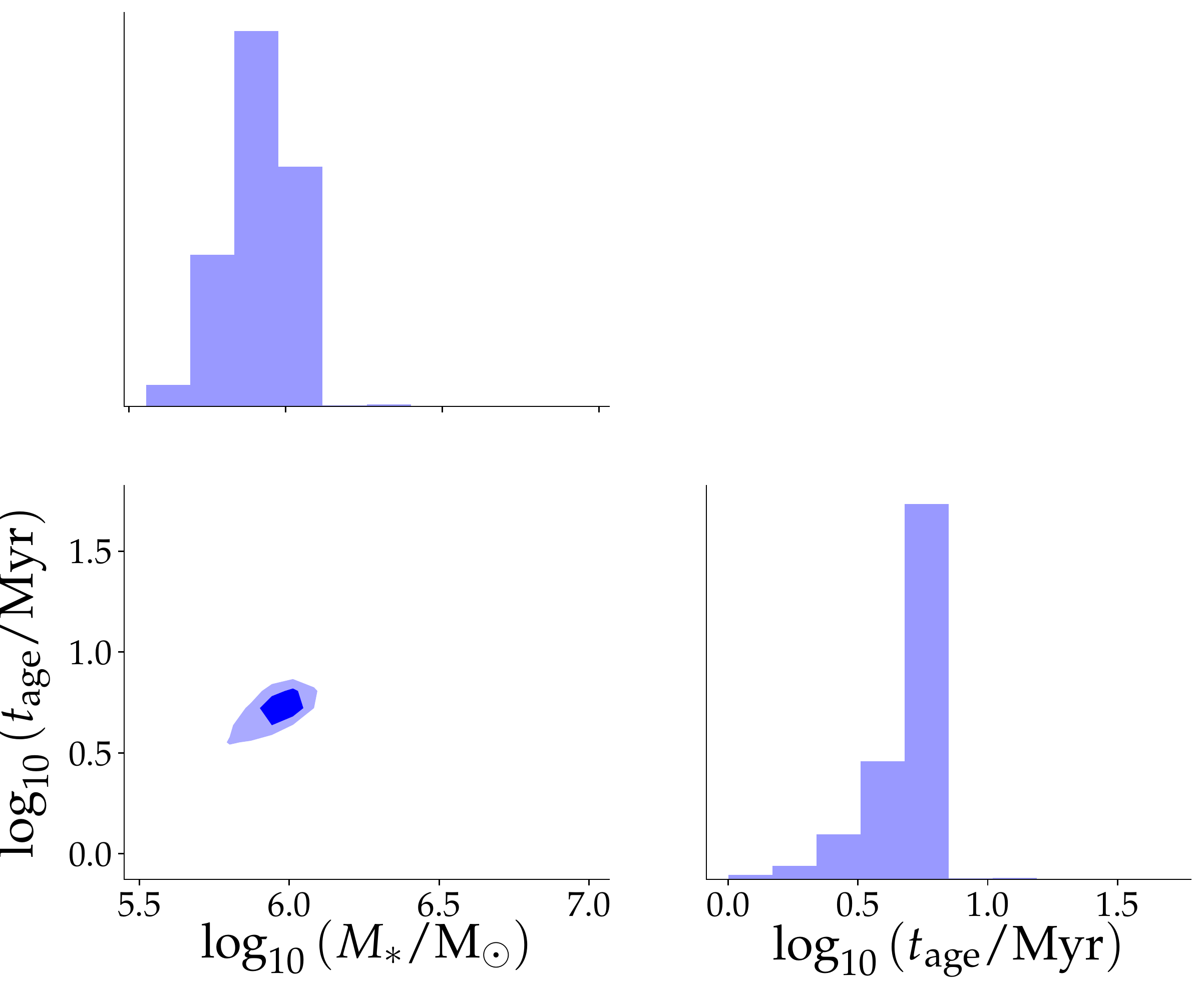}{0.48\textwidth}{}}
\caption{Left panel: Same measured SED as in Fig.\,\ref{fig:ssp-model}(a) and Fig.\,\ref{fig:chisq-plot}(a).  Red curve is best-fit model SED having $Z = \rm Z_\sun$ and nominal parameters in $t_{\rm age}$ and $M_*$ as listed at the upper right corner, and red shaded region the range of model SEDs having $Z = \rm Z_\sun$ encompassed within $\pm 1 \widetilde{\sigma}$ (68\% confidence levels; see text) of the nominal values, inferred based on our MCMC method.  The posterior distribution of the individual parameters, and combined posterior distribution of both parameters (shaded dark blue within $\pm 1 \widetilde{\sigma}$ and light blue within $\pm 2 \widetilde{\sigma}$), are shown in the panels to the right. }
\label{fig:sed-young}
\end{figure*}

To help visualize the output from our MCMC algorithm, in the following we show exemplar results for four of the measured SEDs shown previously in Figures\,\ref{fig:ssp-model}--\ref{fig:chisq-plot}.  Figure\,\ref{fig:sed-young} shows the same measured SED as in Figure\,\ref{fig:ssp-model}$a$ and Figure\,\ref{fig:chisq-plot}$a$.  The model SEDs with $Z = \rm Z_\sun$ provide a reasonable fit ($\chi_{\rm red}^2 \lesssim 2$) to this measured SED over only a very narrow age range of $\sim$2\,Myr centered at $t_{\rm age} \sim 6$\,Myr as shown earlier in Figure\,\ref{fig:chisq-plot}$g$.  In Figure\,\ref{fig:sed-young}, we also plot histograms of the posterior distributions in both $t_{\rm age}$ and $M_*$ as inferred by our MCMC algorithm.  As can be seen, the two posterior distributions are unimodal, although not symmetric.  The contour plots show the range of parameter values bounded at the 68\% (dark blue) and 95\% (light blue) confidence levels.  The red curve superposed on the measured SED is the model SED having the nominal parameter values; i.e., as extracted at the peak of the multi-dimensional likelihood surface.  The nominal parameter values along with their uncertainty bounds thus derived, as well as the $\chi_{\rm red}^2$ of the fit, is listed in the left-most panel.  The red band bounding the measured SEDs indicates the boundary spanned by all the model SEDs combined having ages within the 68\% confidence level of the nominal age (note that different $t_{\rm age}$ have different corresponding $M_*$).  Neither the upper nor lower boundaries of the red band necessarily correspond to a specific model SED, as different SEDs can define either boundaries in different filters.  Notice that the brightnesses in three of the four filters spanning H$\alpha +$[NII] (indicated by red symbols) are in excess of those predicted by the most probable model SED having $t_{\rm age} = 5$\,Myr (with $f_{\rm cov} = 0$); while it is tempting to attribute this excess to line emission from a HII region, as demonstrated above (see Figs.\,\ref{fig:ssp-model}--\ref{fig:chisq-plot}) excess emission in these filters also is seen towards regions that are far too old to exhibit HII regions.

\begin{figure*}[ht]
\centering
\gridline{\fig{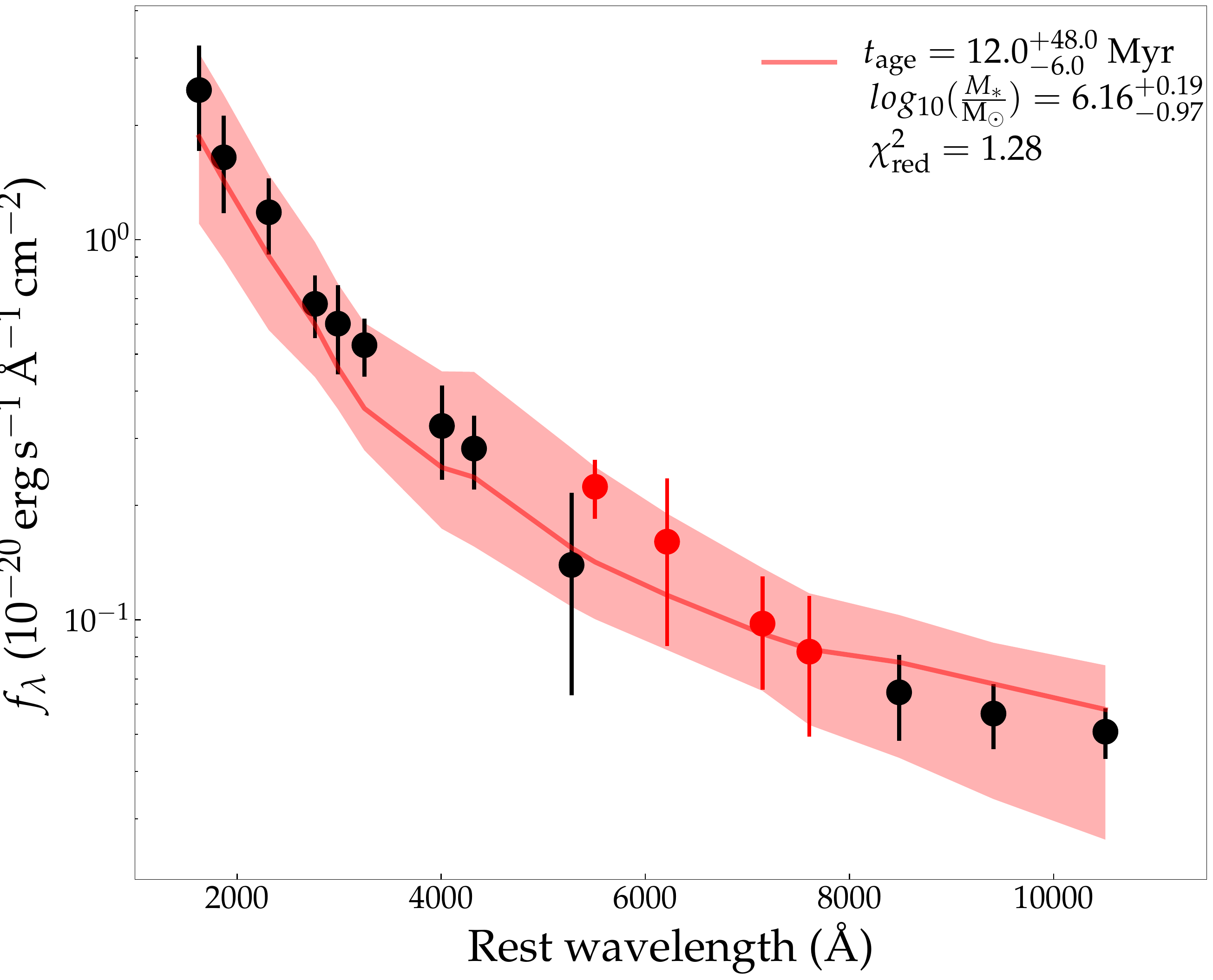}{0.48\textwidth}{}
          \fig{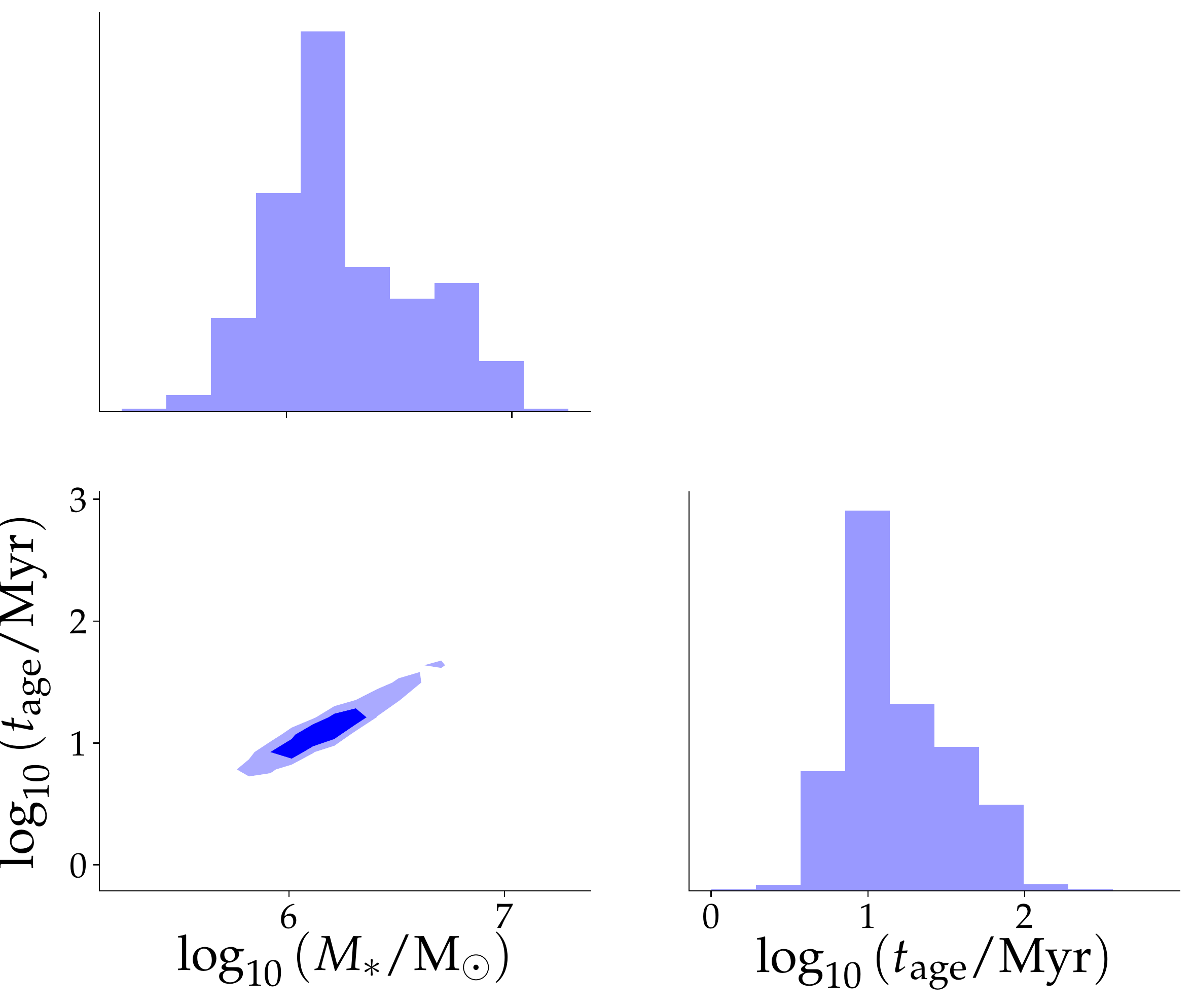}{0.48\textwidth}{}}
\caption{Same as Fig.\,\ref{fig:sed-young}, but for the measured SED in Fig.\,\ref{fig:ssp-model}(b) and Fig.\,\ref{fig:chisq-plot}(b).}
\label{fig:sed-b}
\end{figure*}

Figure\,\ref{fig:sed-b} shows the MCMC results for the same measured SED as in Figure\,\ref{fig:ssp-model}$b$ and Figure\,\ref{fig:chisq-plot}$b$.  The model SEDs with $Z = \rm Z_\sun$ provide a good fit ($\chi_{\rm red}^2 \leq 1$) to this measured SED over the age range $\sim$6-40\,Myr as shown earlier in Figure\,\ref{fig:chisq-plot}$h$.  The relatively broad age range of satisfactory fits reflects the relatively slow evolution in the model SEDs so long as red supergiants, which have lifetimes of up to a few 10s\,Myr, contribute significantly if not dominate the light of the SSP.  As can be seen in Figure\,\ref{fig:sed-b}, the posterior distribution spans a similar decade in ages, with $t_{\rm age} = 12^{+48}_{-6}$\,Myr .

\begin{figure*}[ht]
\centering
\gridline{\fig{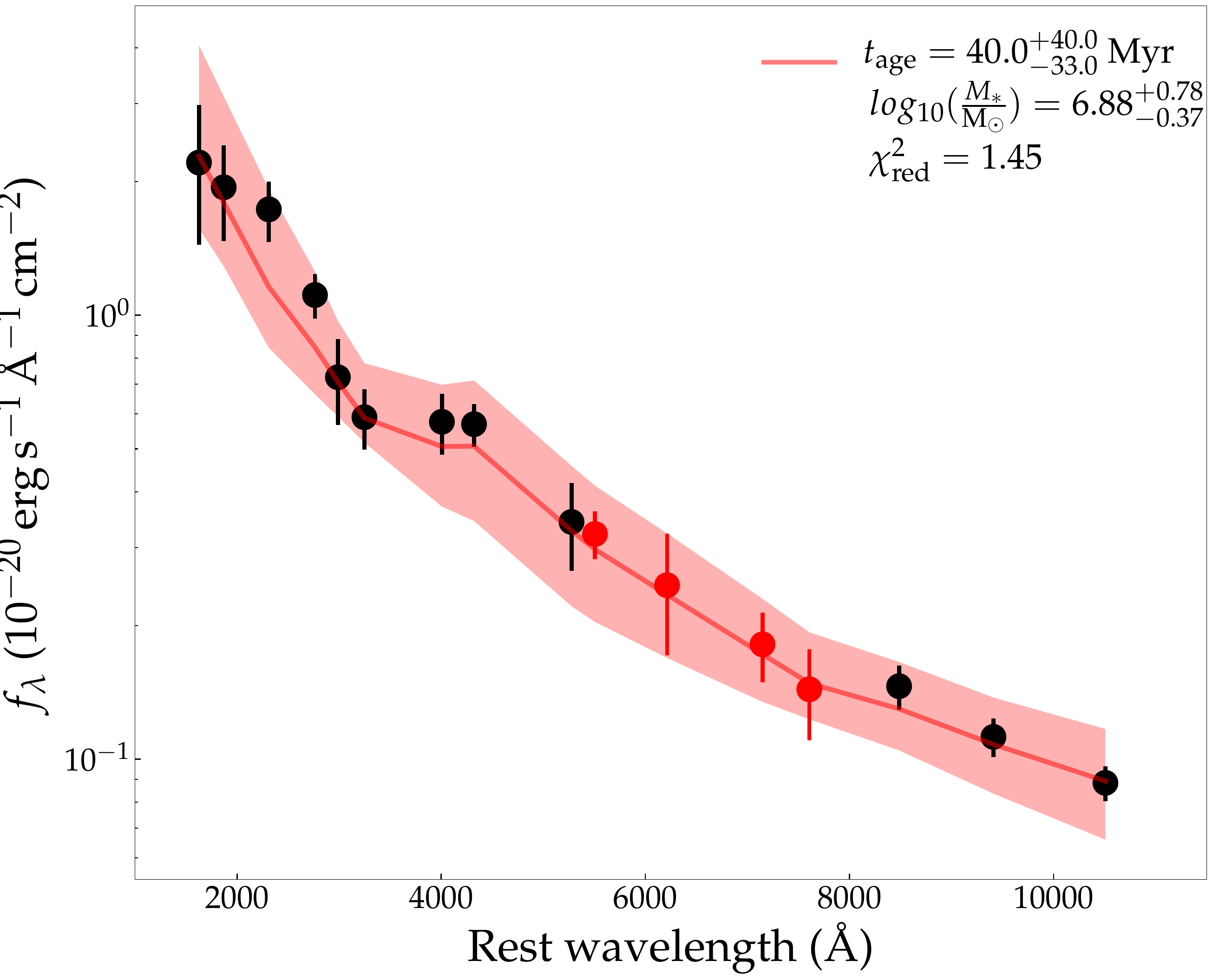}{0.48\textwidth}{}
          \fig{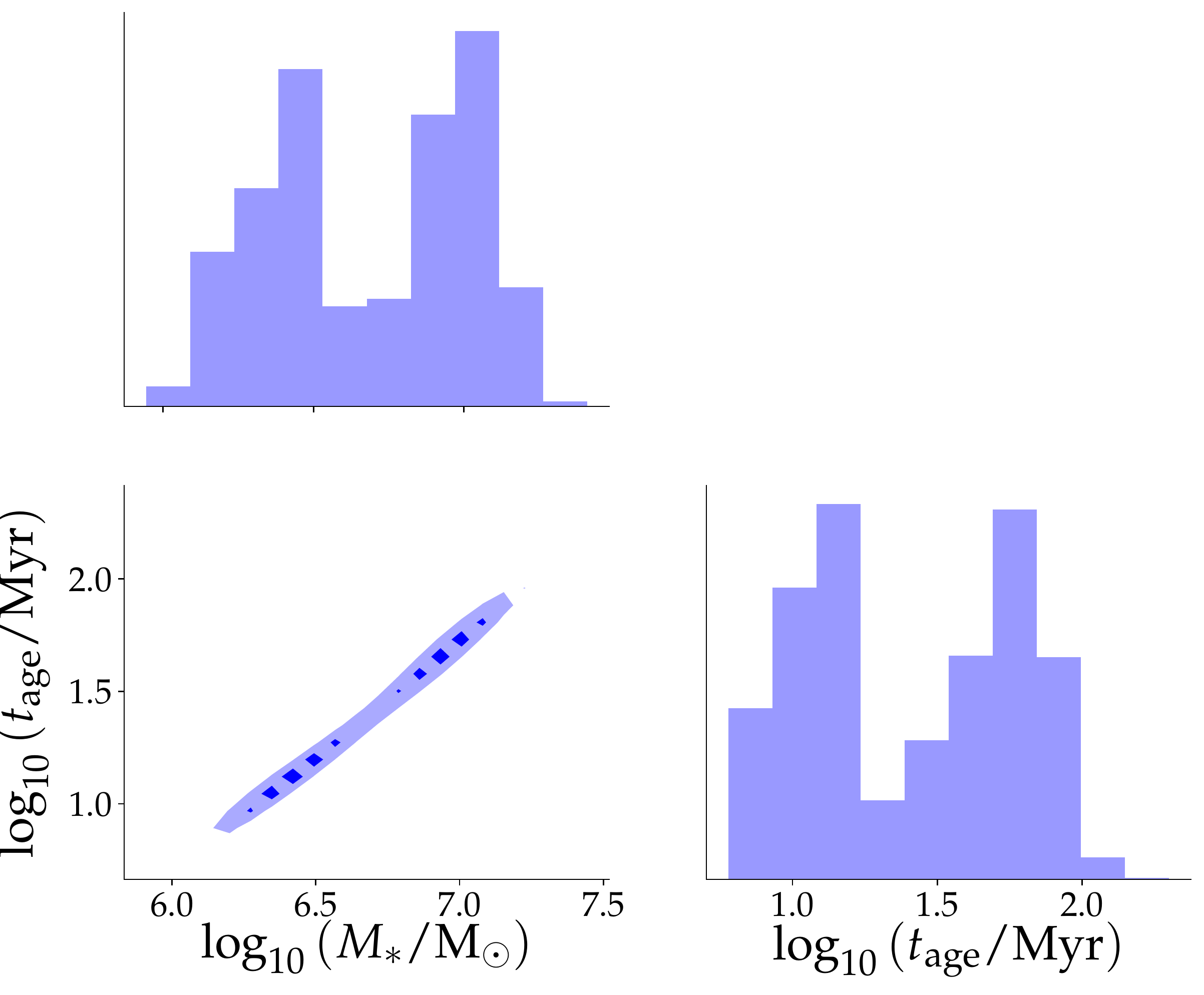}{0.48\textwidth}{}}
\caption{Same as Fig.\,\ref{fig:sed-young}, but for the measured SED in Fig.\,\ref{fig:ssp-model}(c) and Fig.\,\ref{fig:chisq-plot}(c).}
\label{fig:sed-c}
\end{figure*}

Figure\,\ref{fig:sed-c} shows the MCMC results for the same measured SED as in Figure\,\ref{fig:ssp-model}$c$ and Figure\,\ref{fig:chisq-plot}$c$.  The model SEDs with $Z = \rm Z_\sun$ provide a good fit ($\chi_{\rm red}^2 \leq 1$) to this measured SEDs over two distinct age ranges of $\sim$5--8\,Myr and $\sim$30--70\,Myr, separated by a plateau over which $1 < \chi_{\rm red}^2  < 2$, as shown earlier in Figure\,\ref{fig:chisq-plot}$i$.  Similarly, the posterior distribution in age is clearly bimodal with comparable likelihoods for the younger and older age brackets, such that $t_{\rm age} = 40^{+40}_{-33}$\,Myr.

\begin{figure*}[ht]
\centering
\gridline{\fig{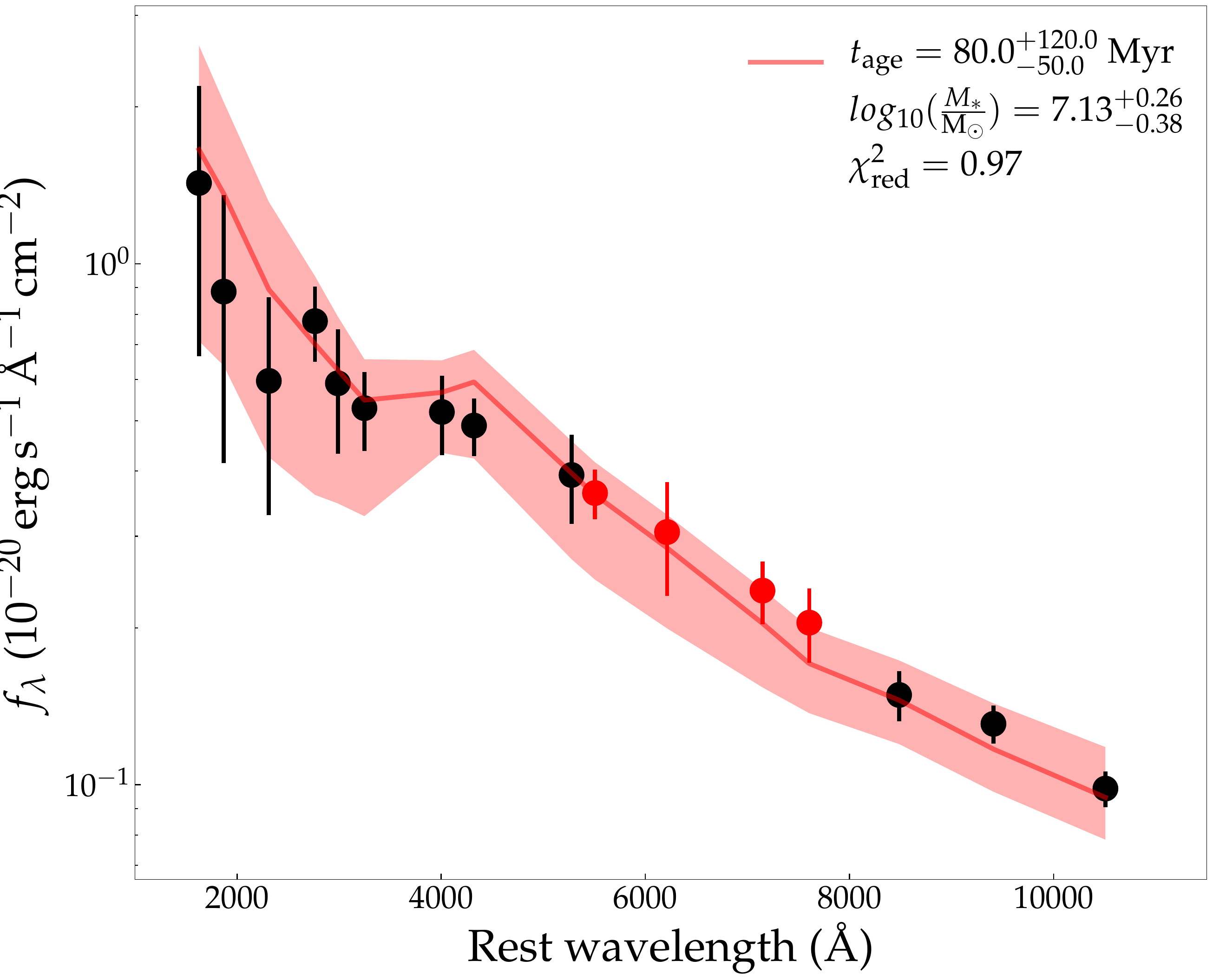}{0.48\textwidth}{}
          \fig{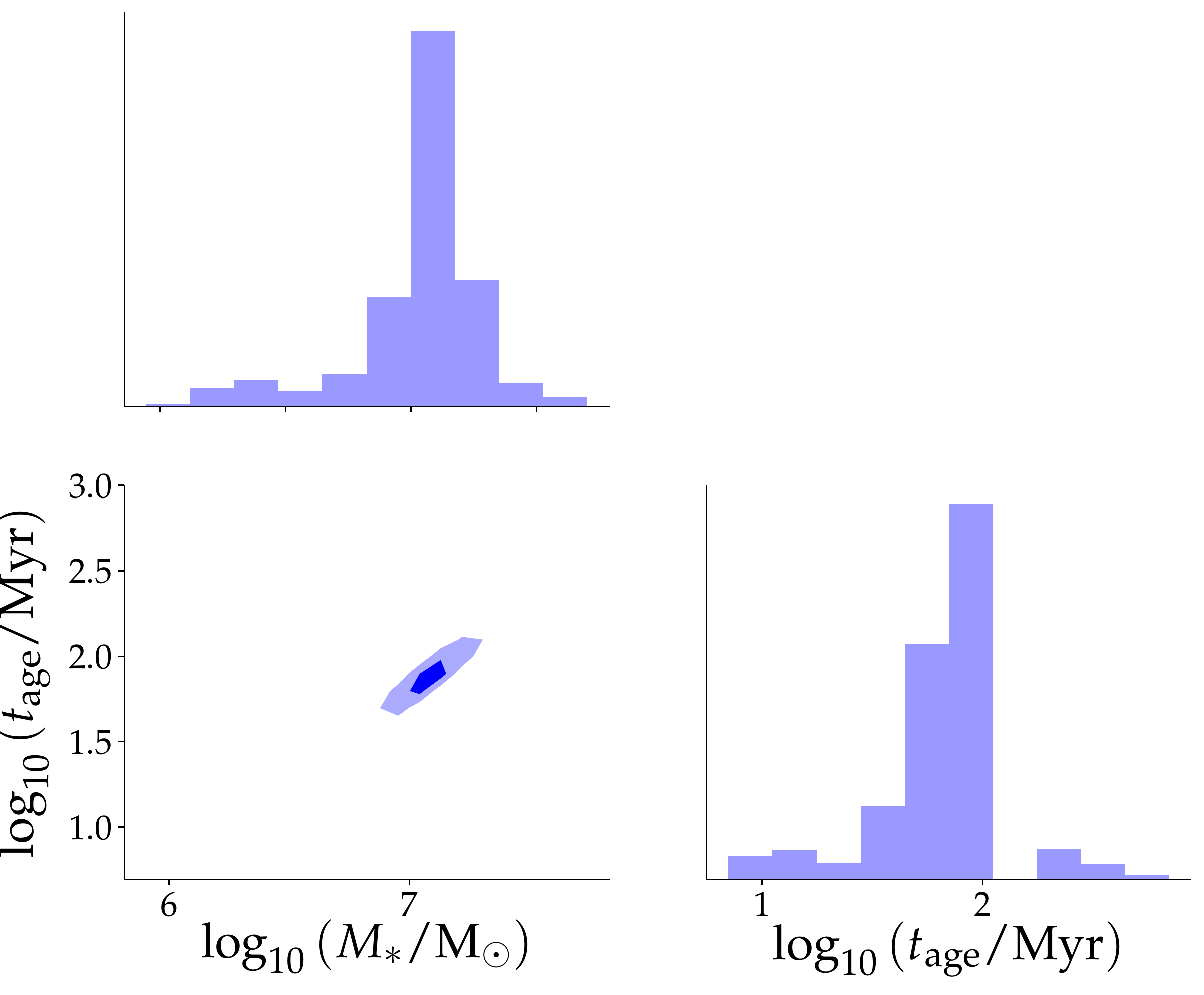}{0.48\textwidth}{}}
\caption{Same as Fig.\,\ref{fig:sed-young}, but for the measured SED in Fig.\,\ref{fig:ssp-model}(f) and Fig.\,\ref{fig:chisq-plot}(f).}
\label{fig:sed-old}
\end{figure*}

In the final example, Figure\,\ref{fig:sed-old} shows the MCMC results for the same measured SED as in Figure\,\ref{fig:ssp-model}$f$ and Figure\,\ref{fig:chisq-plot}$f$.  The model SEDs with $Z = \rm Z_\sun$ provide good fits ($\chi_{\rm red}^2 \leq 1$) to the measured SED over the age range $\sim$60--150\,Myr as shown earlier in Figure\,\ref{fig:chisq-plot}$l$.  Similarly, the posterior distribution in age is strongly peaked (in a logarithmic plot) at $\sim$100\,Myr, such that $t_{\rm age} = 80^{+120}_{-50}$\,Myr.

To see how permitting internal dust extinction might change the results, we also fit model SEDs having $A_V \leq 0.5$ as guided by Figure\,\ref{fig:CC} (top row).  We found that all sightlines with acceptable model SED fits (see Section\,\ref{acceptable fits}) -- which exclude those along which silhouette dust is clearly visible (see Fig.\,\ref{fig:ratios} and Fig.\,\ref{mosaic-residuals}) -- have nominal $A_V$ either considerably lower than 0.5 or consistent with 0 within the uncertainties.  As would be expected, allowing for extinction can expand the age range of acceptable fits owing to an age-extinction degeneracy at optical wavelengths; i.e., increasing either the age (at a fixed dust extinction) or the dust extinction (at a fixed age) both flattens the SED slope at optical wavelengths.  Given the very low if not null extinction found for all sightlines with acceptable model SED fits, in agreement with Figure\,\ref{fig:CC} (top row), we henceforth present results only for fits by model SEDs having no (internal) dust extinction.

\section{Results}\label{sec:results}
We present here the results from our MCMC approach.  First, we explain how we select satisfactory ($\chi_{\rm red}$ of order unity) as well as sensible ($t_{\rm age}$ well constrained) -- the two conditions that must be met for selecting acceptable -- model SED fits to the measured SEDs of the young stellar population.  As will become clear, acceptable fits are found for nearly all sightlines towards the young stellar population except where silhouette dust is visible; on these sightlines, our subtraction of the old stellar population is badly compromised and therefore also the measured SEDs of the young stellar population.  This ubiquity of acceptable model SED fits indicates that, oftentimes, a SSP likely dominates the brightness along a given sightline towards the young stellar population.  Of course, where the S/N ratio is poor, satisfactory fits can always be found, as is the case on sightlines towards the outer bounds of the young stellar population where ages cannot be sensibly constrained.  Furthermore, satisfactory but not sensible fits can always be found towards blank sky where model SEDs are fitted to random noise fluctuations.

Based on the model SEDs that yield acceptable fits to the measured SEDs, we construct maps showing the inferred nominal age and birth mass of the young stellar population along individual sightlines.  We then compute the star formation history -- star formation rate as a function of time -- of this population based on their nominal parameters as well as from the full posterior distribution of the relevant parameters.

\subsection{Acceptable Model SED Fits}\label{acceptable fits}
As we shall demonstrate, the key to selecting acceptable model SED fits to the measured SEDs lies in the posterior distribution in $t_{\rm age}$ given by our MCMC method.  Specifically, unacceptable fits are characterised by a relatively flat and broad posterior distribution in $t_{\rm age}$, indicating instances whereby model SEDs spanning a broad age range provide either: (i) comparably poor (unsatisfactory) fits to the measured SEDs; or (ii) equally satisfactory but not sensible fits, as in instances where the measured SEDs have poor S/N or correspond to blank sky.

\begin{figure*}[htb]
\fig{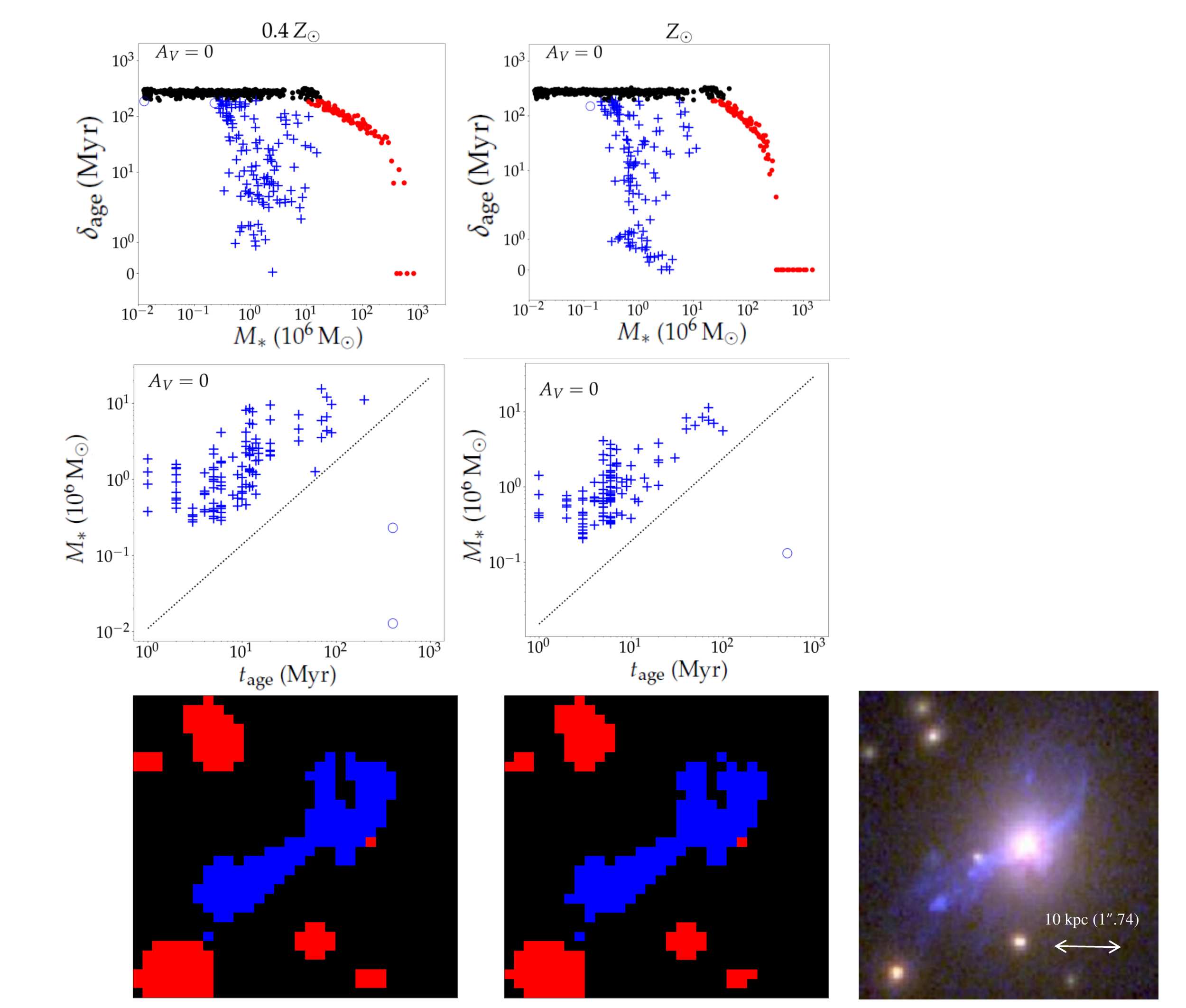}{0.9\linewidth}{}
\caption{
First column: Standard deviation in nominal $t_{\rm age}$, $\delta_{\rm age}$ (see text), versus nominal $M_*$ for model SEDs having $Z=0.4\,\rm Z_{\odot}$ (top row) and $Z = \rm Z_{\odot}$ (mid row) along all sightlines encompassed by the square box indicated in Figures\,\ref{fig:intro} and \ref{mosaic-residuals}.  Second column: nominal $M_*$ versus nominal $t_{\rm age}$ for sightlines colored blue.   Dotted diagonal line is the approximate detection threshold determined in the manner described in Section\,\ref{sec:selection effects} and plotted in Fig.\,\ref{fig:selection-bias}, and is used to select acceptable fits among these sightlines.  Third column: spatial locations of the data points colored black (blank sky), blue (young stellar population), and red (all except one sightline corresponding to cluster members). The RGB image in the last row is the same as in Figure \ref{fig:intro}, zoomed in to the dashed box.}
\label{fig:selection}
\end{figure*}

Figure\,\ref{fig:selection} (first row) shows the standard deviation of the posterior distribution in $t_{\rm age}$, $\delta_{\rm age}$, versus the nominal birth mass, $M_*$, for $Z = 0.4 \, \rm Z_\sun$ (first column) and $Z = \rm Z_\sun$ (second column), in both cases for $A_V = 0$.  The standard deviation of the posterior distribution is defined (following convention) as $\delta_x^2 = \int_{x_{\rm lower}}^{x_{\rm upper}} \, (x - \mu)^2 \, f(x) \, dx$, where $f(x)$ is the probability distribution function in $x$ (in this instance, $t_{\rm age}$) and $\mu = \int x f(x) \, dx$ the weighted average value in $x$.  Three trends are immediately apparent from the distribution of $\delta_{\rm age}$ versus $M_*$: (i) a pileup in $\delta_{\rm age}$ over the range $\sim$200--300\,Myr (black data points); (ii) a grouping of data points having $\delta_{\rm age}$ significantly below this pileup over the range $10^5 \, \rm M_\sun$$\lesssim M_*  \lesssim 10^7$$\,\rm M_\sun$ (blue data points); and (iii) a systematic decrease in $\delta_{\rm age}$ with increasing $M_*$ at $M_* \gtrsim 10^7 \, \rm M_\sun$ (red data points).  The corresponding spatial locations of these three groups of data points are shown in Figure\,\ref{fig:selection} (third row).  As can be seen, the blue data points coincide with the young stellar population, whereas the red data points coincide almost exclusively if not entirely with neighbouring cluster members.  The black data points coincide primarily with regions in between the young stellar population and neighbouring cluster members (i.e., blank sky), although a few coincide with the inner region of the BCG where silhouette dust is clearly visible and where the fits are poor (for the reasons explained above).

The pileup at $\delta_{\rm age} \approx 200$--300\,Myr reflects the relatively flat and broad posterior distribution in $t_{\rm age}$ when model SEDs are fitted to random noise fluctuations (the situation in most cases) or where the fits are poor\footnote{For example, in the case of a log-uniform posterior distribution, $\delta_x^2 = \left[ \frac{ x_{\rm upper}^2 - x_{\rm lower}^2}{2 \, {\rm ln} (x_{\rm upper}/x_{\rm lower}) } \right] - \left[ \frac{x_{\rm upper} - x_{\rm lower}}{{\rm ln} (x_{\rm upper}/x_{\rm lower}) } \right]^2$, so that for $x_{\rm lower} = 1 \, {\rm Myr}$ and $x_{\rm upper} = 1 \, {\rm Gyr}$ (the lower and upper bounds in age range of the model SEDs used in our MCMC computations), $\delta_{\rm age} = 227$\,Myr.}. The blue data points (young stellar population) are formally separated from the black data points (blank sky or unacceptable fits) by a demarkation line that we define at $\delta_{\rm age} = 190$\,Myr, lying just below the pileup in $\delta_{\rm age}$.  Figure\,\ref{fig:selection} (second row) shows $M_*$ plotted as a function of $t_{\rm age}$ for the blue data points.  As is apparent, the data points are strongly concentrated at the upper left corner, with a trend of an increasing lower bound in $M_*$ with increasing $t_{\rm age}$ suggestive of an observational detection threshold.  Indeed, the approximate detection threshold computed in the manner described in Section \ref{sec:selection effects} and indicated by a diagonal line closely defines the lower bound of this main group of data points.  The blue data points plotted as open circles mostly lie well below the detection threshold, and have relatively large $\delta_{\rm age}$ just below the demarkation line at 190\,Myr; they coincide with the outer regions of the young stellar population where the S/N is poor.  The remaining blue data points plotted as crosses ($+$) are therefore selected as satisfactory and sensible (i.e., acceptable) model SEDs fits to the measured SEDs.  Although not plotted in this figure, the red data points are concentrated at the upper bound in age considered in our model SED fits of 1\,Gyr, implying that their stellar populations have older ages (as would be expected for early-type galaxies).

\subsection{Nominal Model Parameters}\label{nominal parameters}
\begin{figure*}[ht]
\centering
\fig{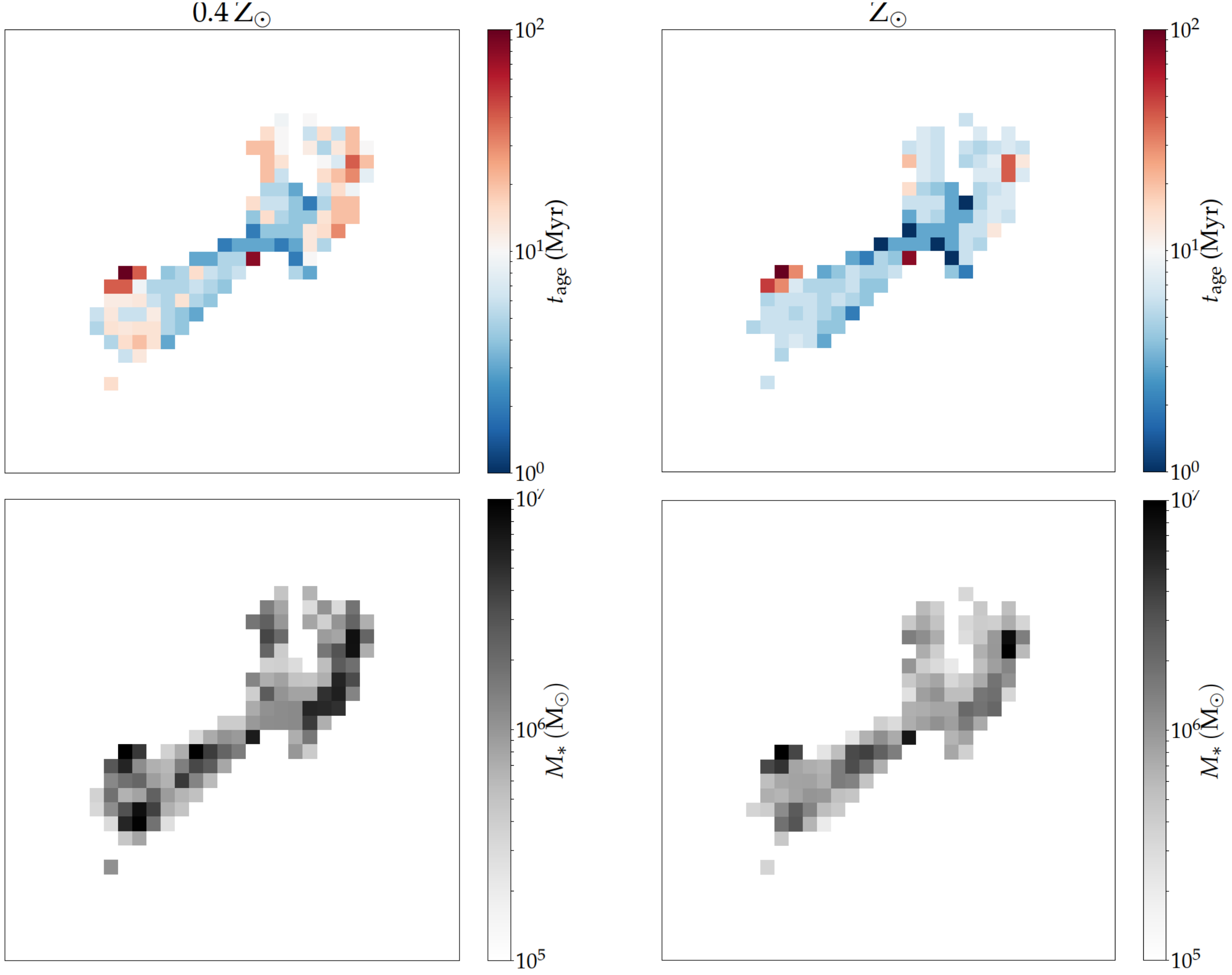}{0.8\textwidth}{}
\caption{Nominal $t_{\rm age}$ (top row) and nominal $M_*$ (bottom row) for $Z=0.4\,\rm Z_{\odot}$ (left column) and $Z=\rm Z_{\odot}$ (right column) inferred from our MCMC method for the young stellar population.}
\label{fig:nominal_parameters} 
\end{figure*}

In Figure\,\ref{fig:nominal_parameters}, we present maps of the nominal parameters for the young stellar population in ages (first row) and birth masses (second row) for $Z = 0.4 \, \rm Z_\sun$ (left column) and $Z = \rm Z_\sun$ (right column).  As can be seen, the nominal ages for $Z = \rm Z_\sun$ (second column) tend to be somewhat younger than for $Z = 0.4\, \rm Z_\sun$ (first column) owing to an age-metallicity degeneracy.  Given the generally small differences in ages between the two metallicities, however, the birth masses are very similar irrespective of whether $Z = 0.4\, \rm Z_\sun$ or $Z = \rm Z_\sun$.  The nominal ages span $\sim$1--100\,Myr, and the nominal birth masses (which, along a given sightline, encompasses an area of $1.5 \times 1.5$\,kpc) span $\sim$$10^5$--$10^7 {\rm \, M_\sun}$.

\subsection{Emission-line Nebula}\label{Nebula}
In observations utilizing filters, emission-line maps are usually constructed by subtracting an image comprising just stellar continuum from an image comprising both line(s) and stellar continuum together.  Before subtraction, the image comprising just stellar continuum is scaled to match the intensity of the stellar continuum in the image comprising both line and stellar continuum.  This procedure was used by \citet{Fogarty2015} to produce $\rm H\alpha + [NII]$ images for a number of the BCGs in the CLASH program, including that studied here.  When scaling for the stellar continuum, \citet{Fogarty2015} adopted the same color throughout the galaxy corresponding to that of its old stellar population.  

The situation for BCGs hosting a young stellar population, as is the case here, is more complicated as the stellar continuum along some sightlines includes a contribution from this population in addition to that from an old stellar population -- in the case of the BCG here, many of the same sightlines that exhibit line emission.  To further complicate matters, the young stellar population has a spatially-varying SED owing to their different ages and birth masses along different sightlines, and therefore contribute different amounts of continuum light at different spatial locations.  In such instances, the key to producing an accurate emission-line map is to accurately model or infer the stellar continuum of both the young and old stellar population.

\begin{figure*}[htb]
\centering
\gridline{\fig{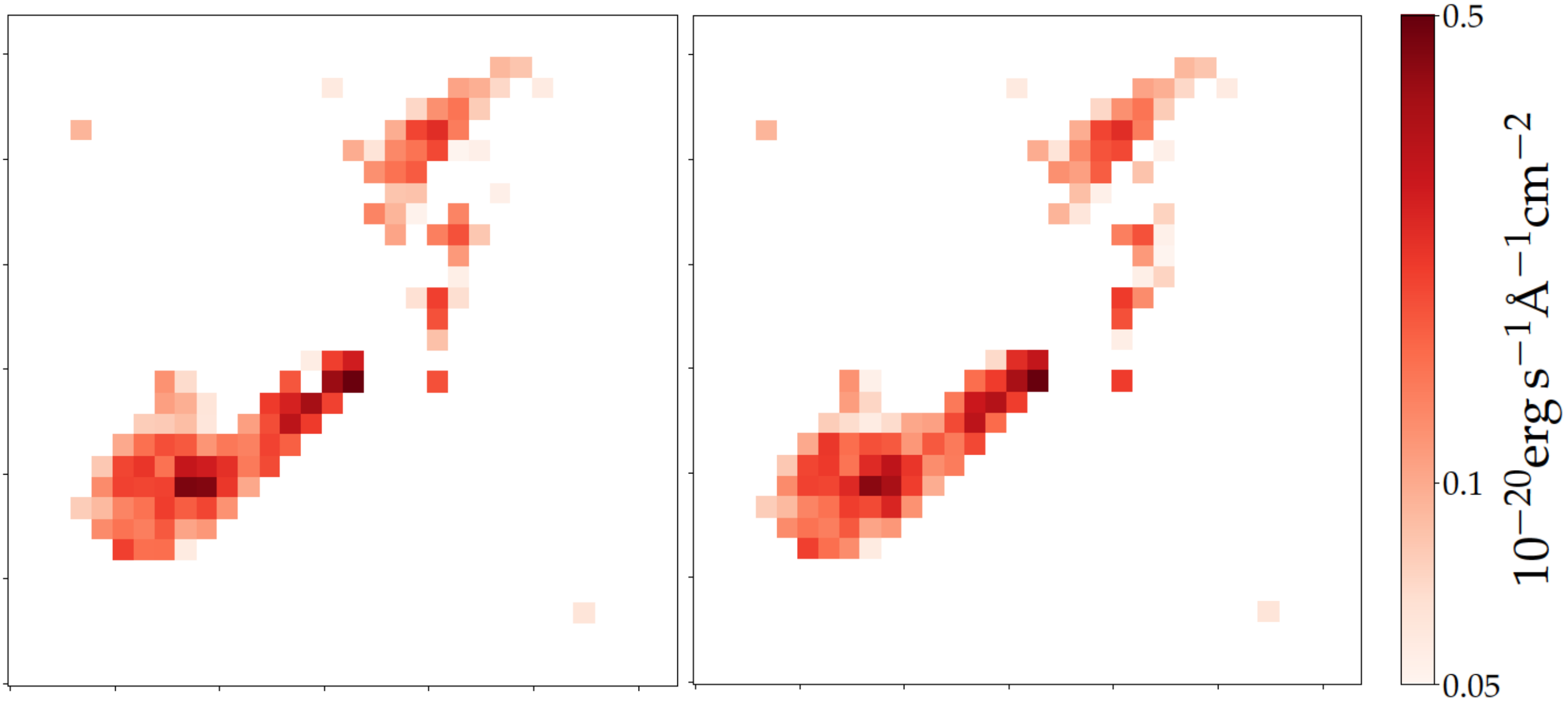}{0.48\textwidth}{(a) Pure H$\alpha$+[NII]}
\fig{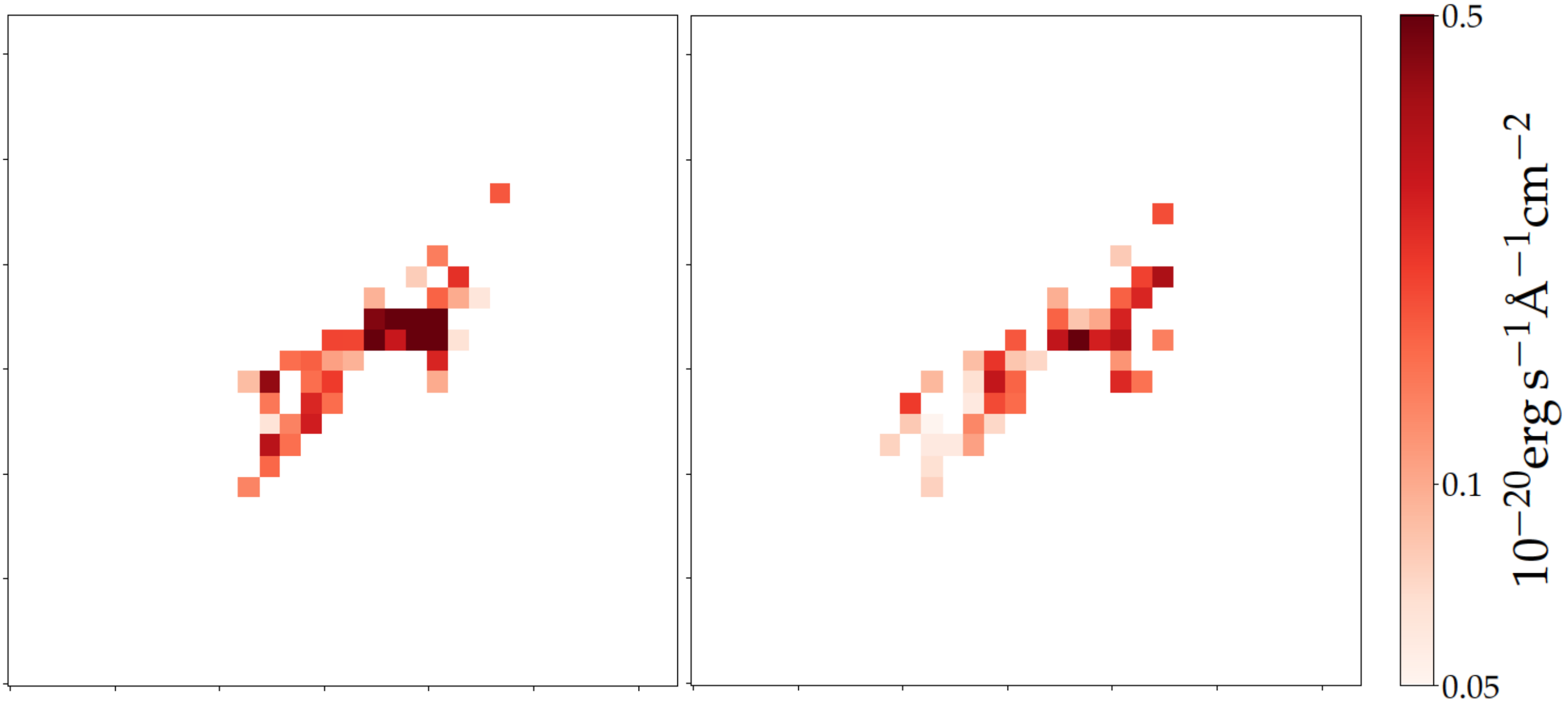}{0.48\textwidth}{(b) Maximal HII regions}}
\gridline{\fig{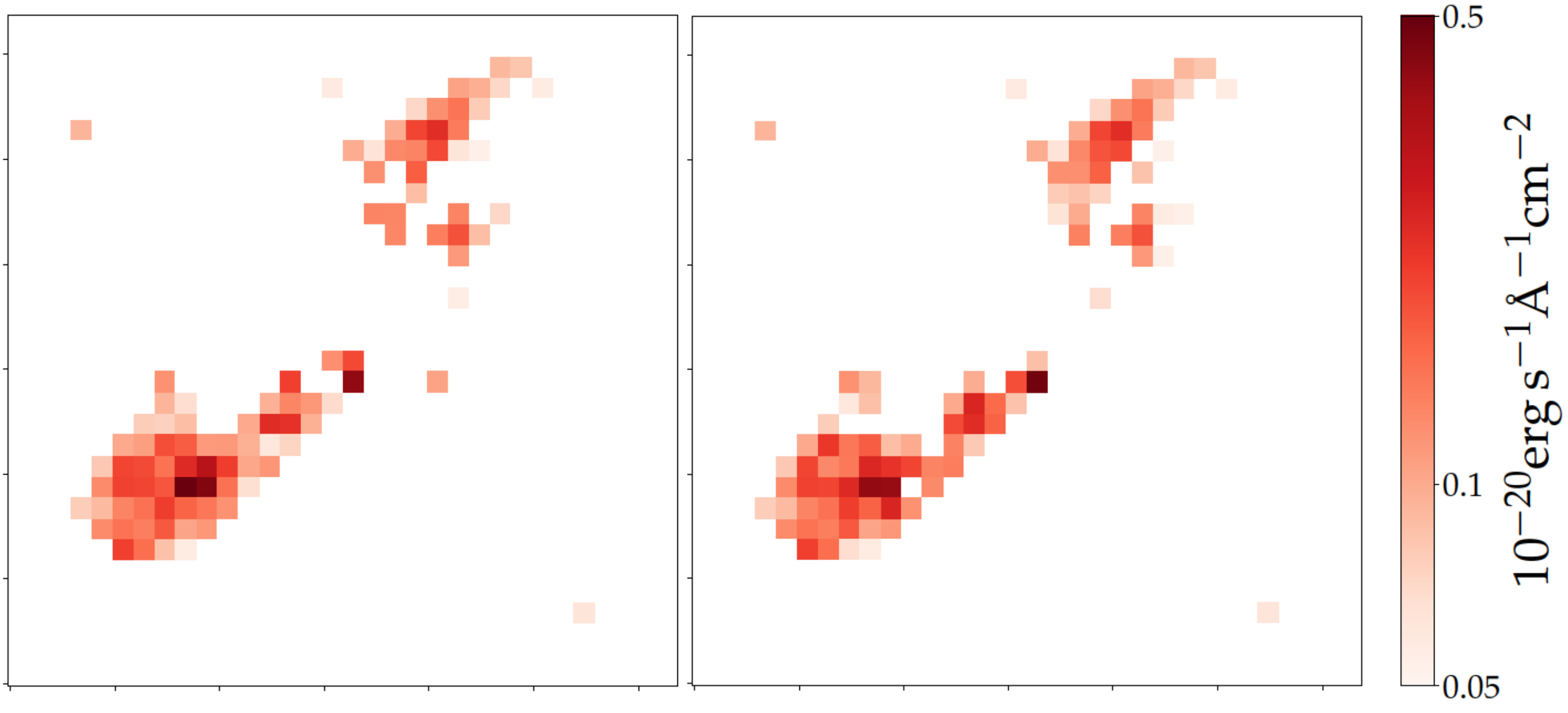}{0.48\textwidth}{(c) Minimal line-emitting nebula}
\fig{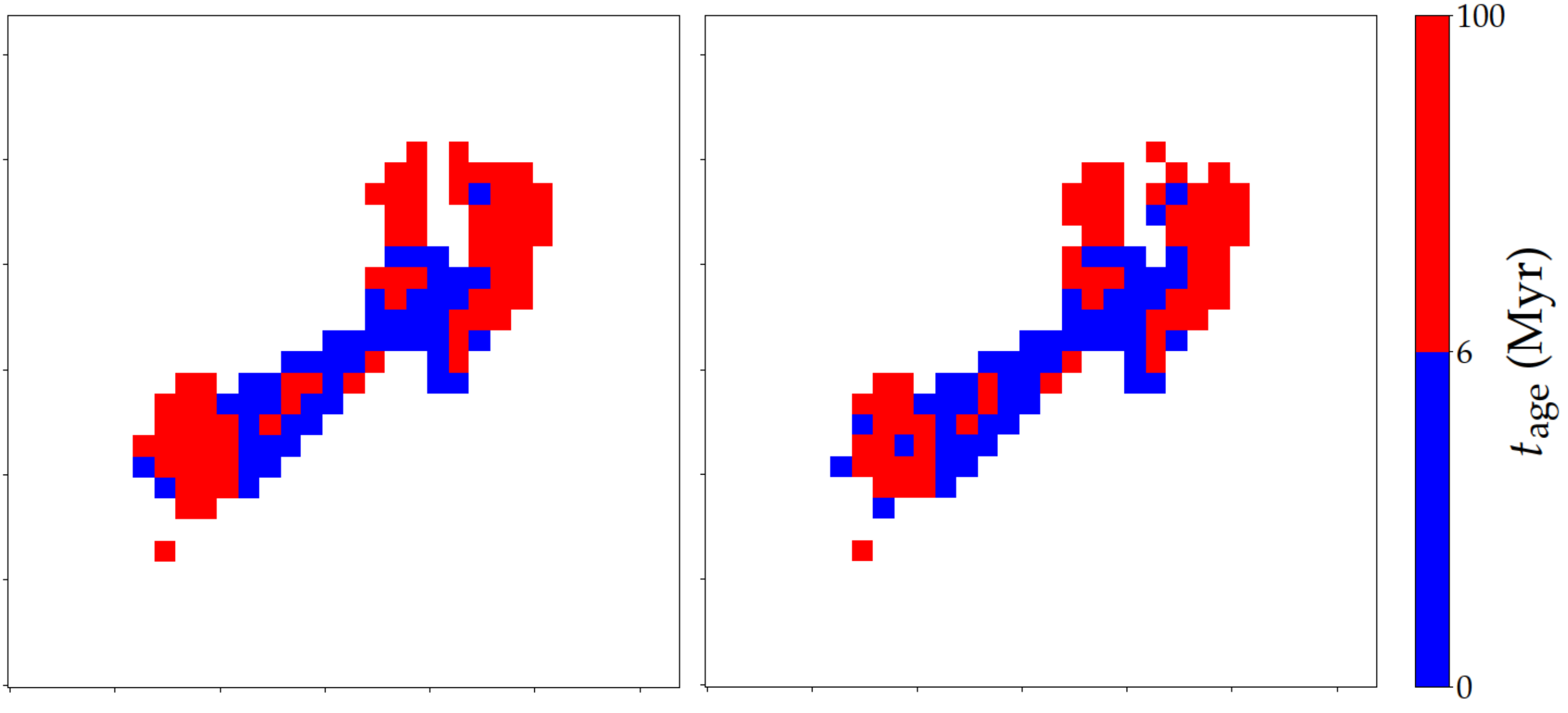}{0.48\textwidth}{(d) Ages of young stellar population}}
\caption{
(a): pure H$\alpha$+[NII] image after subtracting, where necessary, model stellar continuua for the young stellar population from the F850LP image of Fig.\,\ref{mosaic-residuals}, where the stellar continuum from the old stellar population has already been accurately subtracted.  (b): Maximal predicted H$\alpha$+[NII] or relatively weak continuum from HII regions, derived by subtracting model SEDs having $f_{\rm cov} = 0$ from those having $f_{\rm cov} = 1$ based on the nominal $t_{\rm age}$ and $M_*$ along each sightline.  (c): Minimal predicted H$\alpha$+[NII] from an emission-line nebula not associated with HII regions, derived by subtracting the images in second row from those in the first row.  (d): sightlines coded blue where nominal $t_{\rm age} \leq 5$\,Myr and therefore H$\alpha$+[NII] from HII regions potentially detectable, and sightlines coded red where nominal $t_{\rm age} > 5$\,Myr and any H$\alpha$+[NII] from HII regions too weak to be detectable, in the F850LP filter.  
Left column is for $Z=0.4\, \rm Z_{\odot}$ and right column for $Z= \rm Z_{\odot}$.}
\label{fig:Ha}
\end{figure*}

To estimate the stellar continuum of the young stellar population along different sightlines, we computed its model stellar continuum based on its nominal $t_{\rm age}$ and $M_*$ along individual sightlines by setting $f_{\rm cov} = 0$ (i.e., so as not to include any non-stellar contributions from HII regions).  We then subtracted this component from the image in the F850LP filter -- where H$\alpha$+[NII] contribute the largest fractional intensity -- from which the continuum light of the old stellar population had already been accurately removed (Section\,\ref{sec:method}) as shown in Figure\,\ref{mosaic-residuals}.  To isolate regions coincident only with either the young stellar population or emission-line gas (e.g., the filament indicated by an arrow in Fig.\,\ref{fig:ratios}), as well as to help discriminate against noise peaks, we select relatively blue regions in the color maps having intensity ratios $f_{\rm 850LP}/f_{\rm F160W} > 3.0$ and $f_{\rm 814W}/f_{\rm F140W} > 2.0$.  
This procedure produces a pure line image subject only to uncertainties in the inferred parameters ($t_{\rm age}$ and $M_*$) of the young stellar population (and hence its contribution to the stellar continuum) along each sightline (and, in principle, the degree to which the model SEDs actually match the measured SEDs).
The results after subtracting the model continuum light of the young stellar population are shown in Figure\,\ref{fig:Ha} (a) for $Z = 0.4 \rm \, Z_\sun$ (left column) and $Z = \rm Z_\sun$ (right column).  As can be seen, H$\alpha$+[NII] emission is apparent along sightlines even where: (i) the nominal ages of the young stellar population, as shown in Figure\,\ref{fig:Ha} (d), is older than 5\,Myr, at which time the model SEDs show no appreciable contribution from H$\alpha$+[NII] after convolving over the spectral response of the F850LP filter (see Section \ref{model SSPs}); and (ii) there is no detectable young stellar population, most prominently to the north-west of the BCG center (at the location indicated by an arrow in Fig.\,\ref{fig:ratios}).  These results conform with expectations based on the color-color diagrams presented in Figure\,\ref{fig:CC} (middle row) showing H$\alpha$+[NII] emission at levels in excess of those predicted for $f_{\rm cov} = 1$ along many sightlines to the young stellar population, and the color map shown in Figure\,\ref{fig:ratios} (third row, third column) where pure line emission as indicated by an arrow is visible north-west of the BCG center.  Notice that the pure line image sometimes shows no emission along sightlines where the nominal ages of the young stellar population is $\lesssim 5$\,Myr, indicating no detectable HII regions along these sightlines.  In this way, we find an integrated line intensity of $3.8 \times 10^{-15} \, \rm erg \, s^{-1} \, cm^{-2}$ (i.e.,  $L_{\rm H\alpha +\rm [NII]}=4.1 \times 10^{42}\, \rm erg \, s^{-1}$) if the young stellar population has $Z = 0.4 \rm \, Z_\sun$, and an integrated line intensity of $3.7 \times 10^{-15} \, \rm erg \, s^{-1} \, cm^{-2}$ (i.e., $L_{\rm H\alpha +\rm [NII]}=4.0 \times 10^{42}\rm \, erg \, s^{-1}$) if this population has $Z = \rm Z_\sun$.  By comparison, \citet{Fogarty2015} report $L_{\rm H\alpha +\rm [NII]} = (25.1 \pm 2.4) \times 10^{42}\, \rm erg \, s^{-1}$ (after correction for internal dust extinction; see Section\,\ref{sec:star-formation rates}), about six times higher than what we measure.

To estimate the maximal possible line emission from HII regions, we subtracted model images for $f_{\rm cov} = 0$ from those for $f_{\rm cov} = 1$ generated at the same nominal $t_{\rm age}$ and $M_*$ in the F850LP filter.  The results are shown in Figure\,\ref{fig:Ha} (b) for the two respective metallicities.  As would be expected, relatively bright line emission from HII regions is predicted where the nominal $t_{\rm age} \leq 5$\,Myr (note that intensities are color coded on a logarithmic scale, so that the range spanned by light green to yellow is an order of magnitude brighter than the range spanned by black to dark green).  We truncate the scale bar at $0.05 \times 10^{-20} {\rm erg \, s^{-1} \AA^{-1} \, cm^{-2}}$ as emission below this level mostly if not entirely arises, we suspect, from small numerical errors in the YGGDRASIL algorithm used to generate the model SEDs for $f_{\rm cov} = 1$ (see earlier discussion in Section\,\ref{model SSPs}), rather than from genuine HII regions.

By subtracting the maximal model HII-region emission in the second row from the pure line image in the first row of Figure\,\ref{fig:Ha} for the respective metallicities, we generated images showing the minimal line emission from gas not associated with HII regions -- hereafter, emission-line nebula -- as shown in the Figure\,\ref{fig:Ha} (c).  As is apparent, the emission-line nebula overlaps in large part with the young stellar population, consistent with the color-color diagram shown earlier in Figure\,\ref{fig:CC} (b) whereby the contribution from H$\alpha$+[NII] along many sightlines to the young stellar population far exceeds that predicted by the model SEDs even for $f_{\rm cov} = 1$.  Furthermore, the emission-line nebula extends beyond the bounds of the young stellar population (in particular to the north-west), consistent also with the color image shown in Figure\,\ref{fig:ratios} (third column, third row).  From the ratio between the total flux density of the images in the third row and first row of Figure\,\ref{fig:Ha}, we estimate that the emission-line nebula accounts for at least $\sim$60\% (for $Z=\rm Z_\sun$) to $\sim$80\% (for $Z = \rm 0.4 \, Z_\sun$) of the H$\alpha$+[NII] line emission from the BCG.  Attributing the H$\alpha$+[NII] emission predominantly or entirely to star formation would therefore result in an over-estimate of the star-formation rate as computed from the total line luminosity.

\subsection{Star Formation History}\label{SFH}
\begin{figure*}[ht]
    \fig{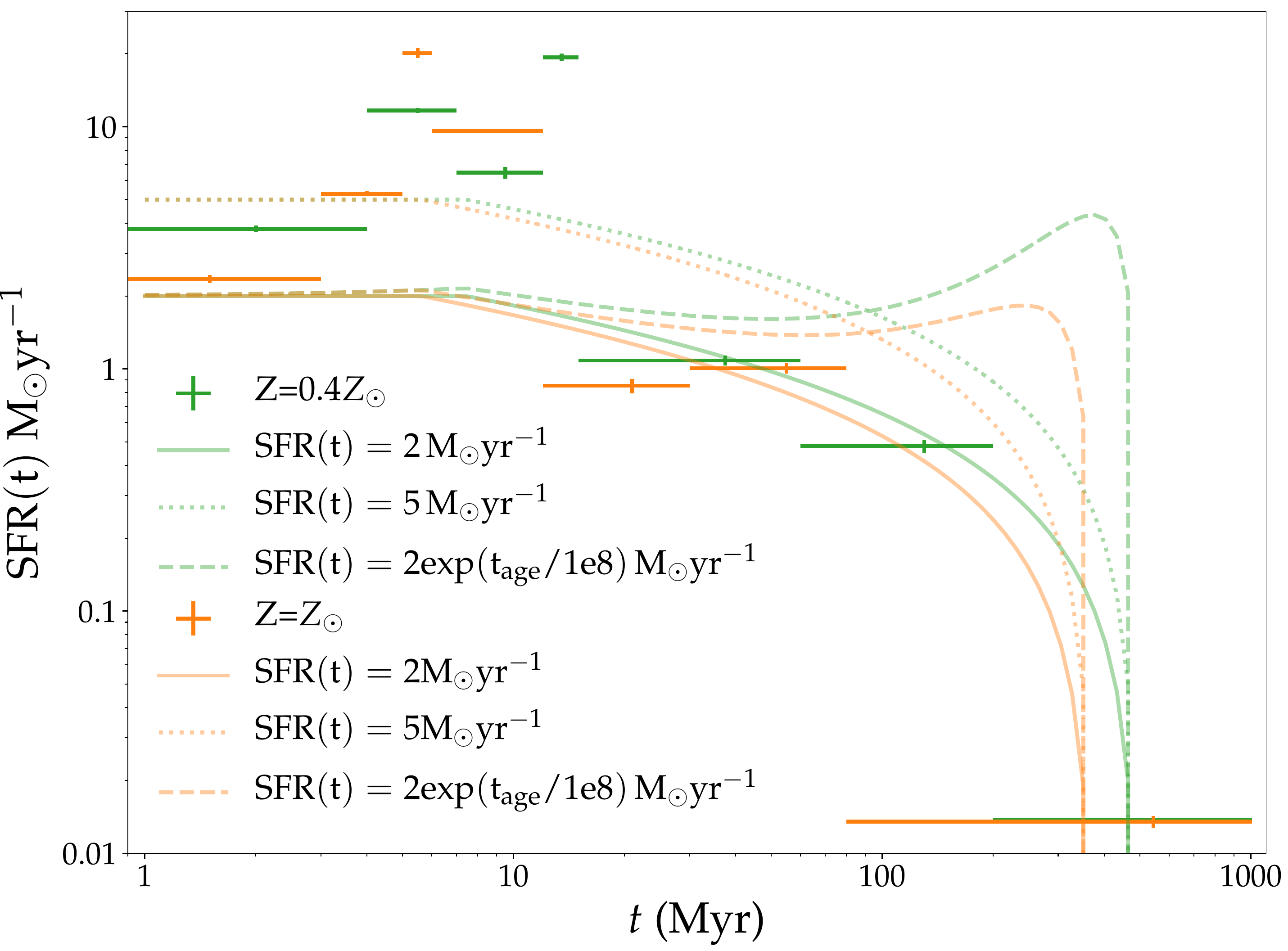}{0.48\linewidth}{}
    \fig{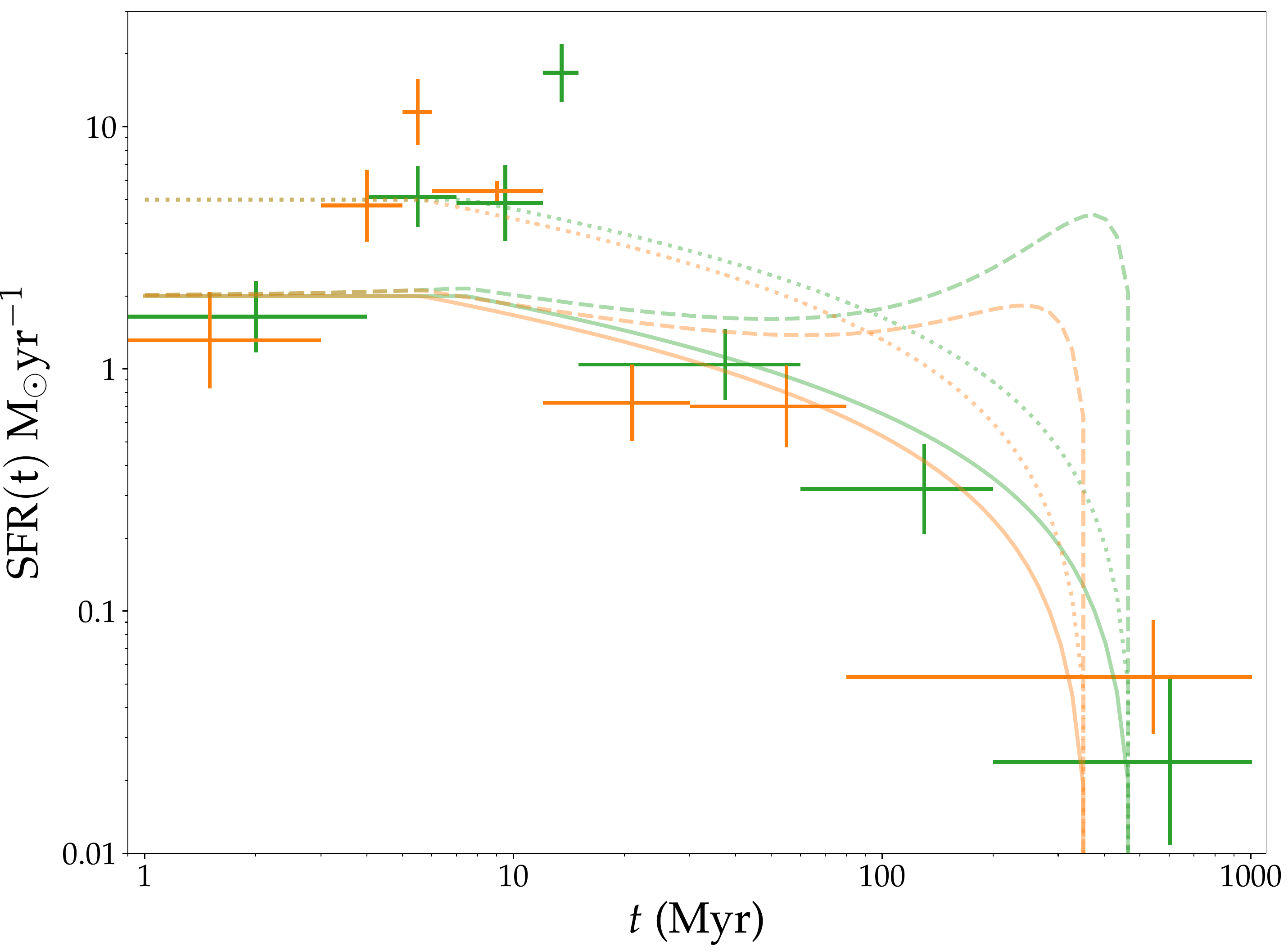}{0.48\linewidth}{}
    \caption{Left panel: star-formation rate, $SFR(t)$, as a function of time, $t$, based on the nominal $t_{\rm age}$ and $M_*$ inferred from our MCMC method.  Right panel: same as left panel, but now utilising the full posterior distribution for $t_{\rm age}$ and $M_*$ (see text).  The bin widths, corresponding to time intervals, were selected such that there are roughly equal number of samples in each bin.  Green color is for $Z=0.4\,\rm Z_{\odot}$ and orange color is for $Z=\rm Z_{\odot}$.  Solid curves are for $SFR(t) = 2 \rm M_\sun \, {\rm \, yr^{-1}}$, dotted curves for $SFR(t) = 5 \rm M_\sun \, {\rm \, yr^{-1}}$, and dashed curves for a starburst that began $\sim$400\,Myr ago and has since decayed exponentially with an e-folding timescale of $100$\,Myr, in all cases after applying selection effect I (an increasing detection threshold in $M_*$ with increasing $t_{\rm age}$) as discussed in the text. }
    \label{fig:no-prob-SFH}
\end{figure*}

Based on the nominal $t_{\rm age}$ and $M_*$ along each sightline towards the young stellar population, we computed the ensemble star formation rate as a function of time, $SFR(t) = \Delta M_* / \Delta t_{\rm age}$, whereby $\Delta M_*$ is the sum of the birth masses along all sightlines spanning a selected range in nominal $t_{\rm age}$ of $\Delta t_{\rm age}$.  The results are shown in Figure\,\ref{fig:no-prob-SFH} (left panel) for $Z = 0.4 \, \rm Z_\sun$ by the green histogram and $Z = \rm Z_\sun$ by the orange histogram.  In this plot, we selected different $\Delta t_{\rm age}$ centered at different $t_{\rm age}$ so as to give a comparable number of samples within each bin.  The difference in $SFR(t)$ between the two metallicities are small except at $t_{\rm age} \sim 10$--100\,Myr, whereby the model SEDs are dominated by red supergiants having evolutionary timescales that depend on metallicity.  

Taking into consideration the uncertainties in $t_{\rm age}$ and $M_*$ along each sightline provided by our MCMC method, Figure\,\ref{fig:no-prob-SFH} (right panel) shows the star-formation history computed using the full posterior distribution of these parameters along individual sightlines.  The mathematical treatment is described in full in Appendix\,\ref{app:sfr}.  
In essence, for all sightlines towards the young stellar population, we summed the products between $M_*$ at a particular $t_{\rm age}$ and the probability given by the posterior distribution for that age bin.  We then computed $SFR(t)$ by choosing an appropriate $\Delta t_{\rm age}$ centered at a given $t_{\rm age}$ so as to give roughly equal number of sightlines in each $\Delta t_{\rm age}$ bin.  

At this stage, we caution that the $SFR(t)$ thus derived should not be taken to reflect the intrinsic $SFR(t)$, as the measurements shown in Figure\,\ref{fig:no-prob-SFH} do not take into account selection effects as we shall describe next.


\section{Discussion}\label{sec:discussion}

\subsection{Selection Effects}\label{sec:selection effects}
Over a given wavelength interval, stellar populations generally fade over time as progressively less massive stars reach the terminal phases of their evolution.  Correspondingly, between ages of $\sim$10\,Myr and $\sim$100\,Myr, our model SSPs fade in brightness  by: (i) just over an order of magnitude (factor of $\sim$20) in the rest-frame UV; and (ii) just under to about an order of magnitude in the rest-frame optical to near-IR.  Between ages of $\sim$1\,Myr and $\sim$10\,Myr, the fading at all these wavelengths is significantly less, corresponding to a factor of $\lesssim 4$.  This fading over time leads to the following selection effects depending on the composition of SSPs (i.e., ensemble of star clusters possibly having different birth masses and/or ages) along individual sightlines: (i) if composed of a singular SSP or multiple SSPs having similar ages, then a limitation on the oldest detectable SSP for a given (total) birth mass (selection effect I); and (ii) if composed of multiple SSPs having a broad range of ages but otherwise similar birth masses, then shifting the inferred age from the mean age towards the age of the youngest (and hence brightest) SSP, as well as overestimating its birth mass as part of the light is actually contributed by SSPs having different ages (selection effect II).  Both these selection effects lead to the same bias in the star-formation history as inferred from the measurements: inevitably underestimating at progressively greater severity the star-formation rate in the more distant past.

\begin{figure*}[ht]
\gridline{\fig{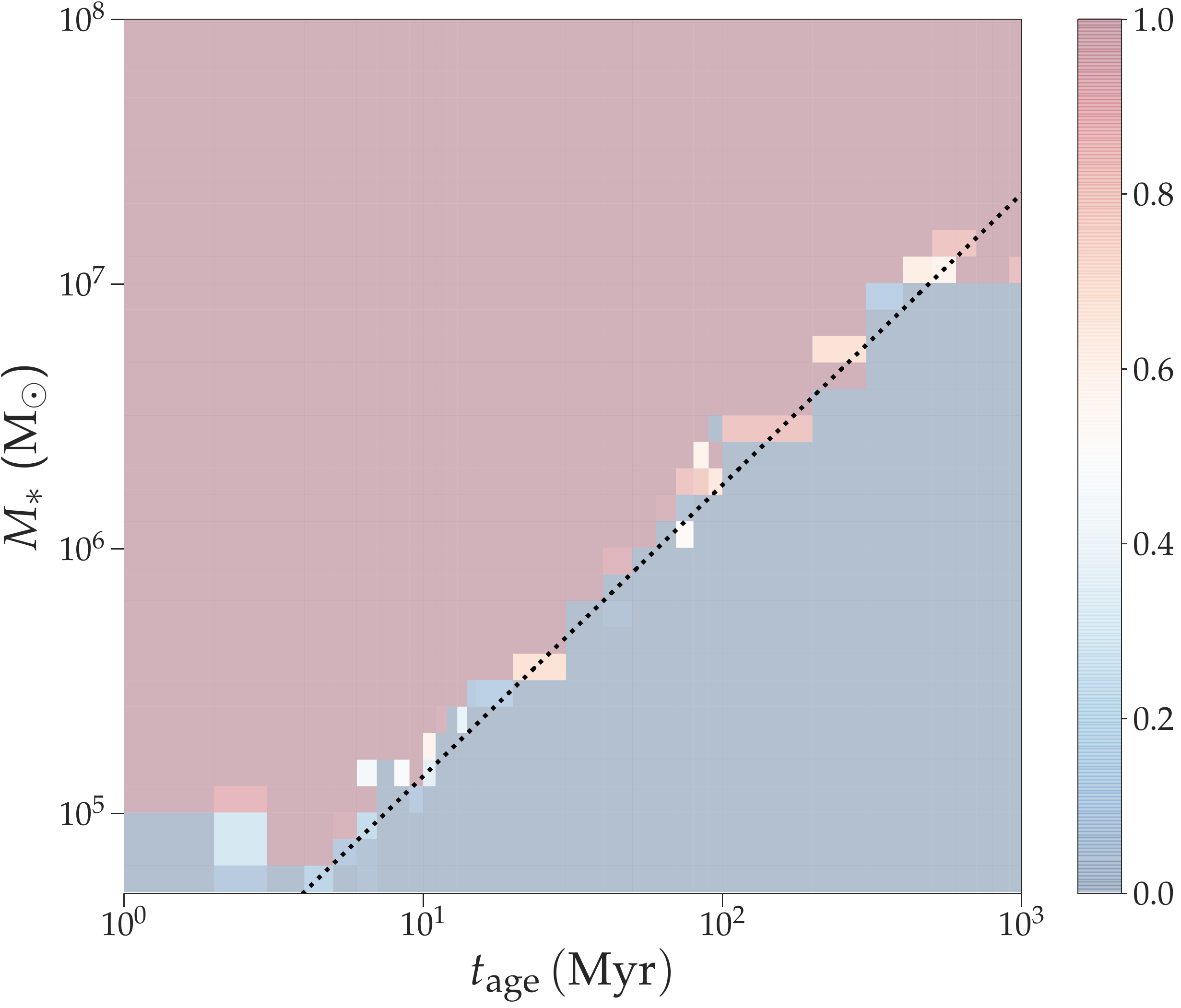}{0.48\textwidth}{}
\fig{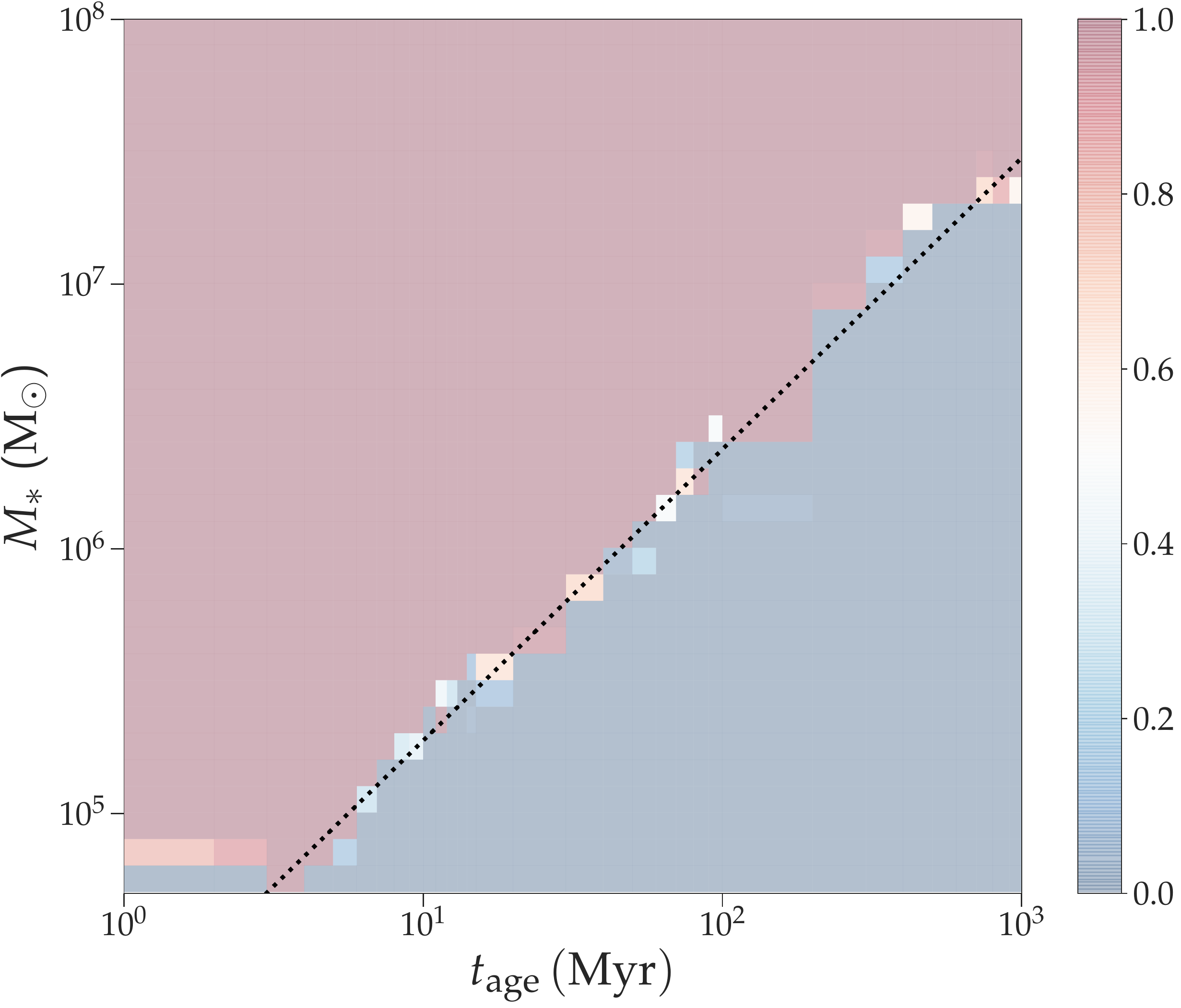}{0.48\textwidth}{}}
\caption{Probability of detecting $M_*$ as a function of $t_{\rm age}$, with blue colors indicating a probability less than 50\% and red colors indicating a probability above 50\%, for $Z=0.4\,\rm Z_{\odot}$ (left panel) and  $Z=\rm Z_{\odot}$ (right panel).  The dividing line between a probability of less than 100\% and a probability of 100\% is quite sharp, as indicated approximately by the black  diagonal dotted lines.  The same diagonal lines are plotted in Fig.\,\ref{fig:selection}.}
\label{fig:selection-bias}
\end{figure*}

To understand the consequences of selection effect I on our results, we generated from our model SEDs for either $Z = 0.4 \, \rm Z_\sun$ or $Z = \rm Z_\sun$, and for $M_*$ spanning the range $10^4-10^8 \, \rm M_\sun$ at uniform logarithmic steps (every one-tenth of a decade), a corresponding suite of mock model SEDs perturbed in each filter by the maximal noise measured for the image in that filter to produce 30 independent realizations of each model SED.  We then fit, using our MCMC approach, these mock noise-perturbed model SEDs by the original noiseless model SEDs having the same corresponding metallicities.  In Figure\,\ref{fig:selection-bias}, we show the probability of the noise-perturbed suite of model SEDs at a given $t_{\rm age}$ and $M_*$ passing our selection criterion of ${\delta}_{\rm age} < 190 \, \rm Myr$, shaded from blue to red in order of increasing probability such that white corresponds to a probability of $\sim$50\%.  The methodology of deriving this probability is explained in full in Appendix\,\ref{app:selection-prob}.  As can be seen, the boundary between a high (red shaded region) and low (blue shaded region) likelihood of detection, indicated by a black dotted line in each panel, is sharp, increasing approximately linearly in $M_*$ with $t_{\rm age}$ in a logarithmic plot.  Notably: (i) at $t_{\rm age} \lesssim 5$\,Myr, SSPs having $M_* \gtrsim 1 \times 10^5 \, \rm M_\sun$ are all detectable; (ii) by $t_{\rm age} = 10$\,Myr, those having only slightly higher $M_* \gtrsim 3 \times 10^5 \, \rm M_\sun$ remain detectable; (iii) at $t_{\rm age} = 50$\,Myr, only SSPs having $M_* \gtrsim 10^6 \, \rm M_\sun$ are detectable; and (iv) by $t_{\rm age} = 100$\,Myr, only those having $M_* \gtrsim 4 \times 10^6 \, \rm M_\sun$ remain detectable.  Because we perturbed the mock model SEDs at each wavelength by the highest noise measured in the image at that corresponding filter, the actual detection threshold in $M_*$ can be somewhat lower at a given $t_{\rm age}$.

As shown earlier in Figure\,\ref{fig:selection} (second column), the nominal $M_*$ inferred along the vast majority of sightlines to the young stellar population lie comfortably above the detection threshold as indicated by the black diagonal dotted lines in this figure (corresponding to the same black diagonal dotted lines in Fig.\,\ref{fig:selection-bias}).  Furthermore, the lower bound in the nominal $M_*$  increases with increasing nominal $t_{\rm age}$, as would be expected given the increasing detection threshold in $M_*$ with $t_{\rm age}$.   A few slightlines to the young stellar population, however, have nominal $M_*$ mostly lying far below the detection threshold, corresponding to the blue data points plotted as open circles in Figure\,\ref{fig:selection} (first column).  These sightlines were excluded from further consideration owing to their poorly constrained fits -- all have ${\delta}_{\rm age}$ close to our selection threshold of 190\,Myr -- as a result of the poor S/N of their measured SEDs.  

The consequences of selection effect II are more difficult to quantify.  As explained above, the light from SSPs spanning a range of ages along a given sightline can boost the $M_*$ computed for the singular SSP deemed to have the best-fit age for that sightline.  This effect can explain why the inferred $M_*$ along all sightlines to the young stellar population lie comfortably above the detection threshold as indicated by the black diagonal dotted lines in Figure\,\ref{fig:selection-bias}.

\subsection{Bias in Star-Formation History}\label{sec: bias SFH}
The inevitable rise in the detection threshold for $M_*$ with increasing $t_{\rm age}$ (selection effect I) biases the measured $SFR(t)$ in two distinct ways: (i) generating an apparent decrease in the measured $SFR(t)$ towards the more distant past even if the intrinsic $SFR(t)$ is constant over time if not higher in the past; and also (ii) producing an apparent cutoff in the measured $SFR(t)$ at a distinct epoch as $M_*$ falls below the detection threshold.  As we shall show, both these effects can be seen in the measured $SFR(t)$ of the BCG in MACS\,J0329.7$-$0211.  

To study the effect of an increasing detection threshold for $M_*$ with increasing $t_{\rm age}$ (selection effect I), we assume that the bulk of stars are formed in star clusters having a mass function of $dN/dM \sim M^{-2.1}$.  This functional dependence is the apparent universal mass function of star clusters (see \citealt{Lim2020}), whether it be open star clusters in our Galaxy, newborn star clusters in interacting or merging galaxies, relatively young and massive star clusters in the BCG of the Perseus cluster, or globular clusters having masses $\gtrsim 10^5 {\rm \, M_\sun}$ (above the peak in their mass function).  Adopting this mass function for different model $SFR(t)$, we then computed how such model $SFR(t)$ would manifest themselves in our measurements following the methodology laid out in Appendix\,\ref{app:bias-sfr}.  The results are shown in Figure\,\ref{fig:no-prob-SFH} for two different underlying model $SFR(t)$: (i) a starburst that began $\sim$400\,Myr ago followed by an exponential decay with an e-folding timescale of $\sim$$10^8 {\rm \, yr}$ (dashed curves), the typical decay timescale of starbursts inferred for local starburst galaxies \citep{Bergvall2016}; and (ii) $SFR(t) = \rm constant$ at either $2 {\rm \, M_\sun \, yr^{-1}}$ (solid curves) or $5 {\rm \, M_\sun \, yr^{-1}}$ (dotted curves).   As can be seen, the model starburst provides a poor description of the measured $SFR(t)$.  Instead, apart for a relatively brief period of elevated $SFR(t) \approx 10 \rm \, M_\sun \, yr^{-1}$ around $10^7$\,yr ago, the measurements are better represented by an approximately constant $SFR(t) = 2 \rm \, M_\sun \, yr^{-1}$ (for either $0.4\,\rm Z_\sun$ or $\rm Z_\sun$) over, at least, the past $\sim$400\,Myr, beyond which any star formation drops below the detection threshold.  We note here that the brief elevation in the $SFR$ about $10^7$\,yr ago makes little difference to the time-averaged $SFR$ over the past 100\,Myr or longer.

\begin{figure}
    \fig{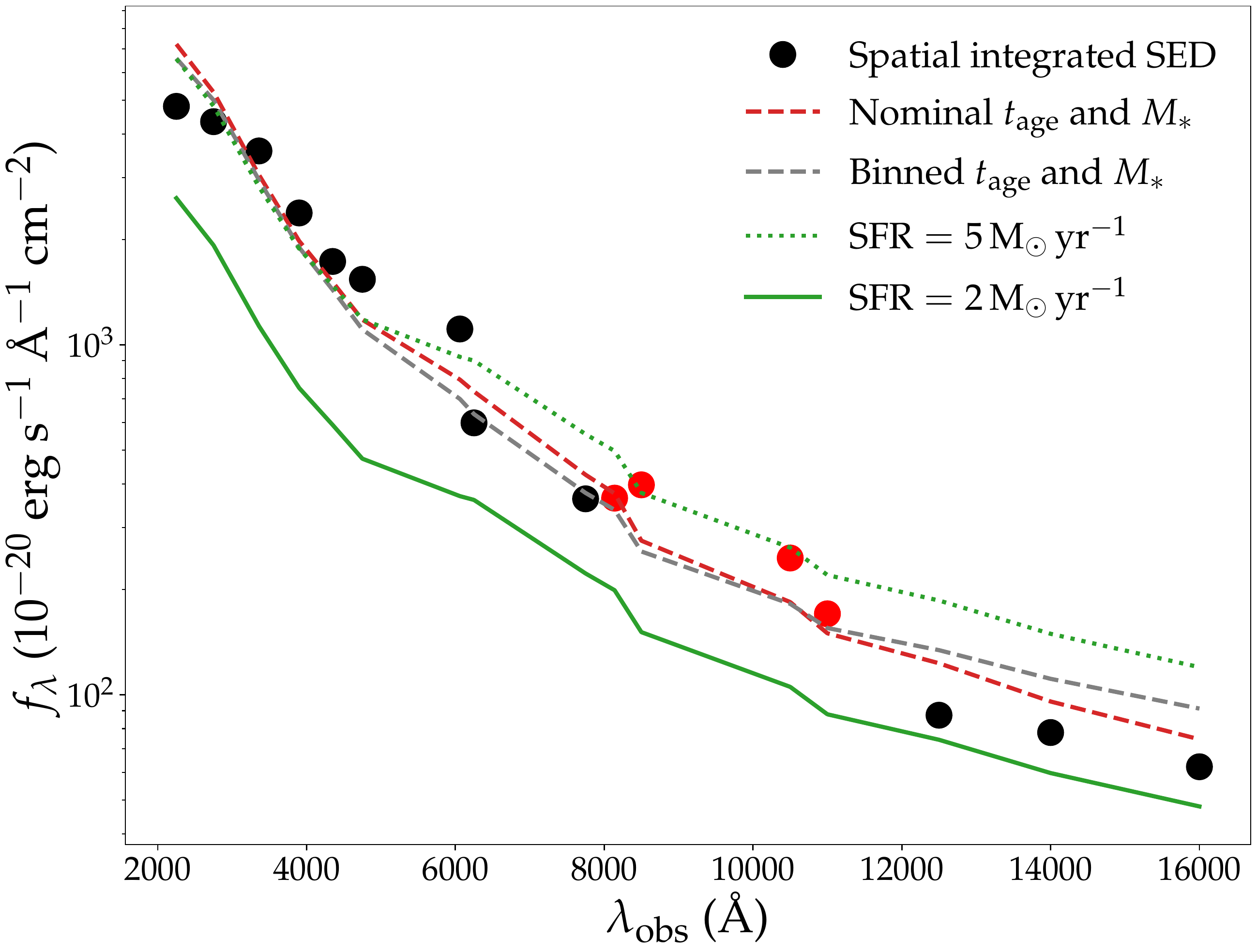}{0.95\linewidth}{}
    \caption{Measured SED integrated over all the selected sightlines towards the young stellar population, where as before the data points colored red correspond to filters spanning the H$\alpha$+[NII] lines.  Overlaid on the measured SEDs are: (i) red dashed curve corresponding to the spatial integration of model SEDs having $Z = 0.4 \, \rm Z_\sun$ and the nominal $t_{\rm age}$ and $M_*$ inferred from our MCMC method along individual sightlines; (ii) gray dashed curve corresponding to the binned star-formation rate over time as in plotted in Figure\,\ref{fig:no-prob-SFH} (right panel) for $Z = 0.4 \, \rm Z_\sun$; (iii) green solid curve corresponding to $SFR(t) =  2 \rm \, M_\sun \, yr^{-1}$ and green dotted curve corresponding to $SFR(t) =  5 \rm \, M_\sun \, yr^{-1}$, both corrected for the increasing threshold in $M_*$ with $t_{\rm age}$.}
    \label{fig:total_SED}
\end{figure}

As a sanity check, Figure\,\ref{fig:total_SED} shows the SED integrated over all the selected sightlines towards the young stellar population as indicated by the black circles.  The red dashed curve corresponds to the spatial integration of model SEDs having $Z = 0.4 \, \rm Z_\sun$ and the nominal $t_{\rm age}$ and $M_*$ inferred along individual sightlines as used for plotting Figure\,\ref{fig:no-prob-SFH} (left panel), and the gray dashed curve to the spatial integration of model SEDs having also $Z = 0.4 \, \rm Z_\sun$ but now for values of $t_{\rm age}$ and $M_*$ that take into account the full posterior distribution of these parameters as plotted in Figure\,\ref{fig:no-prob-SFH} (right panel).  Both these curves are in reasonable agreement with the spatially-integrated SED, providing confidence in our methodology of fitting model SEDs to the measured SEDs along individual sightlines.  The solid green curve correspond to $SFR(t) = 2 \rm \, M_\sun \, yr^{-1}$ and the dotted green curve to $SFR(t) =5 \rm \, M_\sun \, yr^{-1}$, both after correcting for selection effect I using Eq.\,\ref{eq: detected-mass}.  As can be seen, a constant $SFR(t) = 5 \rm \, M_\sun \, yr^{-1}$ provides a good representation of the spatially-integrated SED at the seven shortest wavelength filters, whereas a constant $SFR(t) = 2 \rm \, M_\sun \, yr^{-1}$ presents a better representation of the spatially-integrated SED at the three longest wavelength filters.  Light in the shorter-wavelength filters is dominated by SSPs having relatively young ages, and a constant $SFR(t) = 5 \rm \, M_\sun \, yr^{-1}$ is close to the mean $SFR(t)$ over the past $\sim$$10^7 {\rm \, yr}$.  Light in the longer wavelength filters is less subject to an age bias (selection effect II), and suggests a constant $SFR(t) \approx  2 \rm \, M_\sun \, yr^{-1}$.  The spatially-integrated SED is therefore compatible with the measured $SFR(t)$ as shown in Figure\,\ref{fig:no-prob-SFH}.

\subsection{Comparison with previously inferred Star-Formation Rates}\label{sec:star-formation rates}
\citet{Donahue2015},  \citet{Fogarty2015}, and  \citet{Fogarty2017} have previously inferred star formation rates for the BCGs in the CLASH program.  Using the \citet{Kennicutt1998} relationship between UV luminosity and $SFR$, a relationship that implicitly assumes that $SFR(t) = \rm constant$ over the past $\sim$$10^8 \rm \, yr$ (the overall lifespan of stars that emit significantly in the UV), \citet{Donahue2015} derive $SFR = (25.1 \pm 2.4) \, \rm M_\sun \, yr^{-1}$ for the BCG in MACS\,J0329.7$-$0211, the third highest amongst all the BCGs in the CLASH program.  To estimate the UV luminosity of the young stellar population in the F390W filter, \citet{Donahue2015} corrected for an estimated contribution from evolved stars by adopting a mean color of $\rm (UV-IR) = 5.5$ for these stars, and also corrected for Galactic dust extinction (but not for any dust extinction intrinsic to the BCG).  As we have shown, the $SFR$ for this BCG has not been constant over the past $10^8 \rm \, yr$, such that the $SFR$ inferred by \citet{Donahue2015} is a factor of $\sim$2.5 higher than that even during a brief period of elevated star formation $\sim$$10^7$\,yr ago and over an order of magnitude higher than the time-averaged $SFR$ over the past $10^8 \rm \, yr$.

Repeating the procedure described by \citet{Donahue2015} to determine the UV luminosity of the young stellar population in the F390W filter, we find a luminosity that is nearly exactly a factor of $1 + z = 1.45$ lower than that inferred by \citet{Donahue2015} for the BCG in MACS\,J0329.7$-$0211.  Even then, this UV luminosity remains a factor of $\sim$2.5 higher than that measured from the subtracted image in the F390W filter shown in Figure\,\ref{mosaic-residuals}, indicating that the contribution from the old stellar population is not completely removed using the method adopted by \citet{Donahue2015}.  Using the UV luminosity we cleanly measure for the young stellar population from the subtracted image in the F390W filter, we derive $SFR \approx 7 \rm \, M_\sun \, yr^{-1}$ based on the \citet{Kennicutt1998} relationship.  This $SFR$ is in much better agreement with, albeit still a factor of $\sim$3.5 higher than, the time-averaged $SFR$ over the past $\sim$$10^8 {\rm \, yr}$ that we infer in our work.

Using also the \citet{Kennicutt1998} relationship between UV luminosity and $SFR$, \citet{Fogarty2015} derived $SFR = 42 \pm 2 \, \rm M_\sun \, yr^{-1}$ for the BCG in MACS\,J0329.7$-$0211.  They derived the UV luminosity of the young stellar population in this BCG from images in the four shortest wavelength filters employed in the CLASH program.  Using the same method as \citet{Donahue2015}, they corrected for a contribution to the UV emission from the old stellar population, and corrected for Galactic dust extinction.  Unlike \citet{Donahue2015}, however, \citet{Fogarty2015} inferred and corrected for dust extinction by assuming an intrinsically flat SED -- if flux density is expressed in $f_\nu$ rather than $f_\lambda$ -- over the rest wavelength range 1500--2800\,\AA\ along individual sightlines to the young stellar population, as would be the case if $SFR(t) = \rm constant$ along all such sightlines.  In this way, \citet{Fogarty2015} derived the color excess based on the UV slope of the measured SEDs along individual sightlines, and then used the dust law of \citet{Calzetti2000} to infer spatially-variable dust extinction spanning the range $A_V \sim 0.3$--1.8\,mag.  As we have shown, however, the assumption that $SFR(t) = \rm constant$ along individual sightlines is not justified, nor do we find significant dust extinction along any sightline to the young stellar population (except over a small area, omitted from our work, exhibiting silhouette dust).

As mentioned in Section\,\ref{Nebula}, by subtracting only the estimated stellar continuum of an old stellar population but not that of the young stellar population, \citet{Fogarty2015} derived H$\alpha$+[NII] images for a number of the BCGs in the CLASH program.  Using the \citet{Kennicutt1998} relationship between H$\alpha$ luminosity and $SFR$, a nearly instantaneous measure of the $SFR$ owing to the short lifetimes of massive stars capable of ionizing the surrounding leftover gas from star formation to produce HII regions, \citet{Fogarty2015} derive $SFR = 80 \pm 21 \, \rm M_\sun \, \rm yr^{-1}$ for the BCG in MACS\,J0329.7$-$0211 after correcting for dust extinction inferred in the manner described above.  As we emphasized in Section\,\ref{Nebula}, however, much of the H$\alpha$+[NII] emission associated with this BCG is not related to HII regions, resulting in a overestimate of the $SFR$ (which, in the case of \citet{Fogarty2015}, is further exacerbated by the large extinctions inferred) if the line emission is attributed entirely to HII regions.

Assuming an exponentially decaying star-formation rate to approximate a recent starburst, along also with an exponentially decaying star-formation rate in the distant past for forming an old stellar population, \citet{Fogarty2017} fit model SEDs to the spatially-integrated SED for the BCG in MACS\,J0329.7$-$0211.  They infer $SFR(0) = 39.8^{+20.5}_{-14.8} \, \rm M_\sun \, yr^{-1}$ and therefore a progressively higher $SFR(t)$ into the past, and a burst duration -- defined as the timescale required to form the inferred mass of stars in the burst -- of $1.0^{+1.4}_{-0.6}$\,Gyr.  The $SFR(0)$ they derive is corrected for an inferred dust extinction of $A_V = 0.56^{+0.18}_{-0.19}$.  As shown in Figure\,\ref{fig:no-prob-SFH}, however, there is no evidence to support the argument for an exponential decay in the star-formation rate of this BCG in the recent past.

\subsection{Implications for BCG Growth}\label{BCG Growth}
As explained in Section \ref{sec:intro}, in-situ star formation can pose a double-edged sword for the stellar growth of BCGs.  While increasing their stellar content, star formation that is concentrated in the central regions of BCGs -- as might be expected if cool gas sinks dissipatively into their centers before fuelling vigorous star formation -- causes the galaxy to shrink in response to a central deepening in its gravitational potential well.  By contrast with this picture, however, star formation in the vast majority of BCGs has been found to be spatially extended (see Section \ref{sec:intro}): in the case of the BCG in MACS\,J0329.7$-$0211, the star formation is spatially extended over $\sim$30\,kpc.  For the BCG in the Perseus cluster, newly-formed star clusters are found preferentially towards the outskirts of this BCG \citep{Lim2020} despite the molecular gas as traced in CO being concentrated at its inner regions.  Obviously, star formation at the outskirts rather than the inner regions of a BCG has a different impact on its growth in stellar size.  The manner by which the newly-formed stars contribute to the growth in stellar size of BCGs also depends on whether they form in star clusters, as is the case in nearby star-forming galaxies (whether spiral galaxies or in interacting or merging galaxies) as well as in the BCG of the Perseus cluster.  In such a case, the less massive (and also more numerous) star clusters may be disrupted by strong tidal forces as they free fall into the inner regions of their host galaxies (see discussion in \citealt{Lim2020}), thus disgorging stars along their orbits to promote the growth in stellar sizes of their host galaxies.

The star-formation rates of BCGs as reported in the literature can reach values $\gtrsim 100 \rm \, M_\sun \, yr^{-1}$.  Although the veracity of any reported star-formation rate needs to be scrutinized when derived, especially, from line-emitting gas (which, in BCGs, may not primarily constitute HII regions), star-formation rates derived from UV emission -- which, using the conversion of \citet{Kennicutt1998}, implicitly assumes a constant star-formation rate over the past $\sim$$10^8$\,yr -- may be on somewhat safer grounds, provided that light from the old stellar population is properly subtracted and any dust extinction inferred with utmost care.  High star-formation rates in galaxies have hitherto been regarded as being unsustainable, as the existing gas reservoir is rapidly consumed and, in part, dispersed.  In the case of BCGs, however, the continual replenishment of the gas reservoir from a residual cooling flow can sustain star formation over an indefinite period, as has been found for the BCG in the Perseus cluster \citep{Lim2020} and now demonstrated also for the BCG in MACS\,J0329.7$-$0211.  In apparent agreement with this picture, \citet{Dunne2021} find that the upper mass bound of molecular gas in BCGs as traced in CO remains constant at $\sim$$10^{11} \, \rm M_\sun$ up to $z \sim 1.2$; they argue that star-forming BCGs ``process any accreted molecular gas into stars through means that are agnostic to both their redshift and their cluster mass."  The role of persistent star formation from a residual cooling flow in contributing to the growth in stellar mass {\it and} size of some BCGs -- thus helping explain their broad range of physical properties, as there are multiple pathways that may operate together for their growth -- therefore deserves due consideration.

That said, does sustained star formation fuelled by a residual cooling flow contribute {\it significantly} to the stellar growth of BCGs?  The BCG in the Perseus cluster has formed numerous star clusters (having masses of $\sim$$10^4-10^6 \rm \, M_\sun$) at a relatively steady mass-formation rate of $\sim0.1 {\rm \, M_\sun \, yr^{-1}}$ over the past 1\,Gyr.  Even if sustained since $z \sim 1$--2 (over the past $\sim$8--10\,Gyr), star formation at this rate would have added just $\sim$$10^9 \, \rm M_\sun$ in stars to this galaxy (of course, we cannot rule out that $SFR(t)$ was higher in the more distant past).  Not all these stars may add to the stellar mass or size of the BCG, as a fraction of the stellar mass formed is bound in massive star clusters that may long survive disruption by the tidal field of their host galaxy \citep[see][]{Lim2020}.  By comparison with the BCG in the Perseus cluster, the BCG in MACS\,J0329.7$-$0211 has a persistent star-formation rate of $\sim$$2 {\rm \, M_\sun \, yr^{-1}}$ (albeit with brief excursions to much higher values) that is about 20 times higher.  Over the past $\sim$400\,Myr alone, this BCG has formed $\sim$$8 \times 10^8 {\rm \, M_\sun}$ of new stars, accounting for $\sim$1\% of the original stellar mass of the most massive red nuggets ($\sim$$10^{11} {\rm \, M_\sun}$); if sustained over a $\sim$3\,Gyr (or $\sim$5.6\,Gyr) interval between $z = 0.45$ and $z \sim 1$ (or $z \sim 2$), star formation at this rate could have contributed $\sim$$6 \times 10^{9} \, \rm M_\sun$ (or $\sim$$1.1 \times 10^{10} \, \rm M_\sun$) in stellar mass, and therefore potentially added $\sim$10\% to the original stellar mass of the most massive red nuggets. 

A comparison of the mass of stars potentially formed in a residual cooling flow against the mass of relatively old stars in BCGs provides a gauge of the relative importance of different pathways for the stellar growth of their progenitor red nuggets.  By fitting a model SED having an age of 5.3\,Gyr (i.e., formation at $z \sim 2$) and $Z = \rm Z_\sun$ to the measured SED of the old stellar population in the BCG of MACS\,J0329.7$-$0211, we derive a mass of $\sim$$8 \times 10^{11} \rm \, M_\sun$ for this population.  For comparison, \citet{Burke2015} estimate a stellar mass of $\sim$$4 \times 10^{11} \rm \, M_\sun$ for the same BCG using also images from the CLASH program, but extracting and modelling the measured SEDs in a different way.  We determine an effective radius for the BCG in MACS\,J0329.7$-$0211 based on its old stellar population of $\sim$20\,kpc, at least 20 times the effective radii of red nuggets.  Assuming an order of magnitude growth in stellar mass from $\sim$$8 \times 10^{10} \rm \, M_\sun$ for its progenitor red nugget to that of $\sim$$8 \times 10^{11} \rm \, M_\sun$ at $z = 0.45$, a persistent $SFR(t) \simeq 2 {\rm \, M_\sun \, yr^{-1}}$ from a residual cooling flow since $z \lesssim 2$ could have contributed at most $\sim$1.5\% to the stellar growth of this BCG compared to other pathways such as dry mergers.  
Unless $SFR(t)$ from a residual cooling flow was highly elevated in the past, then this pathway could have played only a minor role in the stellar growth of the BCG in MACS\,J0329.7$-$0211. 

We contrast our results with those of \citet{Fogarty2015} and \citet{Fogarty2017}, from which one might reach a diametrically opposite conclusion.  As mentioned in Section \ref{sec:star-formation rates}, \citet{Fogarty2017} adopted an exponentially decaying starburst to find a current $SFR = 39.8^{+20.5}_{-14.8} \, \rm M_\sun \, yr^{-1}$ (and therefore a higher $SFR(t)$ into the past).  \citet{Fogarty2017} do not report the stellar mass formed in the presumed starburst; nonetheless, based on the parameters provided, the stellar mass formed since the starburst began $\sim$1\,Gyr is well in excess of $4 \times 10^{10} {\rm \, M_\sun}$ (i.e., given a nominal $SFR(0) = 40 \rm \ M_\sun \, yr^{-1}$ and a $SFR(t)$ that increases into the past until the beginning of the starburst).  The stellar mass inferred to have formed over the past 1\,Gyr alone is well over half the original stellar mass of even the most massive red nuggets.  The conclusions one would reach from the results of \citet{Fogarty2015} and \citet{Fogarty2017} for the BCG in RX\,J1532.9+3021 ($z=0.363$), having a factor of just over 2 higher UV luminosity than the BCG in MACS\,J0329.7$-$0211, provide an even greater contrast.  For this galaxy, \citet{Fogarty2015} and \citet{Fogarty2017} fit model SEDs to the spatially-integrated as well as the spatially-resolved SEDs (i.e., over different sightlines) adopting a $SFR(t)$ that has decayed exponentially over time.  In both methods, they derived a stellar mass of $\sim$$1 \times 10^{11} \rm \, M_\sun$ formed by this galaxy during the present starburst of duration $\sim$0.7\,Gyr thus far.  The inferred mass in new stars is comparable with the stellar mass of the most massive red nuggets known at $z \gtrsim 2$.  An even more instructive comparison is provided by contrasting the mass of new stars formed with the mass of old stars in this galaxy of $(5.33 \pm 0.61) \times 10^{11} \rm \, M_\sun$ \citep{Burke2015}.  The present starburst alone, which has lasted $\sim$0.7\,Gyr to date, already constitutes $\sim$20\% by mass of the entire old stellar population in this galaxy!

In brief, at least for the BCGs in the Perseus cluster and MACS\,J0329.7$-$0211 that differ by over an order of magnitude in their star-formation rates, persistent star formation from a residual cooling flow has only contributed in a relatively minor if not negligible manner to their overall stellar growth since $z \sim 1$--2 -- unless their persistent star-formation rates at their respective epochs are much lower than those in their past and, for the BCG in MACS\,J0329.7$-$0211, possibly also future.  Rather than contributing in a major way to the stellar growth of most if not all BCGs, star formation from a residual cooling flow\footnote{Our work does not address stellar growth from star formation owing to wet mergers, which \citet{McDonald2016} argue is the major cause of star formation observed in BCGs at $z \gtrsim$0.6.} may play a more significant role in contributing to the enormous numbers of globular clusters around BCGs.  For example, the BCG in the Perseus cluster has formed, on average, one globular cluster with a mass of $\sim10^5 \rm \, M_\sun$ (at the peak of the mass distribution in globular clusters) every 1\,Myr for the past, at least, 1\,Gyr \citep{Lim2020}.  Extrapolated to the BCG in MACS\,J0329.7$-$0211, it has already formed about ten times as many progenitor globular clusters over a timescale that is a factor of $\sim$2 shorter.  A sustained formation over time does not fit easily into current narratives for the formation of globular clusters (e.g., see a brief review in \citealt{Harris2000}), but deserves due consideration in the case of BCGs.


\section{Summary and Conclusions}\label{sec:conclusions}
We have determined, to the highest degree of accuracy that we believe is possible using data from the CLASH program, the formation history of the young stellar population in the BCG of MACS\,J0329.7$-$0211 at $z=0.45$, observed at a time when the Universe was two-thirds of its present age.  By fitting for and then subtracting the 2-dimensional light distribution of the old stellar population in this BCG, we are able to isolate just the spectral energy distribution (SED) of its young stellar population along individual spatially-resolved sightlines (Section\,\ref{sec:method}).  The subtracted images, which were convolved to a common angular resolution corresponding to the point spread function of the lowest angular resolution image at F110W, reveal that the young stellar population is detectable from the near-UV to the near-IR, as well as the presence of line (H$\alpha$+[NII]) emitting gas not spatially coincident with the young stellar population and a relatively compact dust feature observed in silhouette near the center of the BCG (Section\,\ref{SED young stellar pop}).

We fit model single stellar populations (SSPs; i.e., stars all sharing a common age and metallicity in a given population, and with a Kroupa initial mass function) to the measured SEDs along individual sightlines, each having a cross-sectional area of $1.5 \times 1.5$\,kpc that is comparable to the full-width half-maximum of the common point spread function of the images (Section\,\ref{sec:analysis}).  We find, apart from the region where silhouette dust is visible and hence our subtraction of the old stellar population is badly compromised, no significant dust extinction towards the young stellar population.  A simple comparison between the model and measured SEDs reveal that the young stellar population span ages from a few Myr to a few hundreds of Myr.  To quantify the star-formation history (i.e., star-formation rate as a function of time) of the young stellar population, we employed a Markov Chain Monte Carlo (MCMC) method to fit model SEDs (generated for SSPs) to the measured SEDs along individual sightlines, yielding the full probability distribution in both age and birth mass along each sightline for a given selected metallicity, $Z$, of either $Z = 0.4 \rm Z_\sun$ (approximately that of the intracluster gas) or $Z = \rm Z_\sun$.

We find that H$\alpha$+[NII] line emission not associated with HII regions is detectable along many sightlines to the young stellar population as shown in Figure\,\ref{fig:Ha}, even (and especially) at ages $> 5$\,Myr when any line emission from HII regions are predicted to no longer be detectable over the bandpasses of the filters employed  (Section\,\ref{sec:results}).  This line-emitting gas accounts for at least $\sim$60\% (for $Z = \rm Z_\sun$) to at least $\sim$80\% (for $Z = 0.4 \rm Z_\sun$) of the H$\alpha$+[NII] emission from the BCG.  Attributing all this line emission to that from HII regions would result in a vast over-estimate of the instantaneous star-formation rate.  

The result for the star-formation history of the young stellar population is shown in Figure\,\ref{fig:no-prob-SFH} (Section\,\ref{sec:results}), which presents the central finding of our work.  To properly interpret this result, we need to take into account observational selection effects: in doing so, we find that the formation rate of the young stellar population has, apart from a brief elevation to $\sim$$10 \rm \, M_\sun \, \rm yr^{-1}$ about $10^7$\,yr ago, been approximately constant at $\sim$$2 \rm \, M_\sun \, \rm yr^{-1}$ over the past $\sim$400\,Myr, beyond which any star formation drops below the observational detection threshold (Section\,\ref{sec:discussion}).  Star formation that is extended not only over space (spanning a projected linear dimension of $\sim$30\,kpc in this BCG) but also through time provides further support for the argument that young stellar populations in (some) BCGs are produced by a residual cooling flow that sustains a significant rate of star formation over an indefinite period.  Such persistent star formation from a residual cooling flow can contribute up to $\sim$10\% of the original stellar mass of the BCG in MACS\,J0329.7$-$0211 if its progenitor was among the most massive red nuggets known at $z \sim 2$ having masses $\sim$$1 \times 10^{11} \rm \, M_\sun$, but only a few percent of its overall growth in stellar mass to $\sim$$8 \times 10^{11} \rm \, M_\sun$ at $z=0.45$.  Instead, the prodigious number of star clusters formed from a residual cooling flow may play a more important role in contributing to the enormous numbers of globular clusters around BCGs.

\onecolumngrid

\acknowledgments{
J. Lim acknowledges support from the Research Grant Council of Hong Kong through the grant 17304817, which also supported the MPhil studentship of J. Li. Y.O. acknowledges the support by the Ministry of Science and Technology (MOST) of Taiwan through grants, MOST 109-2112-M-001-021-.}
\facilities{HST (ACS and WFC3)}
\software{Astropy \citep{astropy2013, astropy2018},
            Stsynphot \citep{stsynphot2020}, 
            Imfit \citep{Erwin2015}
            } 

\appendix

\section{MCMC algorithm}\label{app:mcmc}
We adopt a fully Bayesian calculation for extracting physical parameters from the measured SEDs using a Markov Chain Monte Carlo (MCMC) approach.  The similarity between the measured SED and the model SED having a given parameter set $\mathbf{\Theta} \equiv \lbrace Z, f_{cov}, A_v, \log t_{\rm age}, \log M_* \rbrace$, where $Z$ is the adopted metallicity (either $0.4\,\rm Z_\sun$ or 1.0$\,\rm Z_\sun$), $f_{cov}$ the adopted covering factor (either 0 or 1), $A_v$ the dust extinction (either set to 0 or, for the purpose of investigating any dust extinction, restricted to $\leq 0.5$\,mag), $t_{\rm age}$ the age of the stellar population (up to 1\,Gyr), and $M_*$  the birth mass derived by integrating stars having a Kroupa initial mass function over the stellar mass range 0.1--$100 \rm M_\sun$, is evaluated using the standard $\chi^2$ statistic:
\begin{equation}
\chi^2(\mathbf{\Theta}) \equiv \sum_i \frac{\Bar{f}_i - f_i(\mathbf{\Theta})}{\sigma_{noise, i}^2}  \, \, ,  
\end{equation}
where $i$ runs over all the filters used for constructing the SED, $\Bar{f_i}$ is the flux density of the measured SED in the $i$'th filter, $f_i(\mathbf{\Theta})$ the flux density of the model SED in the $i$'th filter, and $\sigma_{noise, i}$ the measurement uncertainty in flux density at the $i$'th filter. 

The probability that a given model SED can reproduce the measured SED is given by $p(\Bar{f}_i | \mathbf{\Theta}) \propto \exp(-\chi^2/2)$, which describes a Gaussian centered at $f(\mathbf{\Theta})$.  The posterior distribution, $p(\mathbf{\Theta} | \Bar{f})$, describing the likelihood that the different model SEDs considered fit the measured SED, is simply obtained from the Bayes Theorem:
\begin{equation} \label{bayesThm}
p(\mathbf{\Theta} | \Bar{f}) \propto p(\Bar{f}_i | \mathbf{\Theta}) p(\mathbf{\Theta}) \, \, ,
\end{equation}
where $p(\mathbf{\Theta})$ is the adopted prior (i.e., initial guess in $\mathbf{\Theta}$, which can be a functional form).  To reflect our ignorance of which model SED can best fit the measured SED, we adopt an uniform prior. 

Equation\,(\ref{bayesThm}) possess an intuitive interpretation: the peak in the posterior distribution corresponds to the model SED having parameters $\mathbf{\Theta}$ at which $\chi^2$ is minimized, and hence that which best agrees with the measured SED.

Equation\,(\ref{bayesThm}) is often numerically intractable because the normalization for the posterior requires expensive numerical integration.  Instead, we generated samples from the posterior distribution $p(\Theta | \Bar{f})$ using a MCMC approach.  The posterior distribution can then be approximated by a kernel density estimation of the samples.  In this work, we use the PyMultiNest algorithm developed by \citet{Buchner2014}, which is a modification of the standard MCMC implementation so as to provide a faster convergence on multi-modal and asymmetric posteriors.  Like for the age parameter $t_{\rm age}$, we used a logarithmic instead of a linear scale for $M_*$ to restrict the number of values explored so as to not far outnumber those explored in $A_V$ (where relevant) and $t_{\rm age}$.  
Otherwise, the algorithm that we wrote for performing the MCMC computation would have been much slower or not converge in practise. 

\section{Computing Star Formation History} \label{app:sfr}
Our goal is to compute the posterior of $SFR(t)$, namely $p(SFR(t)| \lbrace s \rbrace)$, given the posteriors of individual parameters in the set $\Theta_s \equiv \lbrace \log t_{\rm age,s}, \log M_{*,s}\rbrace$ along each sightline, $s$, for a particular adopted metallicity.  To begin, we outline the calculation of the probabilistic average instead of the full probability distribution so as to provide more insight on the full calculation.  In practise, when computing the $SFR$ at a given $t = t_{\rm age}$ having discrete values selected as mentioned in Section \ref{model SSPs}, we choose a time interval, $\Delta t$, equal to the difference between the corresponding adjacent age steps.  Probabilistically, if a particular sightline has the probability $p(t_{\rm age,s}, M_{*,s}|s)$ of having an age $t_{\rm age,s}$ and a birth mass $M_{*,s}$, the expected value in birth mass that this sightline contributes to the age bin $\Delta t^{(i)} \equiv t^{(i+1)}-t^{(i)}$ is:

\begin{equation} \label{eq:expected-stellar-mass}
    \langle M | \Delta t^{(i)} \rangle_s = \int^{t^{(t+1)}}_{t^{(i)}} \int^\infty_0 M_{*,s}\cdot p(t_{\rm age,s}, M_{*,s}|s)) \, dM_{*,s} \, dt_{\rm age,s}.
\end{equation}
Summing the expectation along all sightlines, 
\begin{equation} \label{eq:expected-sfr}
    \langle SFR(t\in \Delta t^{(i)}) \rangle = \frac{1}{\Delta t^{(i)}}\sum_s \langle M | \Delta t^{(i)} \rangle_s = \sum_s \langle \frac{ M}{\Delta t^{(i)}} | \Delta t^{(i)} \rangle_s \,\, ,
\end{equation}
where the bin width $\Delta t^{(i)}$ is a fixed value at a given $t = t_{\rm age}$ as mentioned above.  While the double integral in Eq.\,\ref{eq:expected-stellar-mass} cannot be evaluated analytically, we can estimate the expected value by utilizing the MCMC samples obtained. 

The above formalism on the expectation $\langle SFR(t\in \Delta t^{(i)}) \rangle$ can be extended to calculate the full posterior $p(SFR(t)| \lbrace s \rbrace)$.  With MCMC, we get sample ages, $\Tilde{t}_{\rm age, s}$, and masses, $\Tilde{M}_{*,s}$, from the probability distribution along each sightline, $\Tilde{M}_{*,s}, \Tilde{t}_{\rm age, s} \sim p(t_{\rm age,s}, M_{*,s}|s)$.  As a consequence, Eq.\,\ref{eq:expected-stellar-mass} can be replaced as a Monte Carlo sum, and be promoted to a probability by removing the integration:
\begin{equation}
    p( M_s| \Delta t^{(i)}, s ) = \frac{1}{N_{i,s}} \mathds{1}(M_s = \Tilde{M}_{*,s}) \,\, ,
\end{equation}
where $N_{i,s}$ is the count of the number of samples belonging to the age bin $\Delta t^{(i)}$ for a particular sightline $s$.  The indicator function $\mathds{1}$ simply counts the number of mass samples $\Tilde{M}_{*,s}$ (right-hand side) that matches the argument $M_s$ (left-hand side).  We can promote Eq.\,\ref{eq:expected-sfr} to a full posterior in a similar way.  As $SFR(t\in \Delta^{(i)}) = \sum_s M_{s}/\Delta t^{(i)}$, we therefore have:
\begin{equation}
    p( SFR(t\in \Delta t^{(i)} | \lbrace s \rbrace) = p( \sum_s M_{s}/\Delta t^{(i)} | \lbrace s \rbrace ) = \frac{1}{N_{i,s}} \mathds{1}( SFR=\sum_s \frac{\Tilde{M}_{*,s}}{\Delta t^{(i)}} ) \,\, . 
\end{equation}
Again, the indicator function $\mathds{1}$ is there to match both sides.

In practice, as the indicator function only occupies a point in the entire posterior space, we need to use a kernel density estimation to `smear out' the indicator function according to the density of samples in its neighbourhood.  This procedure is exactly the standard procedure used whenever MCMC samples are converted to probability contours plot.  One can check, upon taking the expectation (i.e., integrating over the probability), that we can get back to Eq.\,\ref{eq:expected-sfr}, as is required.  We have checked that the resulting posteriors obtained from above procedure are, to a good approximation, a log-normal distribution, despite the individual posteriors for each sightlines being much more complex.  Therefore we report in Figure\,\ref{fig:no-prob-SFH} (lower panel) just the mean and standard deviation for $SFR(t)$ calculated using this probabilistic approach.

The age bins over which we plot the SFH as shown in Figure\,\ref{fig:no-prob-SFH} (lower panel) has been designed so as to have approximately equal number of samples in each bin.  Conceptually, we generate many different realizations for $t_{\rm age}$  and $M_*$ over different sightlines, and then selected time intervals over which the number of samples are approximately equal in each time bin.
\\

\section{Selection Effects}\label{app:bias}
\subsection{Completeness Limits}\label{app:selection-prob}
In Section \ref{acceptable fits}, we explained how we selected acceptable -- satisfactory and sensible -- model SEDs fits to the measured SEDs of the young stellar population.  In brief, we used two criteria: (i) a standard deviation in the posterior distribution for age of $\delta_{\rm age} < 190$\,Myr (see Section \ref{acceptable fits} for how $\delta_{\rm age}$ is computed); together with (ii) a nominal birth mass, $M_*$, above a mass threshold that increases with age.  Below, we describe how the adopted threshold of $\delta_{\rm age} < 190$\,Myr sets a completeness limit -- in a probabilistic sense -- for the birth mass at a given age when fitting model SEDs to the measured SEDs using our MCMC approach.  This completeness limit defines the threshold in birth mass as a function of age for acceptable fits.

As mentioned in Section\,\ref{model SSPs} and in Appendix\,\ref{app:mcmc}, for the purpose of the MCMC analyses, we generated model SEDs, $\mathcal F$, at logarithmic intervals in age (except for additional steps of 1\,Myr from 10--15\,Myr ), $t_{\rm age}$, and birth mass, $M_*$, so as to produce an approximately uniform logarithmic grid in these model parameters, $\lbrace \Theta^{(i,j)} \rbrace \equiv \lbrace t_{\rm age}^{(i)}, M_{*}^{(j)} \rbrace$.   The set of model SEDs, $\mathcal F(\Theta^{(i,j)})$, in these parameters alone therefore number $N_{t_{\rm age}} \times N_{M_*} = \sum i \times \sum j$.  To simulate the effect of noise, we perturbed each of these model SEDs by a Gaussian distribution, $\mathcal{G}(0,\sigma_{noise})$, centered at 0 and with a standard derivation of $\sigma_{noise}$; the latter is chosen to be the maximal noise over the image in a given filter (Section \ref{error}), thus corresponding to the upper limit for the detection threshold in that filter.  In this way, we generated a set of $N_k = 30$ noise-perturbed realizations for each model SED, ${\mathcal F}_k(\Theta^{(i,j)})$.  

We treat these noise-perturbed model SEDs as mock data to be fitted by the original noiseless model SEDs using the same MCMC approach as for the measured SEDs (Appendix\,\ref{app:mcmc}), thus yielding the posteriors $p \left( \Theta | {\mathcal F}_k(\Theta^{(i,j)}) \right)$.   At each grid point $(i,j)$, we therefore have $N_k$ posteriors, $p \left(\Theta^{(i,j)} | {\mathcal F}_k(\Theta^{(i,j)})\right)$, and $N_k$ standard deviations for the posterior distribution in ages, $\delta_{t_{\rm age}}^{(i,j,k)} \equiv \left(\left[ p(\tau \in \Theta^{(i,j)} | {\mathcal F}_k(\Theta^{(i,j)}) \right]\right)^{1/2}$.  The probability for a particular parameter combination $\Theta^{(i,j)} = (t_{\rm age}^{(i)}, M_{*}^{(j)})$ to pass the selection criteria $p\left(\delta_{t_{\rm age}}^{(i,j)} < 190 {\rm \, Myr} \middle| \Theta^{ij} = ({t_{\rm age}}^{(i)}, M_{*}^{(j)})\right)$ is given by:
\begin{equation}
p\left({\boldsymbol \sigma}_{t_{\rm age}}^{(i,j)} < 190 {\rm \, Myr} \middle| \Theta^{(i,j)} = {t_{\rm age}^{(i)}}, M{_{*}^{(j)}} \right) \approx \frac{\sum_k \mathds{1}\left({\boldsymbol \sigma}_{t_{\rm age}}^{(i,j,k)} < 190 {\rm \, Myr}  \right)}{N_k} \, \, ,
\end{equation}
whereby $\sum_k \mathds{1} \left({\boldsymbol \sigma}_{t_{\rm age}}^{(i,j,k)}\right)$  is simply the sum of the number of realizations among the noise-perturbed 
model SEDs that individually yield an age posterior with a standard deviation of ${\boldsymbol \sigma}_{t_{\rm age}}^{(i,j,k)} < 190$\,Myr (and so, divided by $N_k$, is simply the fraction of realizations for a particular parameter combination $\Theta^{ij} = ({t_{\rm age}^{(i)}}, M_{*}^{(j)})$ that passes the selection criterion).  
Because the noise-perturbed model SEDs are generated using the maximal noise uncertainty in each filter image, the fraction that passes the selection criterion is a lower limit.  

The results are plotted in Figure\,\ref{fig:selection-bias} for $0.4 \rm \, Z_\sun$ (left panel) and $1.0 \rm \, Z_\sun$ (right panel), where we show the fraction of realizations that satisfy ${\boldsymbol \sigma}_{t_{\rm age}}^{(i,j,k)} < 190$\,Myr as a function of $\log M_*$ versus $\log t_{\rm age}$.  At a given metallicity, the boundary between $p\left(\sigma_{t_{\rm age}}^{(i,j)}\right) = 1$ (red region) and $p\left(\sigma_{t_{\rm age}}^{(i,j)}\right) < 1$ (blue region) increases approximately linearly in $\log M_*$ with $\log t_{\rm age}$ as indicated by a dashed diagonal line, which therefore approximately defines the completeness limit in $M_*$ at a given $t_{\rm age}$.  The same diagonal lines are drawn in Figure\,\ref{fig:selection} (middle column) as one of the two criteria for selecting acceptable model SED fits to the young stellar population in the BCG. 

\subsection{Correcting for Completeness Limits in Star Formation History}
\label{app:bias-sfr}
The increasing lower bound in the birth mass, $M_*$, with age, $t_{\rm age}$, found in our model SED fits to the measured SEDs of the young stellar population as shown in Figure\,\ref{fig:selection} (middle column) is a selection effect as demonstrated in Appendix\,\ref{app:selection-prob}.  This resulting detection threshold is indicated approimately by the dotted diagonal lines in  Figure\,\ref{fig:selection} (middle column), which imposes a bias on the star formation history as computed from $M_*$ and $t_{\rm age}$ for each sightline and integrated over all sightlines.  For example, an intrinsically constant star-formation rate over time ($SFR(t) \propto t^0$) would be inferred from the measurements to be a decreasing star-formation rate into the past; even an increasing star-formation rate into the past could potentially show the opposite behaviour in the inferred star-formation history.   
The completeness limit, $M_{*({\rm min})}(t_{\rm age})$, indicated by the dotted lines in Figure\,\ref{fig:selection-bias} can be approximately described by the functional form:  
  
\begin{equation}
    M_{*({\rm min})}(t_{\rm age}) = M_0 \left( \frac{t_{\rm age}}{\text{Myr}} \right)^\alpha = 
    \begin{cases}
    1.1\times 10^4 \left( \frac{t_{\rm age}}{\text{Myr}}\right)^{1.1} \, \text{M}_\odot,\,  Z=0.4\,\rm Z_{\odot} \\
    1.5\times 10^4 \left( \frac{t_{\rm age}}{\text{Myr}}\right)^{1.1} \, \text{M}_\odot,\,  Z=\rm Z_{\odot}
    \end{cases}\, \,
\end{equation}

to within a scatter of $\lesssim 0.1$ dex irrespective of the adopted metallicities and internal dust extinction of $0 \le A_V \le 0.5$.   The number of sightlines that would be selected, $N_\text{obs}(M_*,t_{\rm age})$, owing to the completeness limit is therefore:
\begin{equation}
N_\text{obs}(M_*,t_{\rm age}) =
\begin{cases}
     N(M_*,t_{\rm age}), & M_* \geq M_{*({\rm min})}(t_{\rm age}) \\
    0, & M_* < M_{*({\rm min})}(t_{\rm age})
\end{cases}\, \, ,
\end{equation}
where $N(M_*,t_{\rm age})$ is the total number of sightlines having $M_*$ and $t_{\rm age}$.

\begin{figure*}[htb]
\gridline{\fig{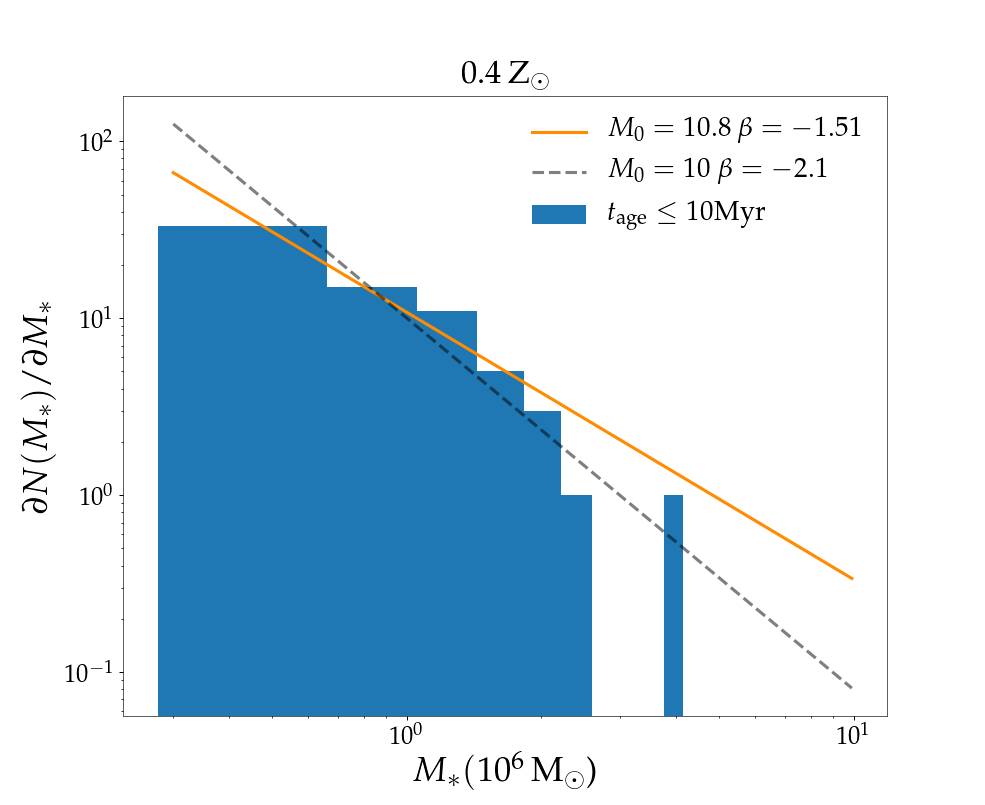}{0.48\linewidth}{}
        \fig{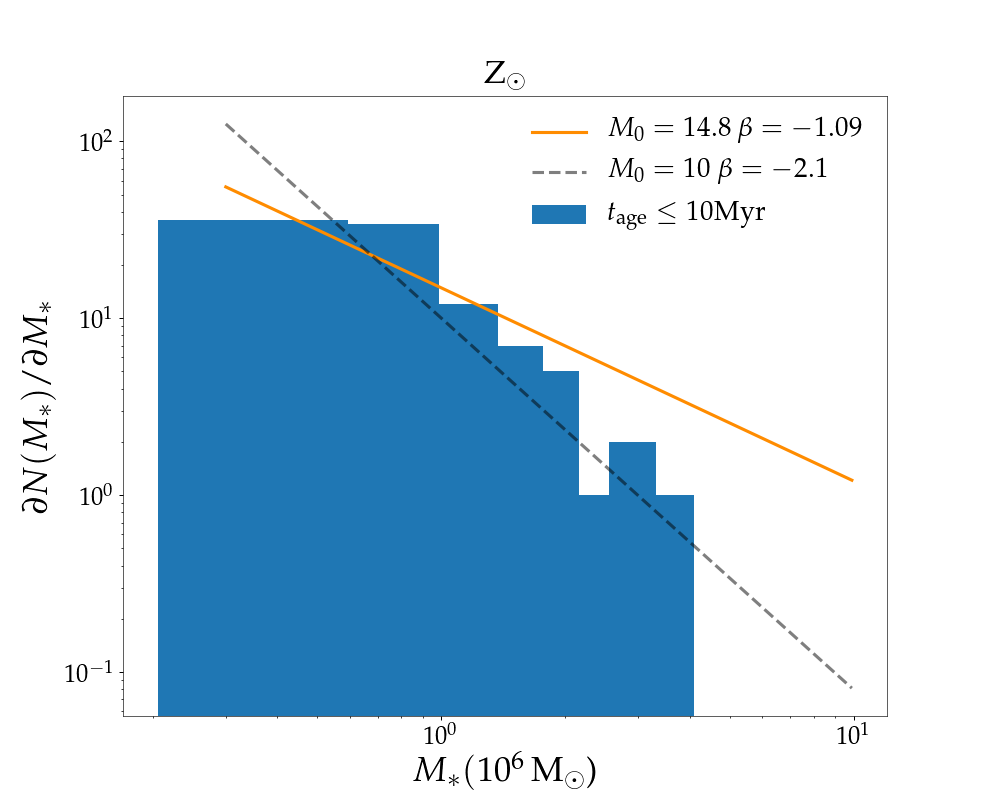}{0.48\linewidth}{}}
\caption{Histograms of inferred masses ($M_*$) with inferred ages ($t_{\rm age} \leq 10\,$Myr), assuming $A_V=0$, $Z=0.4\,\rm Z_{\odot}$ (left) and $Z=\rm Z_{\odot}$ (right). The bestfit power-law functions are plotted as orange lines with the parameters indicated in legends. Power-law with $\beta=-2.1$ is also plotted manually as grey dashed line to make visual comparison.
}
\label{fig:mass-function}
\end{figure*}

At a given $t_{\rm age}$, the detectable total mass, ${\mathbb M}_{\rm obs} (t_{\rm age})$, having $M_* \ge M_{*({\rm min})}(t_{\rm age})$ along individual relevant sightlines compared with the actual total mass, ${\mathbb M}_{\rm total} (t_{\rm age})$, above an intrinsic low-mass cutoff, $M_{*(\rm low)}$, is given by:

\begin{equation} \label{eq: detected-mass}
    \frac{{\mathbb M}_{\rm obs} (t_{\rm age})}{{\mathbb M}_{\rm total} (t_{\rm age})} = \frac{\int^{M_{* \rm (high)}}_{M_{*({\rm min})}(t_{\rm age})} M_* \cdot dN(M_*, t_{\rm age})}{\int^{M_{* \rm (high)}}_{M_{*(\rm low)}} M_* \cdot dN(M_*, t_{\rm age})} \, \, ,
\end{equation}
where $M_{*(\rm high)}$ is the intrinsic high-mass cutoff.  Evaluating Eq.\,\ref{eq: detected-mass} requires knowledge of ${\partial N(M_*)}/{\partial M_*}$, the number of sightlines as a function of $M_*$ at a given $t_{\rm age}$.  Figure\,\ref{fig:mass-function} shows ${\partial N(M_*)}/{\partial M_*}$ versus $M_*$  for all sightlines having $t_{\rm age} \le 10\,{\rm Myr}$ for both $Z = 0.4\,\rm Z_\sun$ (left panel) and $Z =\rm Z_\sun$ (right panel).  The best linear fits to these log-log plots are indicated by the solid lines, and have a power-law index, $\beta$ (whereby ${\partial N(M_*)}/{\partial M_*} \propto M_*^\beta$), of between $-1.2$ and $-1.5$.  As the adopted model SSPs fade by a factor of 4 between ages of 1\,Myr and 10\,Myr, the slopes of the best-fit lines represent upper limits; i.e., the actually slope should be steeper after correcting for the completeness limit as a function of age.   Dashed lines having a power-law index of $-2.1$ are shown also in Figure\,\ref{fig:mass-function}, corresponding to an apparently universal number dependence of star clusters with mass across all mass scales -- whether it be open star clusters in our Galaxy, massive star clusters in interacting or merging galaxies, massive star clusters in the BCG in the Perseus cluster, or globular clusters above the peak in their mass function -- thereby implicating a common underlying mechanism for the formation of star clusters over all mass scales (see \citealt{Lim2020}).  If each sightline to the young stellar population in the BCG of MACS\,J0329.7$-$0211 encompasses multiple star clusters having the same number dependence, then statistically we would expect ${\partial N(M_*)}/{\partial M_*}$ versus $M_*$ to also exhibit a power-law index of $-2.1$, as we shall henceforth adopt.

Adopting also $M_{*(\rm low)} = 10^5 {\rm \, M_\sun}$, approximately the lowest $M_*$ detected, and $M_{*(\rm high)} = 10^7 {\rm \, M_\sun}$, approximately the highest $M_*$ detected, Eq.\,\ref{eq: detected-mass} can then be solved to give:
\begin{equation}
     \frac{{\mathbb M}_{\rm obs} (t_{\rm age})}{{\mathbb M}_{\rm total} (t_{\rm age})} = \frac{M_0^{\beta+2}t_{\rm age}^{\alpha (\beta+2)} - M_{*(\rm high)}^{\beta+2}}{M_{*(\rm low)}^{\beta+2} -M_{*(\rm high)}^{\beta+2} } \, \, .
\label{integral}
\end{equation}
This scaling relation allows a simple correction to be made to the measured $SFR(t)$ so as to derive the actual $SFR(t)$; in practise, owing to the uncertainties associated with the measured $SFR$ at a given $t$, we compare the latter to that we would infer for different star-formation histories given the selection bias introduced by Equation\,\ref{integral}.  The results are shown in  Fig\,\ref{fig:no-prob-SFH} for a constant or, to mimic a short-duration starburst in the local Universe, an exponentially decaying star-formation rate over time.  Although we adopt $\beta = -2.1$ to correct for completeness, motivated by the expectation that the young stellar population in the BCG is composed of an ensemble of star clusters, the correction is only weak sensitive to the exact value of $\beta$ at least between $-1.0$ and $-2.1$.

\bibliography{reference}{}
\bibliographystyle{aasjournal}

\end{document}